\pdfoutput=1

\documentclass{article}

\usepackage{arxiv}

\usepackage[utf8]{inputenc} % allow utf-8 input
\usepackage[T1]{fontenc}    % use 8-bit T1 fonts
\usepackage{hyperref}       % hyperlinks
\usepackage{url}            % simple URL typesetting
\usepackage[normalem]{ulem}
\usepackage{multirow}
\usepackage{booktabs}
\useunder{\uline}{\ul}{}

\usepackage{amsfonts}       % blackboard math symbols
\usepackage{nicefrac}       % compact symbols for 1/2, etc.
\usepackage{microtype}      % microtypography

\usepackage{graphicx}

\usepackage[export]{adjustbox}
\usepackage[caption=false]{subfig}% Support for small, `sub' figures and tables

\usepackage{amssymb}
\usepackage{amsmath}
\DeclareMathAlphabet{\mathpzc}{OT1}{pzc}{m}{it}

\usepackage{mathrsfs} % for \mathscr{T}
\usepackage{afterpage}

\title{Texture Selection for Automatic Music Genre Classification}

\date{May 24, 2019}	% Here you can change the date presented in the paper title
\author{
  Juliano H. Foleiss \thanks{Also as a PhD candidate at the School of Electrical and Comppter Engineering, University of Campinas -- Brazil.}\\
  Department of Computing\\
  Federal University of Technology -- Paraná\\
  Campo Mourão, PR -- Brazil \\
  \texttt{julianofoleiss@utfpr.edu.br} \\
   \And
  Tiago F. Tavares \\
  School of Electrical and Computer Engineering\\
  University of Campinas\\
  Campinas, SP -- Brazil \\
  \texttt{tavares@dca.fee.unicamp.br} \\
}

\begin{document}
\maketitle

\begin{abstract}
Music Genre Classification is the problem of associating genre-related labels to digitized music tracks. It has applications in the organization of commercial and personal music collections. Often, music tracks are described as a set of timbre-inspired sound textures. In shallow-learning systems, the total number of sound textures per track is usually too high, and texture downsampling is necessary to make training tractable. Although previous work has solved this by linear downsampling, no extensive work has been done to evaluate how texture selection benefits genre classification in the context of the bag of frames track descriptions. In this paper, we evaluate the impact of frame selection on automatic music genre classification in a bag of frames scenario. We also present a novel texture selector based on K-Means aimed to identify diverse sound textures within each track. We evaluated texture selection in diverse datasets, four different feature sets, as well as its relationship to a univariate feature selection strategy. The results show that frame selection leads to significant improvement over the single vector baseline on datasets consisting of full-length tracks, regardless of the feature set. Results also indicate that the K-Means texture selector achieves significant improvements over the baseline, using fewer textures per track than the commonly used linear downsampling. The results also suggest that texture selection is complementary to the feature selection strategy evaluated.  Our qualitative analysis indicates that texture variety within classes benefits model generalization. Our analysis shows that selecting specific audio excerpts can improve classification performance, and it can be done automatically.
\end{abstract}

% keywords can be removed

\section{Introduction}
\label{sec:intro}

% Nesta seção: não usar equações, somente discussões e figuras.
% * What is Automatic Genre Classification
% * Weaknesses of Genre labels / uses of genre labels
% * What is a texture / why does it relate to genre
% * How does AGC work: iVector
% * How does AGC work: feature selection do Baniya
% * How does AGC work: ensemble learning do Smaragdis
% * Research questions:
%   * Does frame selection contribute to AGC?
%   * What are effective methods for frame selection?
%   * How does it compare to using feature selection?
%   * Does the frame selection method actually point to meaningful frames?

Genre is a descriptive tag commonly associated with music tracks. It relates to the social groups that participate in the process of producing, marketing and consuming particular music pieces or styles. Music genre also relates to the instruments and techniques used for musical composition, performance, and perception, and this reflects on the spectral patterns that are present in its digital audio recordings. Such patterns can be used to devise mathematical models for sound perception that allow automatically associating genre tags to digital music, that is, automatic Music Genre Classification (MGC) \cite{tzan2002}. This task is also referred to as automatic Music Genre Recognition (MGR).

%A great part of the research on AMGC has been on developing novel feature sets to describe timbre. 
Many MGC systems rely on assuming that spectral patterns that allow predicting genre are typically a few (1 to 5) seconds long. These patterns were described by Tzanetakis and Cook  \cite{tzan2002} as \textit{textures}. They were modelled with low-order statistics of low-level features calculated in short-time (around 20 ms) frames \cite{tzan2002} . %This model maps audio to a perceptually-inspired vector space and can be used for further classification steps.

%Just as a texture describes a short-time audio sample, a digital music track can be described by a sequence of textures.
One possible method to use the sequence of textures in a track consists of summarizing them into a single vector \cite{tzan2002}. This method, which we call \textit{Full Track Statistics} (\texttt{FTS}), results in a compact encoding, but can discard texture information that can be useful for genre classification. %It also assumes that all the textures in the recording are equally representative for genre recognition. 

Another possibility for such is to use a collection of textures to represent each track. 
%The idea is to acknowledge that textures are not homogeneous throughout a given track. 
This allows for richer descriptions of the underlying genre tags because the collections comprise distinct sound textures that are present in each musical piece. Many MGC approaches use collections of textures and disregard their time-domain ordering. These approaches can be seen as variants of the Bag of Frames (BoF) \cite{aucouturier07}. There are many approaches for combining the textures in classification, including statistical modelling \cite{aucouturier07}, dictionary learning and quantization \cite{marques11,yeh13,fu2011}, and independent texture labelling combined by voting procedures \cite{henaff2011,hamel2011,wulfing2012,jeong2016,yandre2017}. Using diverse textures for classification avoids the problem of discarding information, but increase the computational cost for dataset processing and storage.

In this work, we argue that selecting specific, typical textures from each track is an effective way to reduce the computational workload in storage and processing, while preserving classification accuracy. For such, we propose to select textures prototypes from each track using K-Means clustering centroids. These centroids are used for further classification steps.

We compare the results of our texture selection approach to linear downsampling, which is widely used in previous research \cite{henaff2011,hamel2011,wulfing2012,jeong2016,yandre2017}. We also investigate how texture selection compares with feature selection. Our results show that, in the datasets employed in this work, K-Means centroid downsampling leads to higher classification accuracy using less data than linear downsampling, and that texture selection in time has a greater impact in results than feature selection. Also, feature-space analysis indicate that this improvement depends on the presence of texture diversity within the tracks, which leads to worse results in datasets containing short (10s-long or 30s-long) excerpts.

Another concern related to texture selection regards the feature set used to represent each texture. For such, there are two typical approaches: using handcrafted features \cite{tzan2002} and using automatically obtained, data-driven features \cite{yandre2017}. In this work, we evaluate both possibilities, and, additionally, we evaluate using random projections of Mel-spectrograms as features. Our results show that K-Means texture selection has a greater impact in accuracy than changing the feature sets in the evaluated datasets.

%Lastly, we present a study of the feature spaces derived from both texture selection techniques.

This paper is organized as follows: Section \ref{sec:related} presents some relevant related research. Section \ref{sec:method} details our proposed texture selection approach. Section \ref{sec:experiments} details our experimental setup and results. Section \ref{sec:discussion} presents a qualitative discussion on how our approach contrasts with the commonly used linear downsampling.

\section{Related Work}
\label{sec:related}

% Nesta seção, se necessário, usar equações. A discussão é mais técnica que na introdução.
% * Audio classification 
% * Ensemble learning / Dictionary Learning
% * Deep Learning of Features
% * Random Features (ELM)

A significant portion of research in MGC uses a single feature vector to represent each track in their datasets. Work by Tzanetakis and Cook \cite{tzan2002} uses the mean value of each feature as dimensions for their feature vector. Banyia et al. \cite{baniya2015} summarizes the textures using higher-order feature statistics and a feature covariance matrix. More recently, work by Choi et al. \cite{choi17b} uses a CNN to extract features from mel-spectrograms and summarizes the output of each layer using average pooling, leading to a single vector representation.

Other studies represent tracks using a collection of textures. Yandre et al. \cite{yandre2017} represent each track by slicing its spectrogram into a set of 50 non-overlapping \textit{patches} that comprise the middle 60s of each track. Each \textit{patch} is treated independently during both training and testing phases. A final classification is decided by summing the classification probabilities for each \textit{patch}. This linear slicing setup along with the sum voting rule implicitly assumes that each \textit{patch} is equally relevant for genre classification. Other related work have also used variants of this linearly-spaced texture selection method \cite{henaff2011,hamel2011,wulfing2012,jeong2016}.

Aucouturier and Pachet \cite{aucouturier04} presented a representation in which each track is described by a Gaussian Mixture Model (GMM) that models the probability of any frame being associated with the track. Differently from the linear-selection procedures, this representation models the relevance difference between the textures within tracks. Later, Aucouturier et al. \cite{aucouturier07} popularized the term Bag of Frames, highlighting that this description takes into account that collections of textures characterize audio tags, while disregarding long-term temporal behavior.

Later, Marques et al. \cite{marques11} evaluated various descriptors based on the dictionary learning approach. In this work, tracks are represented as a histogram of occurrences of dictionary elements.
%As commonly done in similar approaches \cite{yeh13,fu2011}, KMEANS clustering is used to find typical frames. Clustering is not done in a track-by-track basis, but with with all frames at once.
As a baseline, they build the dictionary from sampling uniformly-distributed random frames. Their results show that using random frames as dictionary entries achieves at least the same classification performance as selecting the most representative frames. 

Lopes et al. \cite{lopes2010} observe that Bag of Frames approaches implicitly assume that all textures in a given track convey relevant information for genre classification. They argue that not every texture is useful for determining the decision boundaries among genres. To solve this problem, they propose a technique for selecting discriminative textures. The proposed method resembles the ideas behind wrapper feature selection strategies \cite{guyon2002}. Their results for texture selection are not significantly better than the baseline with no selection. Nevertheless, their work raised an interesting question regarding the relevance of individual textures to represent music genres.% It also motivated us to investigate how texture selection can improve classification performance.

Bag of Frames approaches acknowledges the fact that music tracks do not contain homogeneous textures. On the contrary, each music genre can be related to several typical textures.
%Moreover, it is reasonable to assume that there are typical sounds that are very likely to appear in some genres. 
For example, we are likely to hear the sound of electric guitar solos and fast drum rolls in heavy metal music, but not in baroque pieces. %This indicates that genres are related to corresponding typical textures.

We propose to identify these typical textures within each track by clustering their vector representations with a K-Means clustering algorithm. This allows identifying the variety of textures within a track regardless of their frequency of occurrence. The K-Means centroids are then yielded to a machine-learning algorithm that performs texture-level classification. After that, the track is classified using a majority-voting procedure along the textures.

This proposal contrasts with codebook approaches \cite{fu2011,yeh13,marques11}, in which K-Means is used to learn dictionaries for quantization and tracks are represented by a histogram of codewords. The codebook approach represents texture variety using occurrence ratios. On the contrary, we propose to directly use the centroids yielded by K-Means as texture representations. This allows typical sounds to be directly represented in the feature space, and the machine learning algorithm groups them into genre-typical textures.

Our method is described in the next section.

\section{Proposed Method}
%\section{KMEANS Texture Selection}
\label{sec:method}

This section describes the proposed texture selection method. Section \ref{sec:linspace} describes the baseline texture linear downsampling procedure. After that, Section \ref{sec:kmeans} describes the proposed K-Means texture selection method. These methods assume that textures are represented as vectors spanned by perceptually-related features. The feature sets used in this work are described along with the experimental setup in Section \ref{sec:experiments}.

\subsection{Linear Texture Downsampling}
\label{sec:linspace}
The main concern presented by previous research regarding using collections of textures to describe music tracks is the amount of computing power needed to train models. Many learning algorithms do not scale well as the number of training vectors increases. Therefore, it is necessary to reduce the number of textures used for training, specially in systems that do not rely on neural networks.

A number of previous research solves this problem by downsampling textures of each track \cite{yandre2017}. Some variations of this method have been used before by many authors such as in \cite{yandre2017, henaff2011, jeong2016, wulfing2012, hamel2011}. 

\begin{figure}[h!]
    \centering
    \includegraphics[scale=0.8]{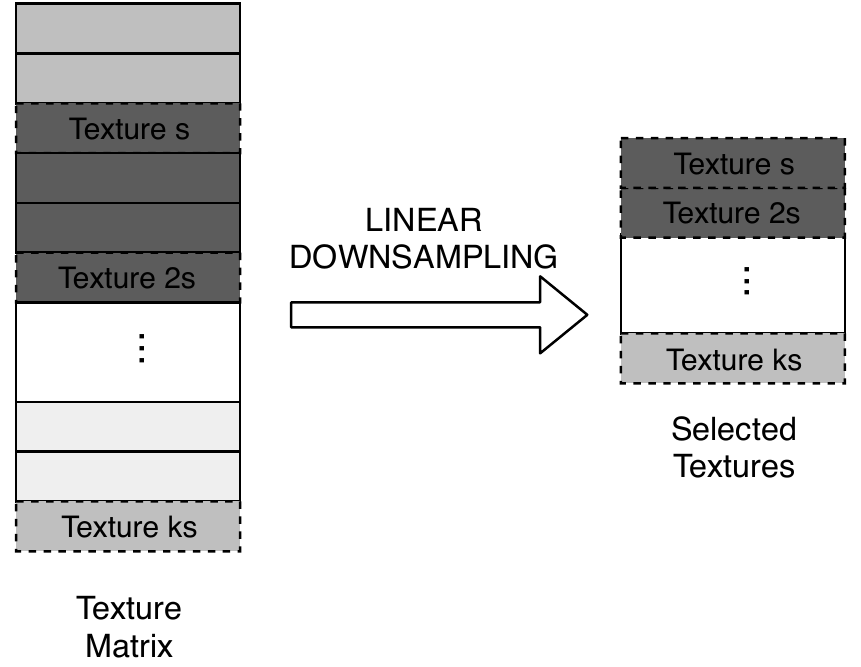}
    \caption{LINSPACE Texture Selection}
    \label{fig:linspace_selection}
\end{figure}

A common strategy is to pick $k$ linearly-spaced textures along the time axis. This procedure is shown in Figure \ref{fig:linspace_selection}. We call this strategy \textbf{LINSPACE Texture Selection}. Mathematically, a track texture matrix with $m$ textures and $n$ features $T \in \mathbb{R}^{m \times n}$, $T = [t_0, t_1, \dots, t_m]^T$, $t_i \in \mathbb{R}^n$, is summarized by LINSPACE yielding a matrix $L \in \mathbb{R}^{k \times n}$ such that $L = [t_s, t_{2s}, \dots, t_{ks}]^T$, where $s = \left \lfloor \frac{m}{k} \right \rfloor$ and $k$ is a parameter that controls the granularity of the downsampling procedure.

LINSPACE does not use content to select representative textures for genre classification. In contrast, the proposed K-Means Texture Selection aims to select appropriate textures based on the premise that there are typical sounds that are likely to appear in tracks of the corresponding genres.

\subsection{K-Means Texture Selection}
\label{sec:kmeans}
%what is kmeans and how it can help us with our problem
K-Means is a well-known clustering algorithm introduced in \cite{macqueen1967}. It works by iteratively estimating points, known as \emph{centroids}, that characterize different trends in a dataset. These points can then be used to query the dataset for other points following the trend. %Clustering algorithms have been widely used in information retrieval for outlier detection, trend extraction, data exploration, and dictionary learning \citep{}.

\begin{figure}[h!]
    \centering
    \includegraphics[scale=0.8]{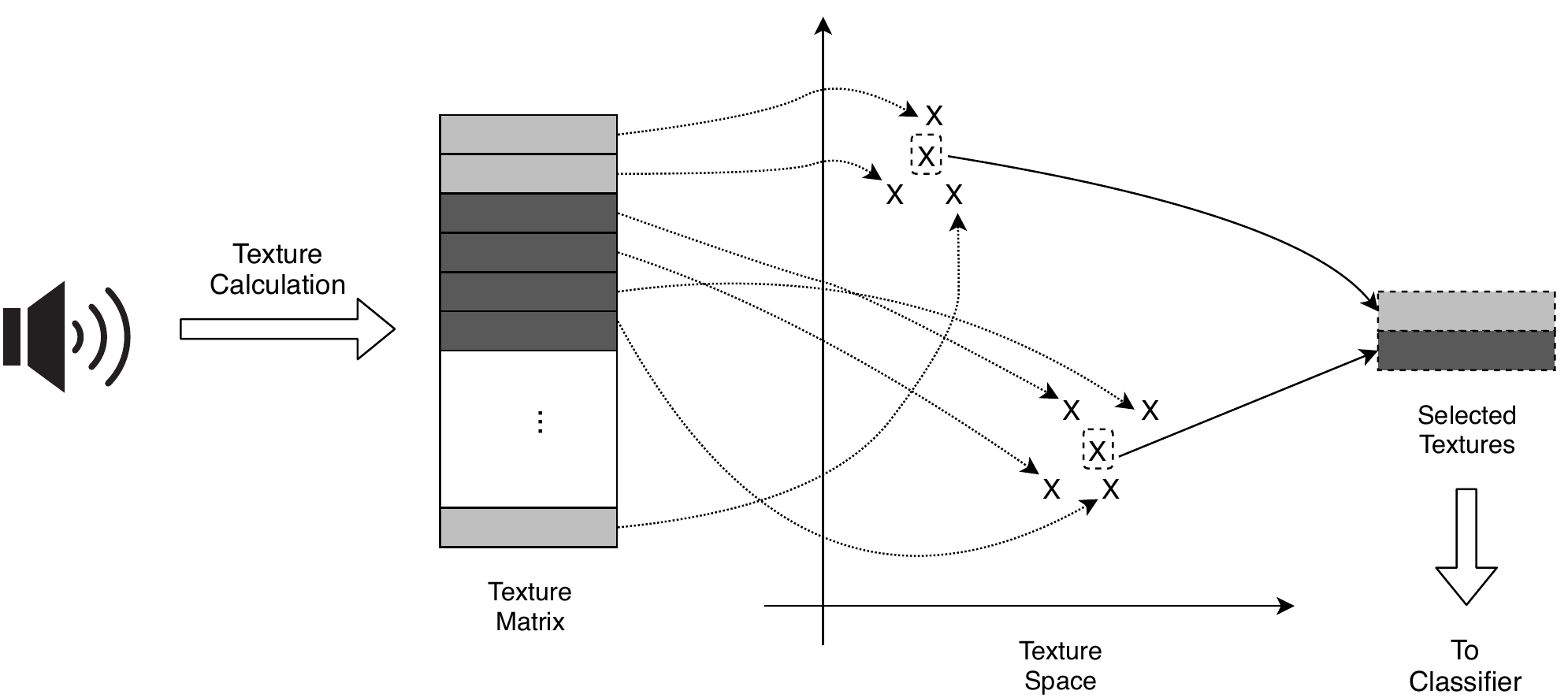}
    \caption{K-Means Texture Selection}
    \label{fig:kmeans_selection}
\end{figure}

%how we use kmeans for texture selection
As stated before, each audio texture is represented by a vector in a perceptually-inspired space $\mathbb{R}^n$, where $n$ is the number of features that span the space. Thus, a music track can be described by a texture matrix $T \in \mathbb{R}^{m \times n}$ where $m$ is the number of textures in the track and $n$ is the number of features. When K-Means is applied to $T$, it estimates a centroid matrix $C \in \mathbb{R}^{k \times n}$, where $k$ is the number of centroids. %, such that $k < n$.
These centroids can be interpreted as vector representations of the acoustic trends in track $T$, and $k$ is the number of trends that are extracted from the audio track. At the end of this process, $T$ is summarized into $C$, and $C$ is yielded to further classification steps. Figure \ref{fig:kmeans_selection} depicts this procedure.

We call this method \textbf{K-Means Texture Selection}. Because K-Means tends to yield centroids that are distant from each other, we can expect that the $C$ matrix contains information that tends to be diverse. As centroids are more likely to be located around denser point clouds, they are also likely to represent typical textures. As a result, the $C$ matrix highlights diverse typical sounds within the track.
%We also highlight that one of the arguments for this track description is that it should support the assumption that music tracks consist of a set of varied sound textures. Therefore, aggregating the centroids into a single texture using low order-statistics invalidates such assumption and may yield an inaccurate description of the whole track. 
To the best of our knowledge, K-Means has not been used before to select representative textures for genre classification. %In contrast to codebook approaches \citep{fu2011,yeh13,marques11}, where KMEANS is used to learn codewords for quantization, we propose using the centroids themselves as instances. Thus, in our approach every track is represented by a set of centroids. To collect a larger variety of textures for the overall genre model, we chose to perform texture selection independently for each track. 

K-Means Texture Selection is compared to Linear Texture Downsampling in music genre classification experiments, as discussed next.

\section{Experiments and Results}
\label{sec:experiments}

In this section we present experiments that aim to investigate how texture selection impacts music genre classification performance. Specifically, we are interested in the effectiveness of selecting representative textures to describe a track with K-Means clustering. To this end, we evaluated four texture selection methods: K-Means clustering (KMEANSC), described in Section \ref{sec:kmeans}, LINSPACE downsampling, described in Section \ref{sec:linspace}, a FTS vector for each track, as proposed by \cite{tzan2002}, and using all textures with no summarization or downsampling (ALL). %, as proposed by \cite{aucouturier07}.
Our main objective was to evaluate how texture selection impacts classification performance in systems where tracks are represented by collections of textures. %Nevertheless, we achieved results comparable to the state of the art in two datasets.

To evaluate the texture selection effectiveness in different scenarios, we designed classification systems taking into consideration two components that are commonly evaluated in MGC systems: the feature sets, and the use of feature selection. Four feature sets of varying abstraction levels, and an univariate correlation filter for feature selection were evaluated along with the texture selection methods.

Our experimental setup is thoroughly described in Section \ref{sec:setup}. We note that our goal with these experiments is to investigate the effects of texture selection in a variety of MGC scenarios. In other words, we aimed at generating insight on the underlying mechanisms behind the result differences. For such, we executed MGC experiments using different texture selection methods, as discussed in Section \ref{sec:selection}, diverse feature sets, as shown in Section \ref{sec:fss}, and datasets with different characteristics, as discussed in Section \ref{sec:datasets}. After that, Section \ref{sec:arch} presents the parameterization of the systems evaluated, and Section \ref{sec:results} shows classification results.

\subsection{Experimental Setup}
\label{sec:setup}

\begin{figure}[h!]
    \centering
    \subfloat[Texture Calculation]{
        \includegraphics[scale=0.42,valign=c]{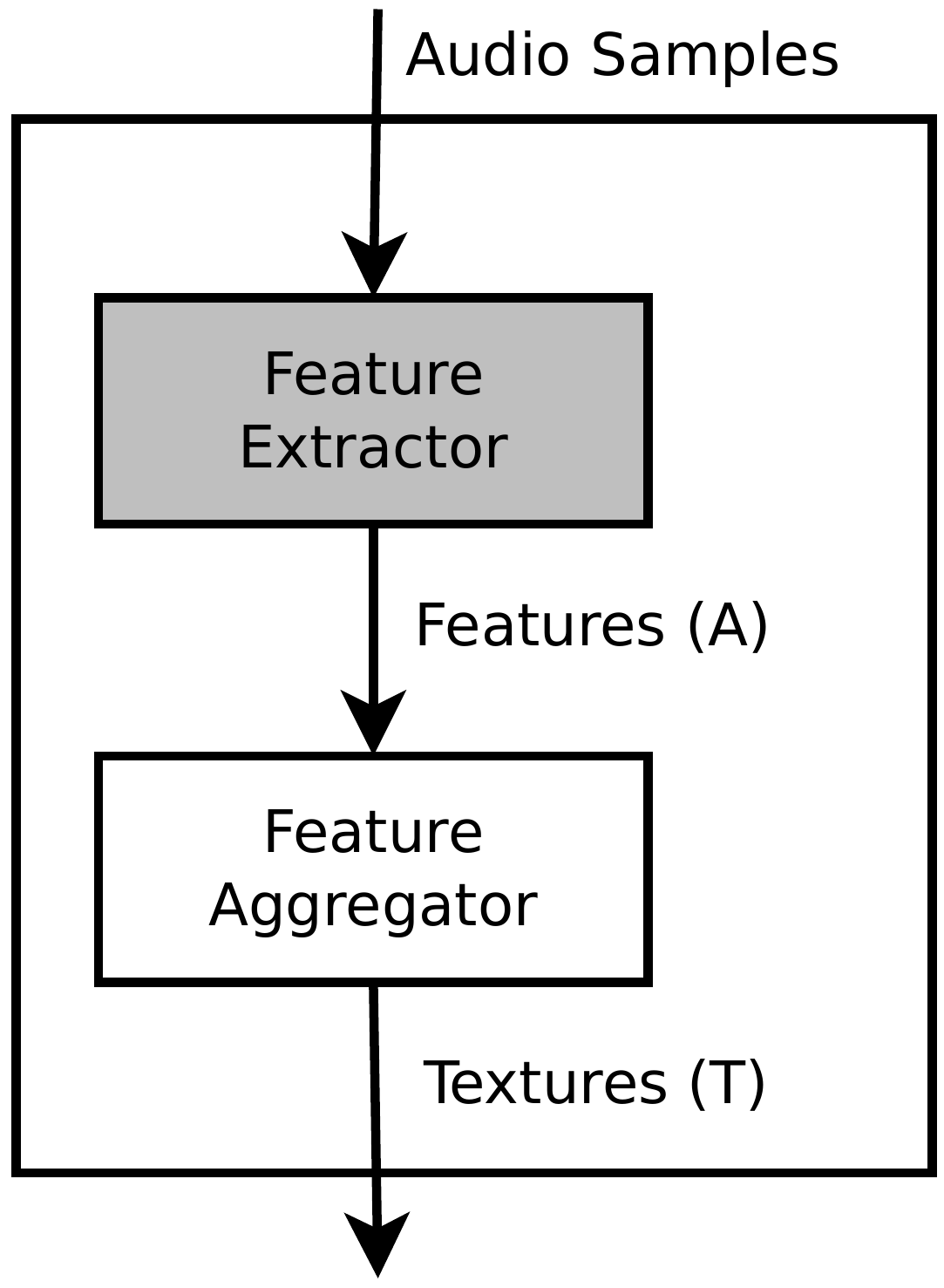}
        \vphantom{\includegraphics[scale=0.42,valign=c]{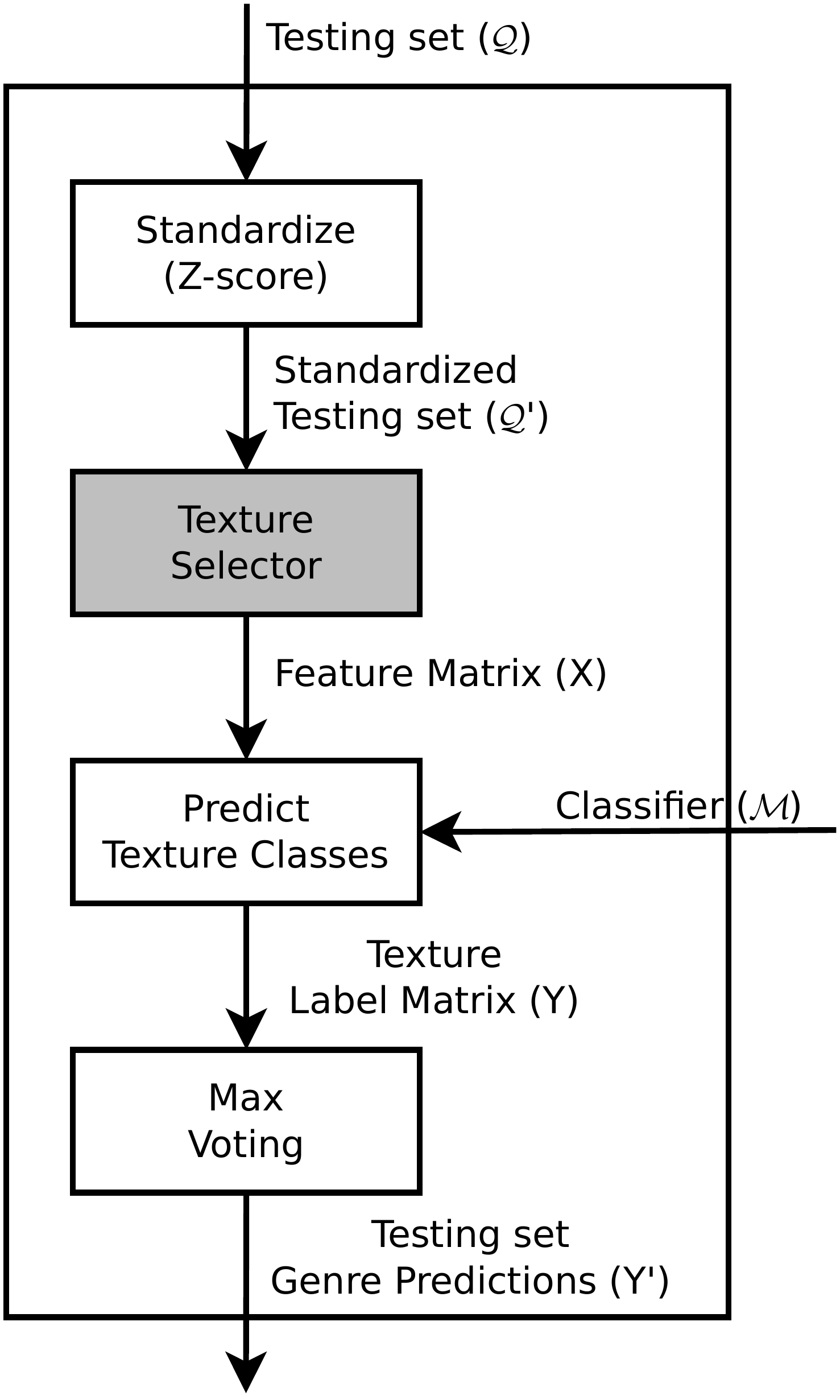}}
        \label{fig:arch_fe}
    }
    \subfloat[Model Training]{
        \includegraphics[scale=0.42,valign=c]{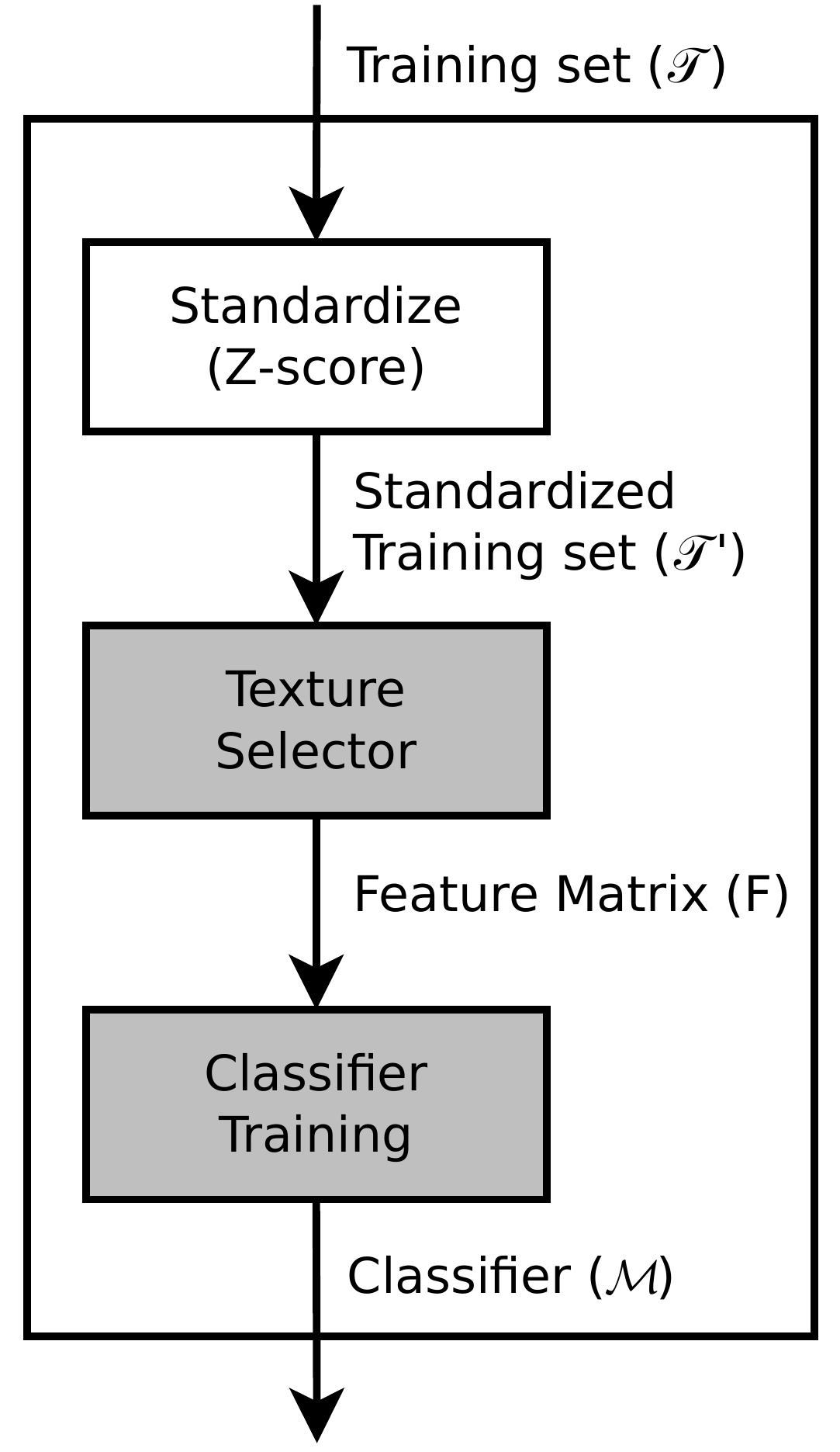}
        \vphantom{\includegraphics[scale=0.42,valign=c]{diagrams/architecture_testing.pdf}}
        \label{fig:arch_training}
    }
    \subfloat[Model Testing]{
        \includegraphics[scale=0.42,valign=c]{diagrams/architecture_testing.pdf}
        \label{fig:arch_testing}
    }
    \caption{System Architecture} \label{fig:architecture}    
\end{figure}

The system architecture is shown in Figure \ref{fig:architecture}. It consists of three stages: Texture Calculation (Figure \ref{fig:arch_fe}), Model Training (Figure \ref{fig:arch_training}), and Model Testing (Figure \ref{fig:arch_testing}). Texture Calculation begins by extracting features over $23$ms frames from music tracks sampled at $44$KHz. For each track, a feature matrix (A) is calculated, consisting of a feature vector for every frame. The features, their first and second-order deltas are calculated and concatenated into a single vector per frame. Textures are then calculated by aggregating 216 consecutive frames, resulting in textures that cover approximately $5s$ of the track. Successive textures are calculated every 10 frames, resulting in a $10$x downsample and around 95\% overlap between textures. We used both average and standard deviation as aggregation functions over every feature. The resulting texture matrix (T) is calculated for every track in the dataset.

The results of the texture calculation are yielded to a machine-learning algorithm. It relies on two different stages: model training and model testing. The model training stage uses a part of the dataset to estimate its latent parameters. The model testing stage uses another, held-out part of the dataset to evaluate the classification performance.

The input to the Model Training stage (Figure \ref{fig:arch_training}) is the training set ($\mathscr{T}$), which is the set of texture matrices corresponding to the training tracks. First, textures are standardised feature-wise to mean $0$, standard deviation $1$. The standardisation parameters are saved to be applied later to the testing set. The resulting standardised training set is given by $\mathscr{T}'$. Texture selection is performed on every texture matrix of the training set independently. The number of desired textures per track is a parameter, which we call $k$. The texture selector outputs a training Feature Matrix (F), which concatenates all selected textures from every track in the training set. Assuming all selected textures are representative of the genre, every texture of the same track are assigned the track label. A classifier is trained with the samples in $F$, along with the corresponding labels. The classifier ($\mathcal{M}$) can be seen as a function that maps each texture to a label%., where $C$ is the set of labels of the dataset.

The Model Testing stage (Figure \ref{fig:arch_testing}) receives a testing set ($\mathcal{Q}$) as input, which is the set of texture matrices corresponding to the testing tracks. The same standardisation parameters applied to the training set $\mathscr{T}$ are applied to $\mathcal{Q}$, resulting in a standardised testing set ($\mathcal{Q'}$).

Similarly to the training samples, texture selection is performed on each track independently for each matrix in the testing set. The same number of textures per track are used for both training and testing. The texture selector outputs a testing matrix ($X$), with every selected texture of each track in the testing set. Then, $\mathcal{M}$ is applied row-wise to $X$, resulting in a Texture Label Matrix ($Y$), with a label prediction for every selected texture of every track in the testing set. A final classification is decided for each track by majority voting. The output for the entire testing set is the prediction matrix ($Y'$).

\subsection{Texture Selection}
\label{sec:selection}
%KMEANS vs LINSPACE in texture selection. ivector. ALL frames.

We used four different texture selection approaches. Both KMEANSC and LINSPACE were described in Section \ref{sec:method}. As stated, they assume that a track is best described by a set of textures, instead of aggregating the description into a single texture. To test this hypothesis, we evaluated the classification performance using FTS as well. In this approach, a single texture is calculated for each track by aggregating all textures with average and variance. %In this case the training matrix in Figure \ref{fig:architecture} becomes $F \in \mathbb{R}^{N \times 2n}$.
%The same approach is used to calculate the FTS vectors for the testing matrix.

We also evaluated how selecting relevant textures impacts classification performance. For this, we evaluated classification performance using ALL textures to represent a track.  %In contrast to our KMEANSC approach, using all textures assumes that all variations of textures are typical of the genre of the track. This may introduce false positives during training, as discussed before. Thus, this approach also allows us to verify the effect of such false positives in classification performance.
%When all textures are used to describe a track, the training matrix in Figure \ref{fig:architecture} becomes $F \in \mathbb{R}^{a \times n}$ where $a = \sum_{i=0}^{N} m_i, i = 1, \dots, N$, and $m_i$ is the number of textures in track $i$ of $\mathscr{T}$. 
Naturally, the higher the number of textures used to describe tracks, the higher the cost of training and testing classification systems. Thus, by comparing the results of texture selection with ALL, we are able to determine the relationship between the number of textures per track and classification performance, as well as how selecting a subset of the textures affects performance.

Because the four texture selection techniques presented have distinct computation requirements, evaluating the resulting classification performance is important. This analysis can lead to guidelines for choosing the most appropriate texture selector given the task size and computing capabilities.

\subsection{Feature Sets}
\label{sec:fss}

We evaluated texture selection performance along with four different feature sets with different underlying assumptions. This allows us to evaluate whether the change in performance due to texture selection is dependent on the feature set describing the textures. It also allows us to draw conclusions on key factors for improving classification performance, in particular: how relevant are the low-level descriptions for music genre classification, and what is the relevance of using more textures to train classification models. %preciso melhorar isso pq essa é a pergunta que começou essa pesquisa.

\subsubsection*{Mel-Scale Spectrograms}

\begin{figure}
    \centering
    \subfloat[MEL-SPEC Feature Set]{
        \includegraphics[scale=1.3]{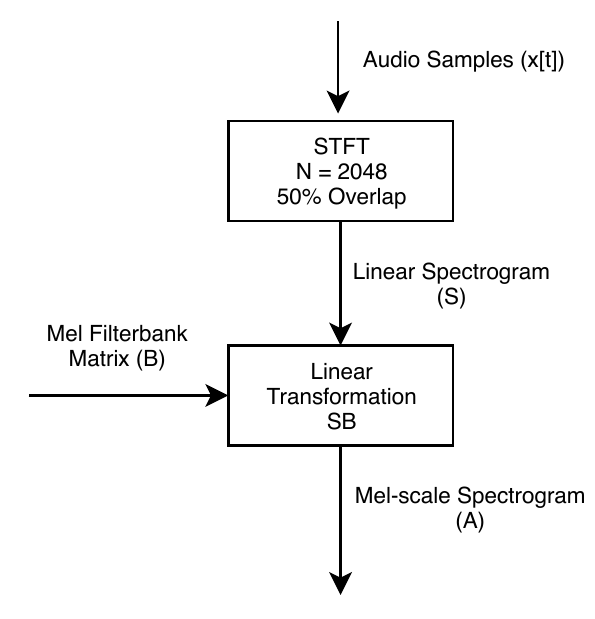}
        \label{fig:MEL-SPEC}
    }
    \hspace{-10pt}
    \subfloat[MEL-RP Feature Set]{
        \includegraphics[scale=1.3]{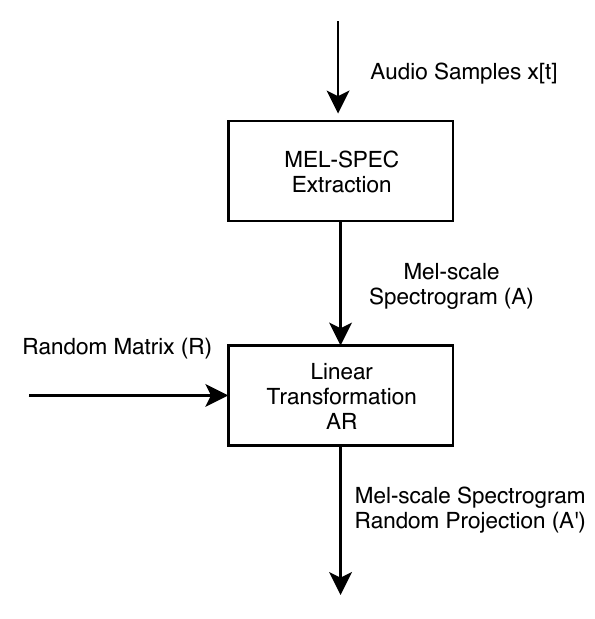}
        \label{fig:MEL-RP}
    }
    \caption{Low-level Feature Sets} 
\end{figure}

Two feature sets derive directly from audio data, and were used as ``raw'' features: MEL-SPEC and MEL-RP. The MEL-SPEC feature extraction is shown in Figure \ref{fig:MEL-SPEC}. It starts with a track signal in the time domain. Then, a STFT is calculated in frames of $2048$ samples, with 50\% overlap. 
%Previous work has shown that the phase content is not relevant for music genre classification \cite{}, thus 
The absolute value of the STFT is calculated, yielding a magnitude matrix ($S$). Then, a 128-bin Mel Filterbank matrix ($B$) was applied to transform $S$ into ($A=SB$). We call $A$ a Mel-Scale spectrogram (MEL-SPEC). This transformation relies on previous research, which has shown that Mel-Scale spectrograms highlight important aspects of the audio spectrum relevant to genre classification \cite{nam17}. %Explicar melhor o pq que escolhemos usar esse feature set (talvez já tá explicado).

Many systems that rely on feature learning have been proposed to work with raw inputs such as MEL-SPEC \cite{choi17b, nam17,dieleman2014}. We used it to evaluate how such texture descriptions can be used in a scenario where MEL-SPEC is used directly as a feature set, not as input to a feature learning mechanism. We assume that such inputs roughly model our timbre perception by highlighting the magnitude of spectral contents. Thus, it is expected that the Euclidean distance of such texture vectors relates to music content similarity. %Texture selection based on KMEANSC is expected to result in clusters representing the typical sounds of each track.

\subsubsection*{Mel-Scale Spectrogram Random Projection}

Another feature set derived directly from data is the random projection of mel spectrograms (MEL-RP). MEL-RP feature extraction is shown in Figure \ref{fig:MEL-RP}. The first step is to calculate the Mel-Scale Spectrogram $A$ for the music track, as shown in Figure \ref{fig:MEL-SPEC}. Then, a linear transformation $R$, is applied to $A$, resulting in $A'=TR$. The number of columns in $R$ is lower than in $A$, which results in a projection into a lower-dimensional space . Because $R$ is a random matrix sampled from a Gaussian distribution with mean $0$ and variance $1$, we call $A'$ the Mel-Scale Spectrogram Random Projection (MEL-RP). MEL-RP aims at transforming each texture in $T$ into a stable embedding, preserving the underlying distance topology. This allows texture classification \cite{davenport2007} in the projected space, instead of the Mel-Scale spectrogram space.

The effectiveness of random projections for dimensionality reduction is well-known in machine learning literature \cite{baraniuk2010}. The Johnson-Lindenstrauss (JL) lemma \cite{johnson1984} states that a random matrix $R \in \mathbb{R}^{N \times M}$, when $N > M$, projects the points in $A \in \mathbb{R}^{T \times N}$ into a stable embedding with high probability if $M = O(\log{(T)} \epsilon^{-2})$, $\epsilon \in (0,1)$. $N$ is the data original dimensionality, $M$ is the target dimensionality, and $T$ is the number of points in $A$. The $\epsilon$ constant quantifies the distortion introduced by the random transformation. It follows that, as more distortion is allowed, the smaller M can be. Thus, $\epsilon$ is a parameter that can be implicitly adjusted according to the application by optimizing $M$ in relation to a classification system performance measure.

Since the dimension of the embedding domain is lower ($M<N$), the computational effort needed for training a machine learning model in $\mathbb{R}^{M}$ is expected to be smaller than a model in $\mathbb{R}^{N}$. Furthermore, for suitable classifiers, the curse of dimensionality can be alleviated by transforming data points into a lower-dimensional space.

To the best of our knowledge, no previous research has employed MEL-RP as a feature set for genre classification. A similar idea, proposed by Choi \cite{choi17b}, consists in setting the weights of a convolutional neural network to random values. The classification results are used as baseline for other approaches. The main difference is that the architecture is made up of convolution operations followed by non-linear activation functions. Choi has reported satisfactory results with this method. Furthermore, other authors have used random projections as a dimensionality reduction tool in further related works \cite{panagakis2009,chang2010,Banitalebi2014}.

\subsubsection*{Handcrafted Features}

Classification systems with handcrafted features were also evaluated. For simplicity sake, we call the chosen subset HANDCRAFTED. The features used are well-known in the music information retrieval literature and some of them are known as discriminative features for some tasks. Handcrafted features are based on specialist knowledge. Therefore, they can be seen as features at a higher abstraction level than the ``raw'' feature sets presented earlier. In the context of our work, we are interested in how handcrafted features work in tandem with texture selection.

The handcrafted features selected for the experiments are calculated for every frame from the STFT magnitude spectrum. First, the audio input signal sampled at 44Khz is centered at mean 0, variance 1. Then, a $2048$ sample STFT is calculated with $50\%$ overlap over consecutive frames. A hamming window is applied to prevent high-frequency artifacts from slicing.

The following equations describes the features in our HANDCRAFTED feature set. $M[f]$ is the magnitude of the FFT of a given frame at frequency bin $f$, and $N$ is the total number of frequency bins. \cite{tzan2002} describe the \textbf{Spectral Centroid} as a measure of spectral brightness. It is calculated from the following equation:

$$
    C = \frac{\sum_{f = 1}^{N} f M[f]  }{\sum_{f = 1}^{N} M[f] }
$$

\textbf{Spectral Rolloff} is a measure of spectral shape \cite{tzan2002}. It corresponds to the value $R$ in the following equation:

$$
    \sum_{f = 1}^{R} M[f] = 0.85 \sum_{f = 1}^{N} M[f]
$$

The \textbf{Spectral Flux}, which is a measure of spectral variation \cite{tzan2002}, can be calculated by:

$$
    F = || M[f] - M[f-1] ||_2
$$

\textbf{Energy} is a measure of signal strength, is calculated by:

$$
    E = \sum_{f = 1}^{N} M[f]^2
$$

\textbf{Spectral Flatness} is a measure of noise in an audio signal \cite{dubnov04}, and it is calculated by:

$$
    L = \frac{ \exp{ \left ( \frac{1}{N} \sum_{f = 1}^{N} \ln{(M[f])} \right )} } { \frac{1}{N} \sum_{f = 1}^{N} M[f] }
$$

\textbf{Zero Crossing Rate} is another measure of noise in an audio signal. It is calculated over the time-domain signal, $x[i], | \, i = \{ 1, 2, \dots, T \}$:

$$
    Z =  \frac{1}{2T} \sum_{t=1}^{T} \left | \text{sign}(x[t+1]) - \text{sign}(x[t])  \right |
$$

\noindent
where $\text{sign}(k) = 1$ if $k \geq 0 $, otherwise $0$. 

MFCCs (Mel-Frequency Cepstral Coefficients) \cite{hunt1980} are also part of the HANDCRAFTED feature set. MFCCs are widely used by the speech recognition and music information retrieval research communities. In MIR, they are commonly used as timbre descriptors.

In total, there are 26 features in the HANCRAFTED feature set: Spectral Centroid, Spectral Rolloff, Spectral Flux, Spectral Flatness, Energy, Zero Crossing Rate and the first 20 MFCC coefficients. 
%There have been many different combinations of these features used in previous music genre classification work. 
We are interested in how much these rather simple features work along our proposed texture selection. Specifically, we want to verify how the classification results compare to more sophisticated feature sets, such as an autoencoder representation, discussed next.

\subsubsection*{Mel-Scale Autoencoder}

An autoencoder is a neural network that learns a mapping from the input to the input itself \cite{goodfellow16}. A well-known architecture is based on a series of fully connected neuron layers. The middle layer is known as the \emph{bottleneck}, is also known as the latent representation, and usually has a lower dimensionality than the other layers. The idea is that the bottleneck represents a compressed form of the input signal, hence transforming the input into a lower dimensionality vector. %As the autoencoder is trained with a 

\begin{figure}
    \centering
    \includegraphics[scale=0.75]{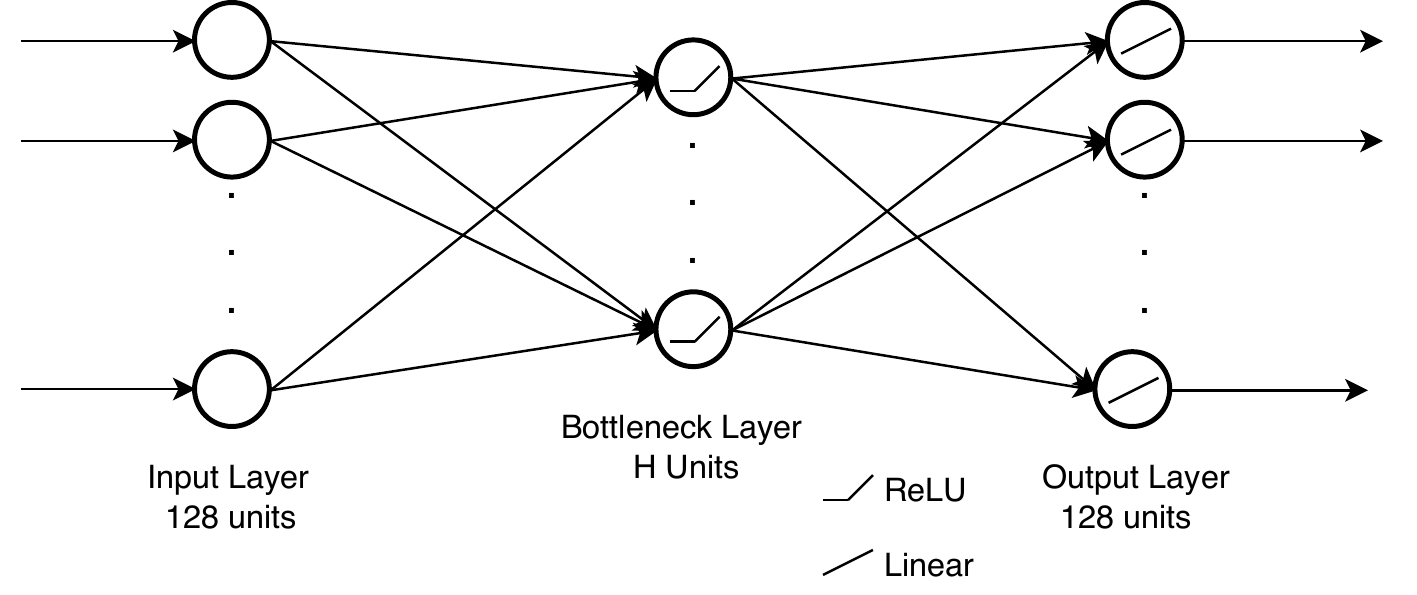}
    \caption{Autoencoder Architecture}
    \label{fig:autoencoder}
\end{figure}

In contrast to the other feature sets presented above, autoencoders provide features learned directly from data. Figure \ref{fig:autoencoder} shows the architecture of the autoencoder used in this paper. Mel Spectrograms are the input to the autoencoder, hence its output as well. The number of neurons in the bottleneck is a parameter ($H$). A non-linear ReLU activation function is used in the bottleneck layer, forcing a non-linear projection of the input. The output layer uses a linear activation function, thus forcing a linear reconstruction of the input signal in the output space. This is an attempt to promote linear separability among different clusters of similar points. The bottleneck layer activations are used as features. We call this feature set MEL-Autoencoder (MEL-AE). 

In preliminary tests we experimented with different autoencoder architectures. The final mean squared error was evaluated for different architectures and the results differed within small error margins. Thus, by Occam's principle, we chose the simplest architecture in terms of number of neurons and number of layers.

\subsection{Datasets}
\label{sec:datasets}

All classification systems were evaluated with four different publicly available datasets: GTZAN \cite{tzan2002}, ISMIR \cite{ismir2004}, LMD \cite{silla2008}, and HOMBURG \cite{homburg2005}. We chose these datasets because each one presents particular challenges. Evaluating system performance on datasets with different challenges leads to a better understanding of how effective the proposed methods are under different situations. They are also well-known in the music information retrieval community. Table \ref{tab:datasets} summarizes the datasets used in the experiments and shows their diversity regarding genres, number of tracks, track length and number of textures.
\begin{table}[h!]
\centering
\caption{Datasets used in the experiments}
\label{tab:datasets}
~\\
\begin{tabular}{@{}lccccc@{}}
\multicolumn{1}{c}{\textbf{Datasets}} & \textbf{Genres} & \textbf{Tracks} & \textbf{Textures} & \textbf{Track Length} & \textbf{\begin{tabular}[c]{@{}c@{}}Avg. Textures\\ per Track\end{tabular}} \\ \midrule
GTZAN                                 & 10              & 1000            & 107K              & 30s                   & 107                                                                               \\
ISMIR                                 & 6               & 1458            & 1,5M              & FULL                  & 1046                                                                              \\
ISMIR (10s)                           & 6               & 1458            & 31K               & 10s                   & 20                                                                                \\
LMD                                   & 10              & 1300            & 1,2M              & FULL                  & 940                                                                               \\
LMD (10s)                             & 10              & 1300            & 27K               & 10s                   & 21                                                                                \\
HOMBURG                               & 9               & 1886            & 40K               & 10s                   & 21                                                                                \\ \bottomrule
\end{tabular}
\end{table}

GTZAN \cite{tzan2002} was one of the first widely available datasets for MGC and it is widely known in the research community. This dataset consists of 1000 music clips of 10 western music genres. GTZAN is balanced among its 10 genres (blues, classical, country, disco, hip-hop, jazz, metal, pop, reggae, rock), with 100 clips each. Each music clip is 30s long. Its flaws are well-known by the research community. The main faults are: repeated tracks, multiple tracks from the same artist, cover tracks, mislabelings and extreme distortions \cite{sturm2013}. To address these faults, we created a 3-fold dataset split following all the recommendations in \cite{sturm2013}. We call this split GTZAN-ARTF in the results section. Although Sturm shows that there are still some unidentified excerpts in GTZAN, the proposed changes fixes most of the identified faults. The ARTF split we used is available online\footnote{\url{https://github.com/julianofoleiss/gtzan_sturm_filter_3folds_stratified/}}.

This 3-fold split is artist-filtered, which means that tracks from same artist and genre do not appear in both training and testing sets. It has been shown that artist-filtered splits makes the genre classification problem significantly harder \cite{pampalk2005,flexer2007}. As a consequence, the classifier performances are lower. One of the reasons for this is that the classifier can learn patterns that relate to specific artists, such as the voice timbre of the lead singer, in contrast to patterns relating to genre, regardless of artist-specific characteristics. Not rarely, an artist works solely on a single genre, thus all of their tracks are labelled with the same genre. Hence, a classification system may correctly predict a genre to a track based on artist-specific patterns, instead of genre-defining patterns. This causes songs from the same artist in the test set to be easier to classify, but this does not mean that the classifier has generalized well to other artists performing on the same genre.

GTZAN is a widely used dataset for evaluating genre classification systems \cite{sturm2012}. Often, a random 10-fold split is used, hence ignoring the artist filtering problem. For comparison purposes, we also included the results with this random split. We call this split GTZAN-RANDOM.

LMD (Latin Music Dataset) \cite{silla2008} is another well known dataset that consists of popular Latin-American music genres. This dataset was originally labelled according to dancing patterns, specially for deciding between genres that are similar regarding timbre, such as ax\'e/pagode and ga\'ucha/sertaneja. We used an artist-filtered subset of LMD in our experiments, which avoids the same problems present in the GTZAN dataset and allows a balanced 3-fold split. We used the same split as in \cite{nanni2016,yandre2017}. We call this split LMD-ARTF. This subset consists of 1300 full-length music tracks across 10 different genres, each with 130 tracks. The LMD-ARTF split we used is available online\footnote{\url{https://github.com/julianofoleiss/lmd_3f_stratified_artist_filter/}}. 

The ISMIR 2004 \cite{ismir2004} dataset consists of 1458 full-length music tracks across 6 western genres. We used the train/test split supplied in the ISMIR 2004 challenge homepage, with 729 tracks in both the training and testing sets. No artist filter was applied to this dataset because the artist information is not available for all tracks, only for tracks in the training set. %\cite{flexer2007}. 
This dataset is not balanced, thus the number of tracks is not the same for all classes. The majority of the tracks in the training set are \textit{classical} (320), followed by \textit{electronic} (115), \textit{jazz\_blues} (26), \textit{metal\_punk} (45), \textit{rock\_pop} (101) and, \textit{world} (122). The number of tracks per genre in the testing set is similar to the training set.

HOMBURG \cite{homburg2005} is a lesser known dataset that also consists of western music genres. It comprises 1886 music clips across 9 genres. This dataset is interesting for evaluating texture selection techniques, since the tracks are only 10s long. We used a standard 10-fold random split for cross validation. We did not apply an artist filter to HOMBURG because the number of tracks per artist is too low, as shown in Table \ref{tab:homburg}.
\begin{table}[h!]
\centering
\caption{Number of Tracks and Artists in the HOMBURG dataset}
\label{tab:homburg}
~\\
\begin{tabular}{@{}lccc@{}}
\multicolumn{1}{c}{\textbf{Genre}} & \textbf{Tracks} & \textbf{Artists} & \textbf{Avg. Tracks per Artist} \\ \midrule
Alternative                        & 145                   & 121                    & 1.20                       \\
Blues                              & 120                   & 80                     & 1.50                       \\
Electronic                         & 113                   & 97                     & 1.16                       \\
Folk/Country                       & 222                   & 177                    & 1.25                       \\
Funk/Soul/RnB                      & 47                    & 39                     & 1.21                       \\
Jazz                               & 319                   & 214                    & 1.49                       \\
Pop                                & 116                   & 106                    & 1.09                       \\
Raphiphop                          & 300                   & 210                    & 1.43                       \\
Rock                               & 504                   & 447                    & 1.13                       \\ \midrule
TOTAL                              & 1886                  & 1491                   &                            \\ \midrule
Weighed Avg.                       &                       &                        & 1.28                       \\ \bottomrule
\end{tabular}
\end{table}

\subsection{Configurations}
\label{sec:arch}

As presented earlier, Figure \ref{fig:architecture} shows the general architecture of the classification systems evaluated. The highlighted \textbf{Feature Extractor}, \textbf{Texture Selector}, \textbf{Classifier Training} boxes represent the components we experimented with different hyperparameter settings. 

The Feature Extractor parameters evaluated are shown in Table \ref{tab:feat_ex}. The four feature sets evaluated were presented in Section \ref{sec:fss}. A single feature set was evaluated at a time. The MEL-RP and MEL-AE input spectrograms are calculated with the same parameters as MEL-SPEC. HANDCRAFTED features are calculated from the STFT instead.

Four target dimension sizes for MEL-RP were chosen linearly between 25 and 100. 26 was used instead of 25 because it allows us to compare the results in terms of dimensionality directly to HANDCRAFTED features, which also has 26 features. To get an idea of how the system performed at a very low dimensionality, we also evaluated the system with only 8 random projection features. For MEL-AE, five different bottleneck sizes are evaluated. They represent compression rates of $\{8, 4, 2, 1, 0.5 \}$ times, respectively.

The number of features in each feature set is shown in Table \ref{tab:n_features}. Column \textit{Texture Size} shows the actual number of features in the texture vector after the aggregation procedures. As described earlier, frame-level features are aggregated with both first and second-order deltas, as well as the original features. Then, frame-level features are aggregated into textures using both the average and standard deviation. This aggregation strategy increases the number of features by a factor of 6.

\begin{table}[]
\centering
\caption{Feature Extraction Parameters}
\label{tab:feat_ex}
~\\
\begin{tabular}{@{}clc@{}}
\textbf{Feature Sets}         & \multicolumn{1}{c}{\textbf{Parameters}} & \textbf{Values}          \\ \midrule
\multirow{4}{*}{MEL-SPEC}    & Mel Bins                               & 128                      \\
                             & FFT Size                               & 2048                     \\
                             & Hop Size                               & 1024                     \\
                             & Window                                 & Hanning                  \\ \midrule
MEL-RP                       & Target Dim (M)                         & \{8, 26, 51, 75, 100\}   \\ \midrule
MEL-AE                       & Bottleneck Dim (H)                     & \{16, 32, 64, 128, 256\} \\ \midrule
\multirow{4}{*}{HANDCRAFTED} & FFT Size                               & 2048                     \\
                             & Hop Size                               & 1024                     \\
                             & MFCC Coeff.                            & 20                       \\
                             & Window                                 & Hanning                  \\ \bottomrule
\end{tabular}
\end{table}

\begin{table}[]
\centering
\caption{Number of Features and Texture Size per Feature Set}
\label{tab:n_features}
~\\
\begin{tabular}{@{}ccc@{}}
\textbf{Feature Sets}   & \textbf{Features} & \textbf{Texture Size} \\ \midrule
HANDCRAFTED             & 26                      & 156                   \\ \midrule
\multirow{5}{*}{MEL-RP} & 8                       & 48                    \\
                        & 26                      & 156                   \\
                        & 51                      & 306                   \\
                        & 75                      & 450                   \\
                        & 100                     & 600                   \\ \midrule
MEL-SPEC                & 128                     & 768                   \\ \midrule
\multirow{5}{*}{MEL-AE} & 16                      & 96                    \\
                        & 32                      & 192                   \\
                        & 64                      & 384                   \\
                        & 128                     & 768                   \\
                        & 256                     & 1536                  \\ \bottomrule
\end{tabular}
\end{table}

The Texture Selector parameters are shown in Table \ref{tab:tsparams}. Two baselines are used: FTS vectors and ALL textures. Since FTS implicitly assumes that the music track can be described by a roughly homogeneous texture, we expect it to be the lower baseline for all experiments. We also present the results using ALL textures to describe each track. This elevates computing costs, specially for the training phase of the SVM learning algorithm. In these cases, we used the ThunderSVM implementation \cite{wenthundersvm18} to train the SVM using a GPU. The ALL baseline gives us in idea of how effective our other two texture selection strategies are. However, we do not consider ALL to be the ceiling for texture selection classification performance. This is related to how some textures within a track can be outliers. %This is related to the same problem LINSPACE faces regarding false positive samples, since LINSPACE is just a downsampled version of ALL.

A single parameter is evaluated for both KMEANSC and LINSPACE: the number of textures used to describe each track. A low number such as 5 textures is used to evaluate which selector achieves good classification performance with lower computing power requirements. A good classification with few textures also gives us an idea of the relevance of the selected textures with respect to the target genre. We also evaluate the systems with an increasing number of textures. This allows us to determine the importance of the number of textures for genre classification. %será que falta falar alguma coisa aqui?

\begin{table}[]
\centering
\caption{Texture Selection Parameters}
\label{tab:tsparams}
~\\
\begin{tabular}{@{}clc@{}}
\textbf{Texture Selectors} & \multicolumn{1}{c}{\textbf{Parameters}} & \textbf{Values}                \\ \midrule
KMEANSC                   & \multirow{2}{*}{\# of Textures (K)}                     & \multirow{2}{*}{\{5, 20, 40\}} \\
LINSPACE                  &                                        &                                \\ \midrule
ALL                       & \multicolumn{1}{c}{--}                 & --                             \\ \midrule
FTS                   & Aggregation                            & MEAN+STDEV                     \\ \bottomrule
\end{tabular}
\end{table}

The Classifier Training procedure consists of feature selection and training the classifier. We chose to use a simple univariate correlation filter for feature selection. A Pearson correlation coefficient is calculated for each feature in relation to the corresponding training labels. Only the features corresponding to the highest scoring coefficients are used for training. This is known as ANOVA feature selection. The number of features to keep is a parameter that should be evaluated. We chose to keep a fraction of the total number of features. The fractions evaluated are shown in Table \ref{tab:anova}. The same features are then used for model testing later. Its important to keep in mind that our goal is to understand the effects of texture selection, not necessarily achieve the best possible results. With that in perspective, this feature selection filter is used because it is intuitive and well-known, thus simplifying analysis. % o fim não ficou tão bom, mas a idéia é não levar pau por ser um método simples de seleção de características.

\begin{table}[]
\centering
\caption{ANOVA Feature Selection Parameters}
\label{tab:anova}
~\\
\begin{tabular}{@{}clc@{}}
\multicolumn{1}{c}{\textbf{Parameters}} & \textbf{Values}     \\ \midrule
Fraction to Keep                     & \{0.2, 0.4, 0.6, 0.8\} \\ \bottomrule
\end{tabular}
\end{table}

Two learning algorithms are used for mapping textures into genres. The Support Vector Machine (SVM) is a well-known learning algorithm used in various machine learning music applications. It relies on mapping points into a higher dimensionality space where they become linearly separable by a hyperplane \cite{vapnik98}. Thus, by design, SVM is not only robust to high-dimensional data, but projecting data into high-dimensional spaces is part of its strategy to solve the pattern classification problem. Two regularization parameters responsible for softening the decision boundary are usually optimized, namely $C$ and $\gamma$. Table \ref{tab:svm_knn} shows the SVM parameters evaluated in the experiments. The main drawback of the SVM is the time complexity of its training algorithm, which depends on the number of training samples, the dimensionality of the training vectors and the type of kernel.

\begin{table}[]
\centering
\caption{Learning Algorithm Parameters}
\label{tab:svm_knn}
~\\
\begin{tabular}{@{}clc@{}}
\textbf{Learning Algorithm} & \multicolumn{1}{c}{\textbf{Parameter}} & \textbf{Values}               \\ \midrule
\multirow{3}{*}{SVM}        & C                                      & \{1, 10, 100, 1000\}          \\
                            & $\gamma$                               & $\frac{1}{n_\text{features}}$ \\
                            & Kernel                                 & rbf                           \\ \midrule
KNN                         & K                                      & \{1, 3, 5, 7, 9\}             \\ \bottomrule
\end{tabular}
\end{table}

The K-Nearest Neighbors (KNN) learning algorithm is another well-known learning algorithm. It uses a distance metric to measure similarity between points, in which similarity is inversely proportional to distance. Assuming that the feature sets capture relevant information, similar-sounding music excerpts yield points that are close in feature space. A commonly used distance metric is the Euclidean distance, which assumes that the weight of each feature is the same. KNN classifies an unknown point by calculating its $K$ nearest points. Then, a majority vote with the corresponding labels is used make the final decision. Table \ref{tab:svm_knn} shows the values for $K$ we evaluate in our experiments.

The main advantage of KNN is that it can be implemented efficiently. A distance tree can lower the nearest neighbor computation to a logarithmic time with respect to the number of training samples, thus making predictions very efficient. Thus, training consists of building the distance tree, which can be done efficiently. Another advantage is related to its interpretability, because euclidean distance can be immediately related to similarity. However, interpretability is conditioned on the interpretability of the individual features of the dataset. A drawback of KNN is that it suffers from the curse of dimensionality, resulting in degraded classification performance in high-dimensional spaces. 

\subsection{Results}
\label{sec:results}
% * Evaluation measures
%cross-validation. explain AE training/prediction protocol. (parameters, splits)
% we chose f1 scores weighed by class support as the evaluation measures for classification

The results presented in this section are F1-scores weighed by class support. We report average F1-scores and standard deviations across folds. The number of folds varies depending on the dataset, as explained in Section \ref{sec:datasets}. For each fold, 80\% of the training set textures are randomly selected for model training and 20\% are for validation of hyperparameters. The parameters evaluated for feature selection and for each learning algorithm are shown in Tables \ref{tab:anova} and \ref{tab:svm_knn}, respectively. Both the learning algorithm parameters and feature selection parameters were optimized at the same time. For both SVM and KNN, $N \times M$ models were evaluated for each (texture selection $\times$ feature set) combination, where $N$ is the number of classifier parameter combinations and $M$ is the number of feature selection parameter combinations.

To increase readability, we use the word ``significant'' exclusively to mean ``statistically significant''. Unless otherwise stated, we used a two-tailed paired Student's T-test to test for statistical significance between any two sample measurements. We used the p-value threshold of 5\% for rejecting the null-hypothesis. Also regarding statistical wording, we use the word ``comparable'' to mean ``a difference lower than two standard deviations''.

Autoencoders were trained for each fold independently. This was done to keep a consistent separation between training and testing data, just as training and testing data are separated to evaluate the classification system as a whole. The number of features are parameters in both MEL-RP and MEL-AE. For brevity sake we present the results for $M=75$ and $H=128$. In most cases, results improve as the number of features increases up until $M=75$ and $H=128$. After that, performance saturates and does not significantly increase any further.

Considering the stochastic behavior in the KMEANSC texture selector resulting from the randomness of the initial centroid candidates, all the experiments were performed three times. The classification performance difference was not statistically significant in any case, thus we report the median results among all three runs. 

To increase readability, we use the \texttt{St} notation to refer to a texture selector \texttt{S} selecting \texttt{t} textures per track. For instance LINSPACE5 indicates that LINSPACE is being used to select \texttt{5} textures. Another example is KMEANSC20+, which refers to the texture selection setups with KMEANSC as the texture selector with 20 textures or more.

% * Show classification results

%pq separamos entre músicas inteiras e trechos?
    %esperamos que em musicas inteiras as texturas não sejam homogêneas entre si.

Preliminary experiments showed us that there are considerable improvements on classification performance for full-length music tracks with texture selection. To determine whether the length of the tracks is relevant for improvement with texture selection regardless of the dataset, we also experimented on 10s samples with both ISMIR and LMD datasets, which are available as full-length tracks. This is the same length of the tracks in the notoriously difficult HOMBURG dataset. The 10s samples for both ISMIR and LMD are from the middle of each song. We call the sliced datasets ISMIR-10s and LMD-10s. The folds used to evaluate these datasets are the same as the folds used to evaluate their full-length equivalents, making comparison straightforward. For clarity, we present the results using full-length tracks first, followed by the results on short-length tracks.
    
%A) para músicas inteiras: (ISMIR + LMD)

%1)4(c,d), 5(c,d) Usar mais que uma textura por música é melhor que ivector. (exceto linspace 5 e alguns kmeansc 5) (de forma geral, kmeansc é significativamente melhor que linspace 5)
    % O efeito se replica nos 4 feature sets

\subsubsection*{Full-length track Datasets}

\begin{figure}[h!]
    \centering
    %\vspace{-10pt}
    \subfloat[SVM]{
        \includegraphics[scale=0.37]{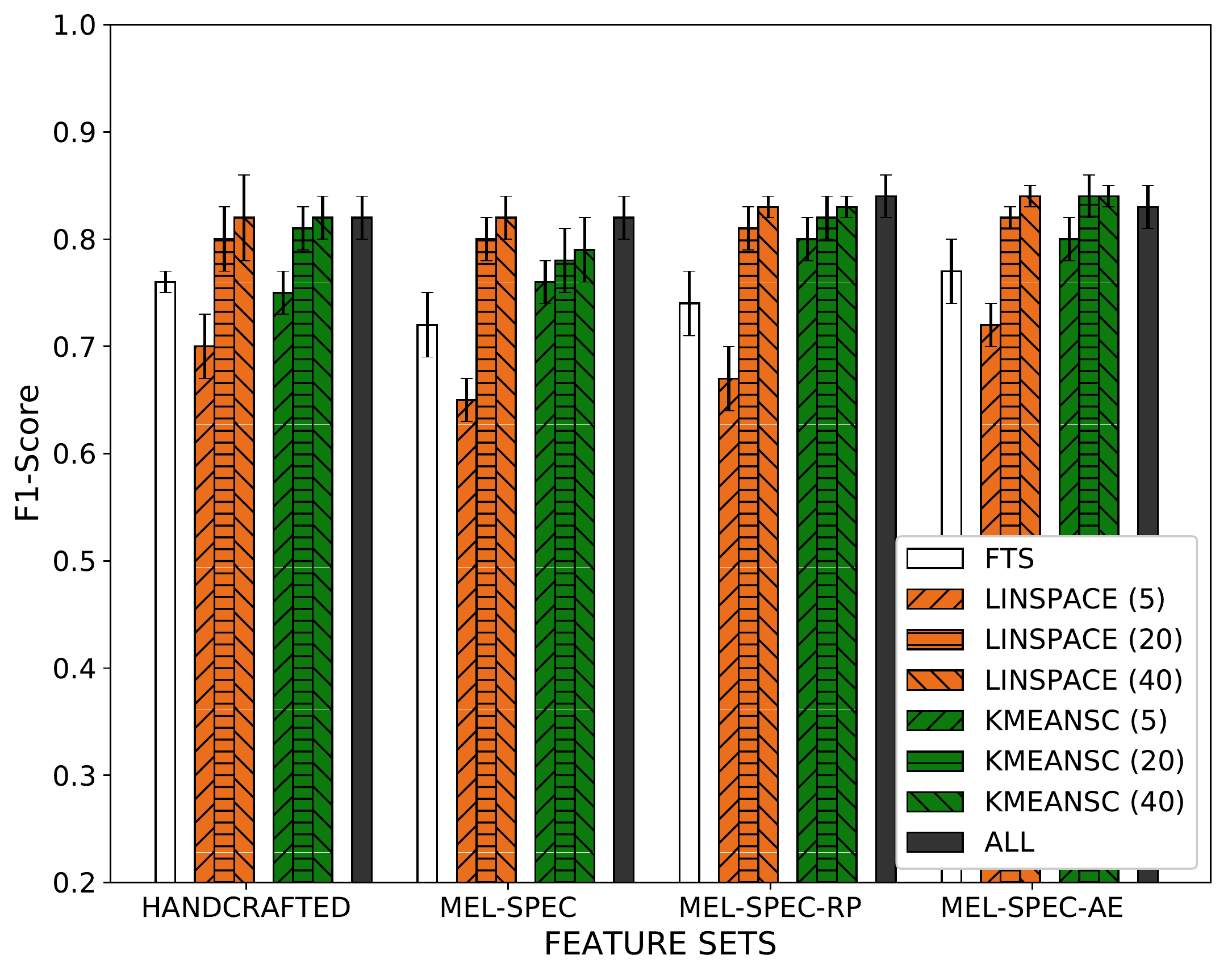}
    }
    \hspace{-10pt}
    \subfloat[SVM+ANOVA]{
        \includegraphics[scale=0.37]{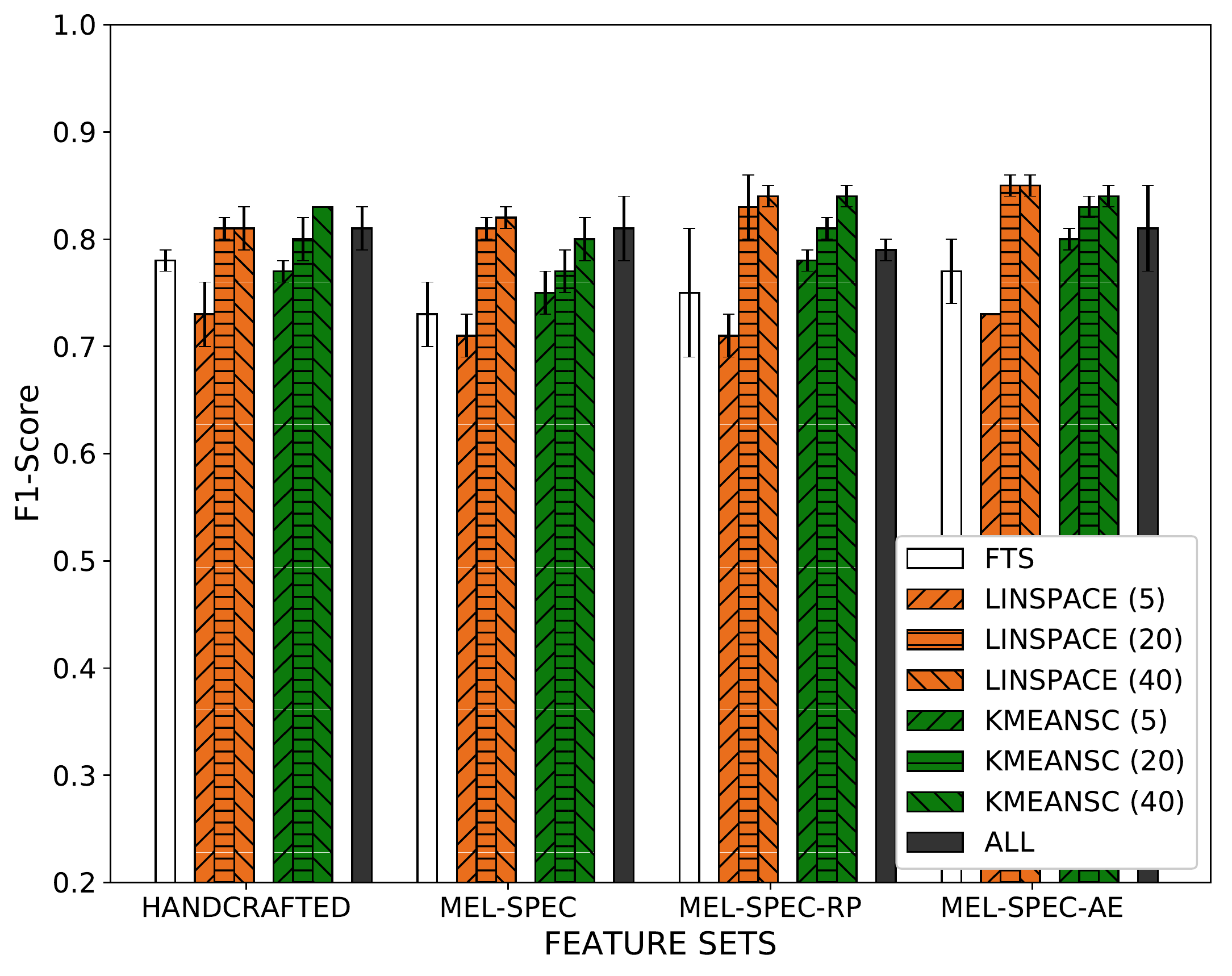}
    }
    \caption{Average F1-score with SVM across 3 Artist-Filtered Folds for the LMD Dataset} \label{fig:lmd_svm}
\end{figure}

\begin{figure}[h!]
    \centering
    %\vspace{-10pt}
    \subfloat[SVM]{
        \includegraphics[scale=0.37]{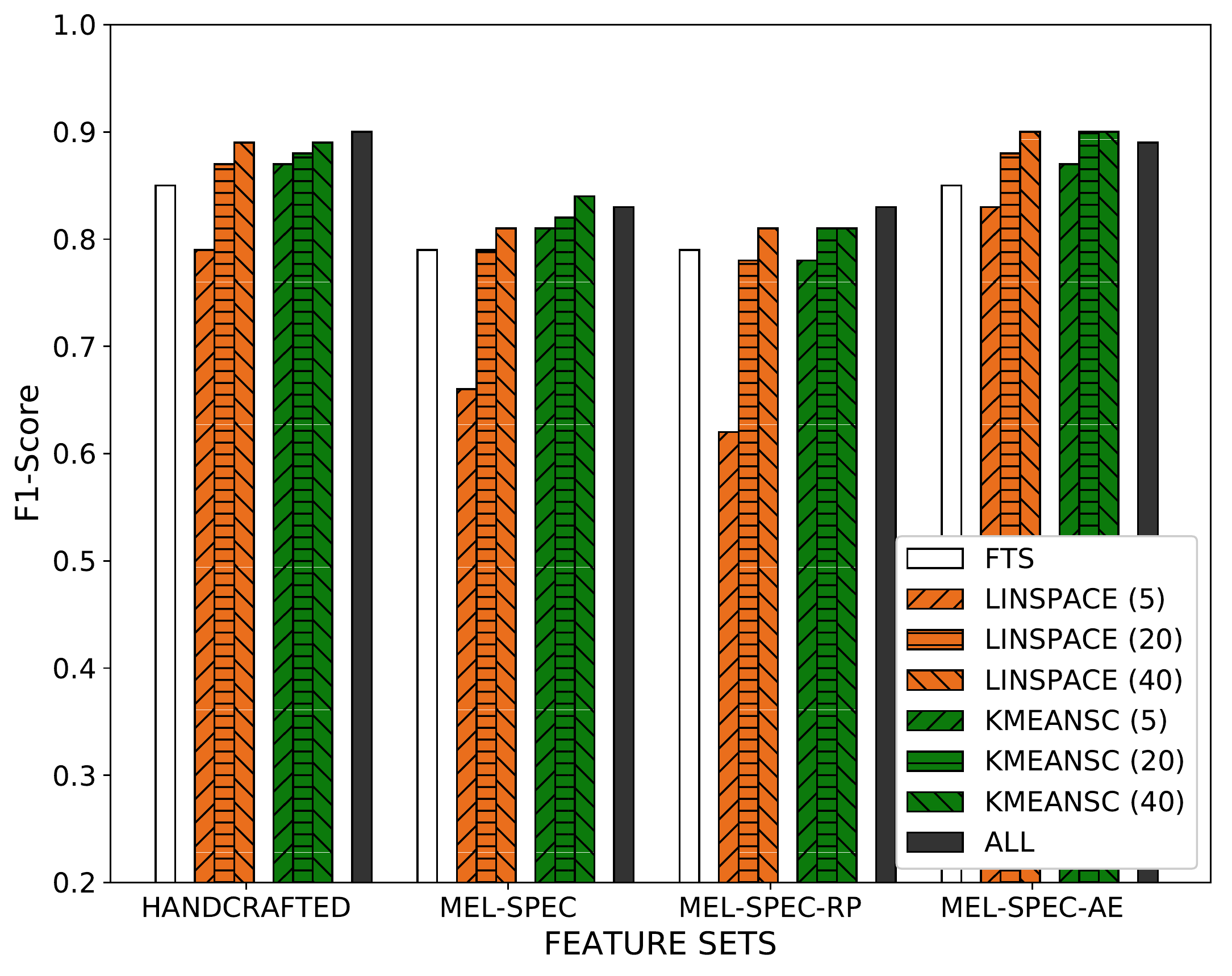}
    }
    \hspace{-10pt}
    \subfloat[SVM+ANOVA]{
        \includegraphics[scale=0.37]{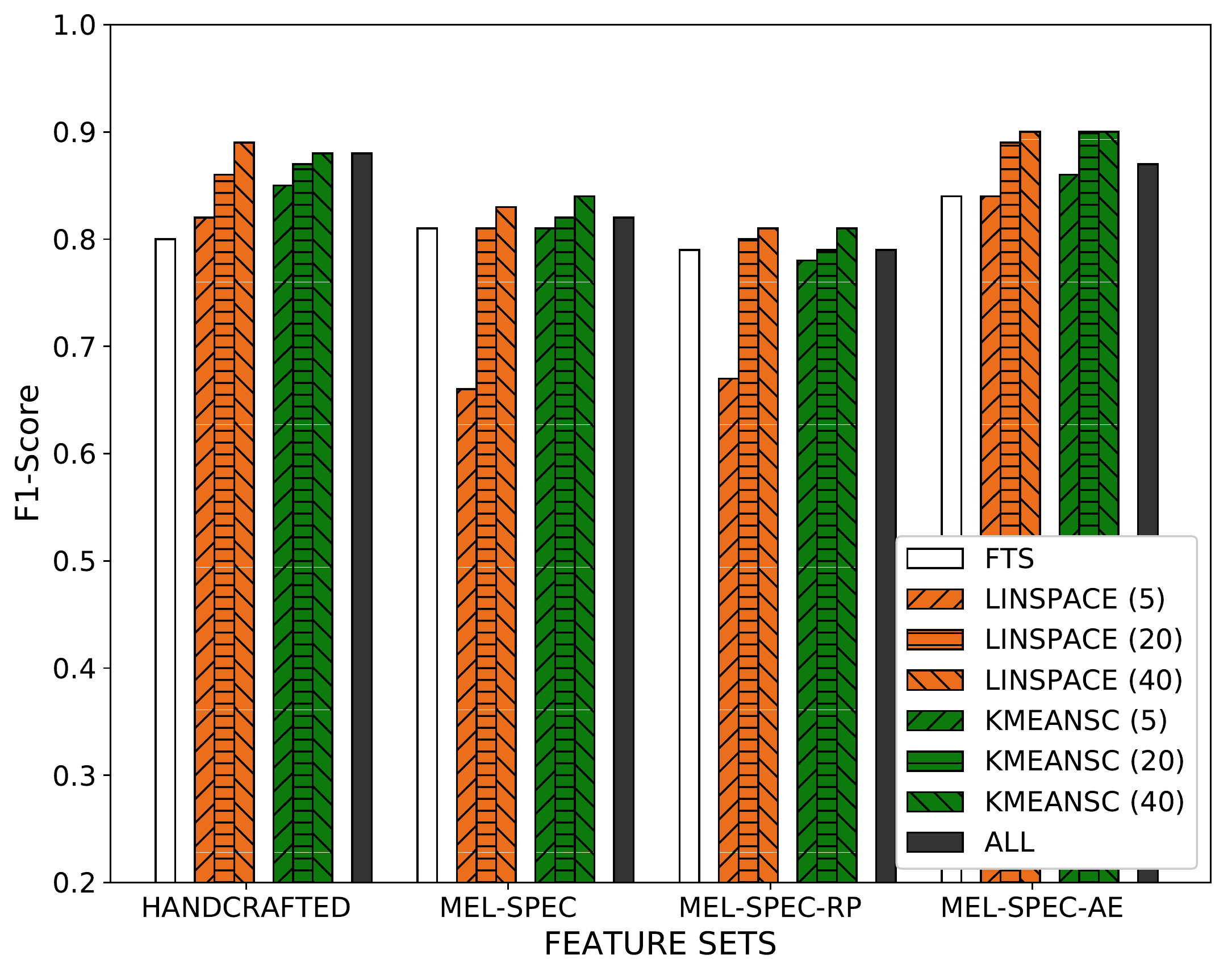}
    }
    \caption{F1-score with SVM for the ISMIR Dataset (ISMIR2004 contest train/test split)} 
    \label{fig:ismir_svm}
\end{figure}

The results with SVM for the datasets with full-length tracks, that is, LMD and ISMIR, are shown in Figures \ref{fig:lmd_svm} and \ref{fig:ismir_svm}, respectively. For both LMD and ISMIR, the results for KMEANSC and LINSPACE using more than 5 textures per track are significantly better than using FTS. This result applies for all feature sets, and the best results are similar across feature sets. 

For KMEANSC5 textures the results are comparable to the results with more than $5$ textures. On the other hand, the results with LINSPACE5 are significantly worse than FTS in almost all cases. This indicates that using more textures per track alone is not enough to improve classification performance. We discuss some possible reasons for this in Section \ref{sec:discussion}.

%2) Escolher texturas tem o mesmo resultado que usar todas as texturas. (tbm se replica em todos os FS)
    %(ISMIR: tem casos que é até melhor 5(b,c,d)-ae)
    %(LMD: melhor que all 4a-ae)

Figures \ref{fig:lmd_svm} and \ref{fig:ismir_svm} also show the results with ALL textures. For both KMEANSC20+ and LINSPACE20+, results are statistically the same as for ALL textures. Because training a SVM model with all textures is very expensive in terms of computing power, obtaining the same result with a fraction of textures is desirable. This also shows that it is possible to undersample tracks preserving essential information for classification. %As pointed out before, using all textures may feed the classifier with false positive samples. This may be the cause for it not leading to statistically better results than the other texture selection methods. 

The best results shown for both LMD and ISMIR are comparable to the state-of-the-art \cite{pohle2009,lim2012,seyerlehner2010,yandre2017,nanni2016}, suggesting that the results reached a glass ceiling. We could not observe statistically significant differences between the results yielded by KMEANSC20+, LINSPACE20+, and ALL.

It is worth highlighting that no significant difference was observed between the results yielded by KMEANSC20+ and LINSPACE20+ in both full-length datasets. Moreover, performance improves as the number of textures per track increases in both cases. However, there is a clear trade-off between computing power required for training and classification performance. In most cases, KMEANSC5 achieves results comparable to ALL in both datasets and for every feature set. Thus, it is reasonable to highlight its performance tradeoff advantage when compared to KMEANSC20+ or LINSPACE20+.

%3) vs seleção de features: (praticamente o mesmo efeito em todos os FS)
    %4(c,d) (LMD: ANOVA improves results further, specially with linspace. 
        % even lispace 5, although its still worse than ivector. 
        % Also improves ivector in all stances, but does not outperform using more than one texture (linspace20+, kmeansc20+)

    %5(c,d) (ISMIR: ANOVA  does not improve much, even for linspace. no improvement for ivector. )

Feature selection is commonly used in MGC systems. Therefore, we also evaluated the relationship between feature selection and texture selection regarding genre classification performance. Figures \ref{fig:lmd_svm}a and \ref{fig:ismir_svm}a shows the results for no feature selection (using all features), while \ref{fig:lmd_svm}b and \ref{fig:ismir_svm}b present the results with ANOVA feature selection. For LMD, feature selection seems to improve results for all texture selection strategies and feature sets. Specifically, feature selection significantly improves results for LINSPACE5, although they are still statistically worse than the results obtained with KMEANSC20+, LINSPACE20+ or ALL. Results for FTS are also improved in some cases. However, the feature selection improvements on FTS are not statistically significant. This suggests that the result improvement due to texture selection is greater than that obtained by using ANOVA feature selection in the LMD dataset.

For ISMIR, feature selection led to positive improvements with LINSPACE5. However, none of the LINSPACE5 results are comparable to the ones achieved with the remaining texture selectors. This behavior is observed both for the results with and without feature selection.

Even in the worst cases feature selection does not significantly decrease performance compared to the cases where no feature selection takes place. There are also some cases where feature selection improves results along texture selection. This suggests that texture and feature selection are complementary towards classification in both full-length datasets evaluated.

%4) Os resultados 1-3 se repetem de forma geral para o classificador KNN (ver apêndice).

Results with the KNN learning algorithm for both LMD and ISMIR are shown in Figures \ref{fig:lmd_f1} and \ref{fig:ismir_f1}, respectively. As with SVM, results improve as the number of textures per track increases for both KMEANSC and LINSPACE with all four feature sets. LINSPACE5 does not improve results when compared to FTS, as with SVM. 

Without feature selection (Figure \ref{fig:lmd_f1}a), in contrast to SVM, the results with KMEANSC20+ for LMD are better than using ALL textures. Specifically, KMEANSC20+ is significantly better than ALL with the MEL-SPEC-AE feature set. The KNN algorithm is known for being less robust to outliers than SVM. As argued before, using ALL textures may include textures that are not typical of the genre. These untypical textures can be considered outliers, degrading KNN performance. We explore this outlier problem further in Section \ref{sec:discussion}.

Overall, there is a clear improvement in all results with KNN and ANOVA feature selection for the LMD dataset (Figure \ref{fig:lmd_f1}b). Apart from selecting features that are more linearly correlated to the output of the training set, ANOVA also lowers the texture dimensionality. The KNN algorithm is known for performing poorly with high dimensional data. In high dimensionality, the density of point clouds becomes too low, with many points becoming equidistant, thus degenerating euclidean distance. Thus, lowering texture dimensionality with a feature selection algorithm such as ANOVA can improve KNN performance.

For the ISMIR dataset, performance improvements due to feature selection are more modest than with LMD. There is a clear improvement in LINSPACE5, although it still does not achieve the results of the other texture selectors. However, the results with feature selection are all comparable to the results with no feature selection. Thus, feature selection had no significant for this dataset.

In general, the results with SVM are better than with KNN in both LMD and ISMIR. More importantly, the improvement patterns related to the number of textures and texture selectors are the same for both classifiers. In other words, for both KNN and SVM, the results improve as the number of textures used to describe each track increases, except for LINSPACE5+. The results for KMEANSC5+ are superior to LINSPACE5+ for both learning algorithms. Lastly, the results for both LINSPACE20+ and KMEANSC20+ are better or statistically the same as using ALL. This suggests that the improvement from texture selection in full-length datasets is not tied to a particular classifier and can also happen with other classifiers.

\subsubsection*{Short-length track Datasets}

%B) para trechos:

%1) HOMBURG
    % não há melhora em usar mais que uma textura por faixa (ivector-> multiplas texturas) em qualquer dataset
    % não há melhora significativa em usar seleção de features (exceto svm ivector mel, mel-rp)
    % discutir que como o clipe é muito curto, a textura é praticamente homogênea, logo não tem muita diferença escolher texturas ou fazer ivector
    
Results with both KNN and SVM for the HOMBURG dataset are shown in Figure \ref{fig:homburg_f1}. In contrast to ISMIR and LMD, which are full-length tracks, the tracks in HOMBURG are only 10s long. The results show that there is no improvement with multiple textures when compared to FTS. Also, we could not observe significant result differences between LINSPACE and KMEANSC in any case, for all feature sets evaluated. We could also observe no significant improvement resulting from feature selection. Therefore, the trends in classification improvement due to texture selection in the full-length datasets are not present in the HOMBURG dataset.

Because the tracks in the HOMBURG dataset are only 10s long, the corresponding textures are more likely to be homogeneous throughout each track. One of the main assumptions with bag of frames approaches is that music tracks are made of a variety of heterogeneous textures. Multiple textures per track are able to describe the varied nature of music. Furthermore, genre models inferred from a greater variety of textures are more likely to represent a more accurate collection of typical textures per genre. With short tracks this heterogeneity assumption is less likely to be true.

%tbm testamos ISMIR e LMD com trechos de 10s. Isso nos permite avaliar se é o comprimento dos clipes ou o conteúdo.
%2) pra mostrar o ponto anterior, também usamos ISMIR e LMD, so que agora só com trechos de 10s do meio das músicas.

We tested this hypothesis by building two additional datasets from ISMIR and LMD, in which we usedonly 10s of the middle of each track. We called the resulting datasets ISMIR-10s and LMD-10s. The results for these shortened datasets allows us determine whether track length is a relevant factor that deters improvement with multiple textures. Because HOMBURG is only available in the 10s form, we are not able to do evaluate it with full tracks. %Thus, evaluating ISMIR-10s and LMD-10s is important to verify the texture selection improvement reliance on longer tracks.

    %os resultados médios caem tanto no LMD quanto no ISMIR.
    
    %LMD: não há diferença significativa entre usar ivector e seleção de texturas! seleção de feats com KNN parece melhorar (melhora só por conta da maldição da dim?).
    %LMD: o resultado com MEL-SPEC-RP não é tão bom quanto os demais. Também, o resultado com ALL é muito melhor rs. (propriedades de media e variancia não são estaveis em RP. no LMD isso é mais importante)
    
    %ISMIR: tbm não há diferença significativa entre usar ivector e seleção de texturas. não há diferença em fazer seleção de features tbm.
    
    %ou seja, de forma geral, HOMBURG, LMD-10s, ISMIR-10s todos enfrentam o problema de não melhorar o baseline.    
    
Figures \ref{fig:lmd_f1}e-h and \ref{fig:ismir_f1}e-h shows the results for LMD-10s and ISMIR-10s, respectively. When compared to the cases where the full-length tracks were used, shown again in \ref{fig:lmd_f1}a-d and \ref{fig:ismir_f1}a-d, the results for both LMD-10s and ISMIR-10s are significantly lower. Furthermore, with SVM in both LMD-10s and ISMIR-10s there is no improvement in F1-score as the number of textures increases.There is also no clear improvement with ANOVA feature selection, except for KNN in LMD-10s. This improvement is most likely related to dimensionality reduction. The results with SVM are significantly better than KNN in all cases for both LMD-10s and ISMIR-10s.

Overall, there was no improvement due to texture selection for any of the three datasets with short tracks. There is also no improvement due to ANOVA feature selection. In fact, none of the improvements shown with full-length tracks are present in the shortened datasets. This suggests that the improvement due to texture selection is dependent on track length. The fact that we evaluated texture selection systems with full-length and 10s tracks of both LMD and ISMIR further provides evidence that texture selection works best with longer tracks.

%C) Trechos de 30s:

We also evaluated texture selection on the GTZAN dataset. This dataset is comprised of 30s tracks. The results for GTZAN without an artist filter (GTZAN-RANDOM) are shown in Figure \ref{fig:gtzan_random_f1}. Similar to the case where full-length tracks were used, the results with SVM (Figure \ref{fig:gtzan_random_f1}c) improve as the number of textures increases. This is true for all feature sets. All feature sets reach comparable results. However, contrasting with the full-length tracks, the difference between LINSPACE5 and KMEANSC5 is no longer significant in every case. However, the average result is still better with KMEANSC5.

The results for GTZAN-RANDOM with ANOVA and SVM are shown in Figure \ref{fig:gtzan_random_f1}d. ANOVA seems to significantly improve results, mostly for FTS and LINSPACE5. Specifically, LINSPACE5 became significantly better than FTS for both MEL-SPEC and MEL-SPEC-AE. In most cases FTS performs significantly worse than the remaining texture selectors, except for HANDCRAFTED and MEL-SPEC-RP.

The results for GTZAN-RANDOM with KNN are shown in Figures \ref{fig:gtzan_random_f1}a,b. In general the results are significantly worse than SVM for every experiment. However, the same trend relating an increase in performance along with the increase on the number of textures is also present. Once again, this suggests that the effect of texture selection may not be dependent on the classifier. Furthermore, the effects of feature selection are more pronounced with KNN. This, along with the improved seen in some cases with SVM, shows that texture selection can be used along feature selection and may bring further improvements to classification performance. 

As stated previously, we have also evaluated texture selection with an artist-filtered version of the GTZAN dataset (GTZAN-ART). Figure \ref{fig:gtzan_artf_f1} shows the results for GTZAN-ART. The F1-scores are lower than GTZAN-RANDOM in all cases for both SVM and KNN. However, the trend of improvement in classification performance as the number of textures increases is shown in Figure \ref{fig:gtzan_artf_f1}c. However, the results with texture selection are not significantly higher than the FTS baseline. A higher variance compared to GTZAN-RANDOM across folds in all experiments shows that the artist filter makes generalization more difficult in GTZAN. This higher variance makes the texture selection results in GTZAN-ART not significantly better than the baseline, although the average results are better.

The results for GTZAN-ART with KNN are shown in Figures \ref{fig:gtzan_artf_f1}a,b. Just as with GTZAN-RANDOM, the results with SVM are superior to KNN. Again, the classification performance improvements are also present as the number of textures increases. Feature selection improves KNN results more than SVM, although the improvements are not significant in most cases.

The positive result for texture selection in both GTZAN-RANDOM and GTZAN-ART is interesting because it shows that it may lead to gains in cases where the entire track is not available. In the case of GTZAN-RANDOM, the gains were significant compared to the FTS representation.

%conclusão: trechos vs musica inteira. Seleção de texturas em músicas inteiras dá resultados superiores do que seleção de features, embora as 2 coisas pareçam ser complementares. Por isso, propomos incorporar seleção de textures no processo de classificação. 

The presented results shows that the effectiveness of texture selection is dependent on the length of the tracks in the training dataset. When track representations are heterogeneous with respect to their respective textures, they are more likely to capture different variations of typical genre sounds. This can lead to classification models that capable of greater generalization. 

Texture selection had the largest positive impact on datasets made up of full-length tracks. The improvement due to texture selection increases as the number of textures per track increases. This improvement was observed in four different feature sets of varying complexity. In most cases, KMEANSC5 lead to results not significantly different than, or comparable with the results obtained with more than 5 textures. Furthermore, we also showed that in most cases LINSPACE5 does not perform significantly the same as KMEANSC5, and did not improve results over the FTS baseline. These two factors suggest that our proposed K-Means based texture selection technique is a promising candidate for texture selection. It offers a good compromise between additional computing power needed for training models and classification improvement.

We have also shown the interaction between texture selection and univariate feature selection based on ANOVA. Feature selection and texture selection worked in a complementary fashion in most cases. This suggests that using both techniques together can improve results in other full-length datasets.

Considering the positive results achieved with full-length track datasets, we propose evaluating texture selection in future music genre classification systems. Just as feature selection has been shown to greatly improve music classification systems, texture selection may also provide classification improvement at a relatively low extra cost during system training and testing.

In the next section we present qualitative results and discussions aimed towards a better understanding on the effects of texture selection for music genre classification. 

\section{Qualitative analysis}
\label{sec:discussion}

% * This is for qualitative research questions (does frame selection point to meaningful frames)?
% * PCA/T-SNE figures here
% * Audio examples are here, too!
% * Discussion: are the selected frames, in fact, meaningful?
% * Paragraph discussing how this work impacts the current state of the art (no citations here!)
% * Discussion: works better because it is ensemble learning or because it is a good frame selection method?

%O que o KMEANS de fato agrupa? Paletas. Mostrar audios. (exemplo de sumarização)
This section analyzes the feature selection processes in an example audio excerpt. First, we use Principal Component Analysis (PCA) to identify the effects of different texture selection processes in the representation of a single track, as shown in Section \ref{sec:projPCA}. After that, we use t-SNE unsupervised manifold learning to identify the effects of texture selection in full datasets, as discussed in Section \ref{sec:tsne}.

\subsection{Single Track Texture Selection}
\label{sec:projPCA}
\begin{figure}[h!]
    \centering
    \subfloat[ALL]{
        \includegraphics[scale=0.43]{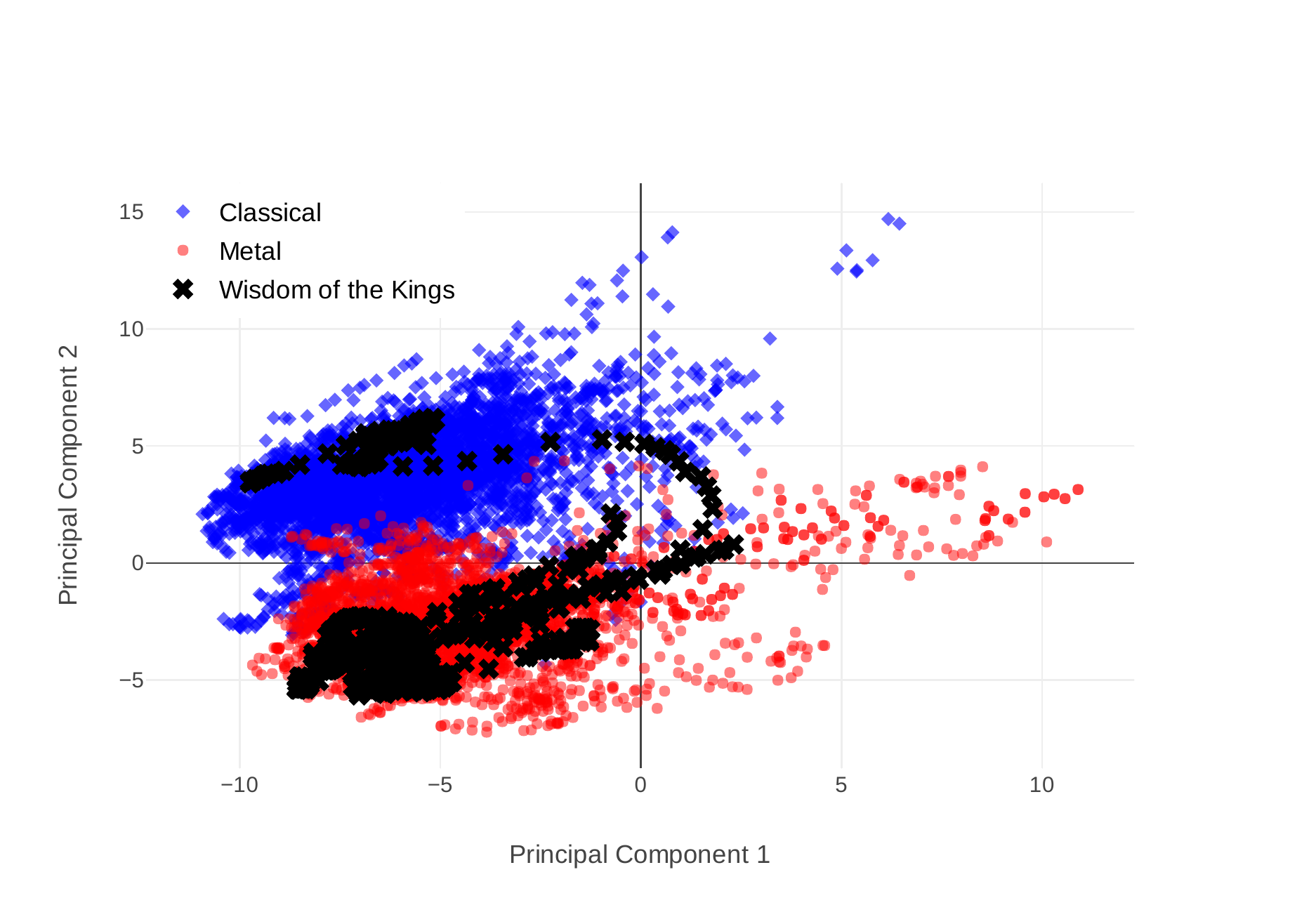}
        \label{fig:pca_all}
    }
    \hspace{-10pt}
    \subfloat[FTS]{
        \includegraphics[scale=0.43]{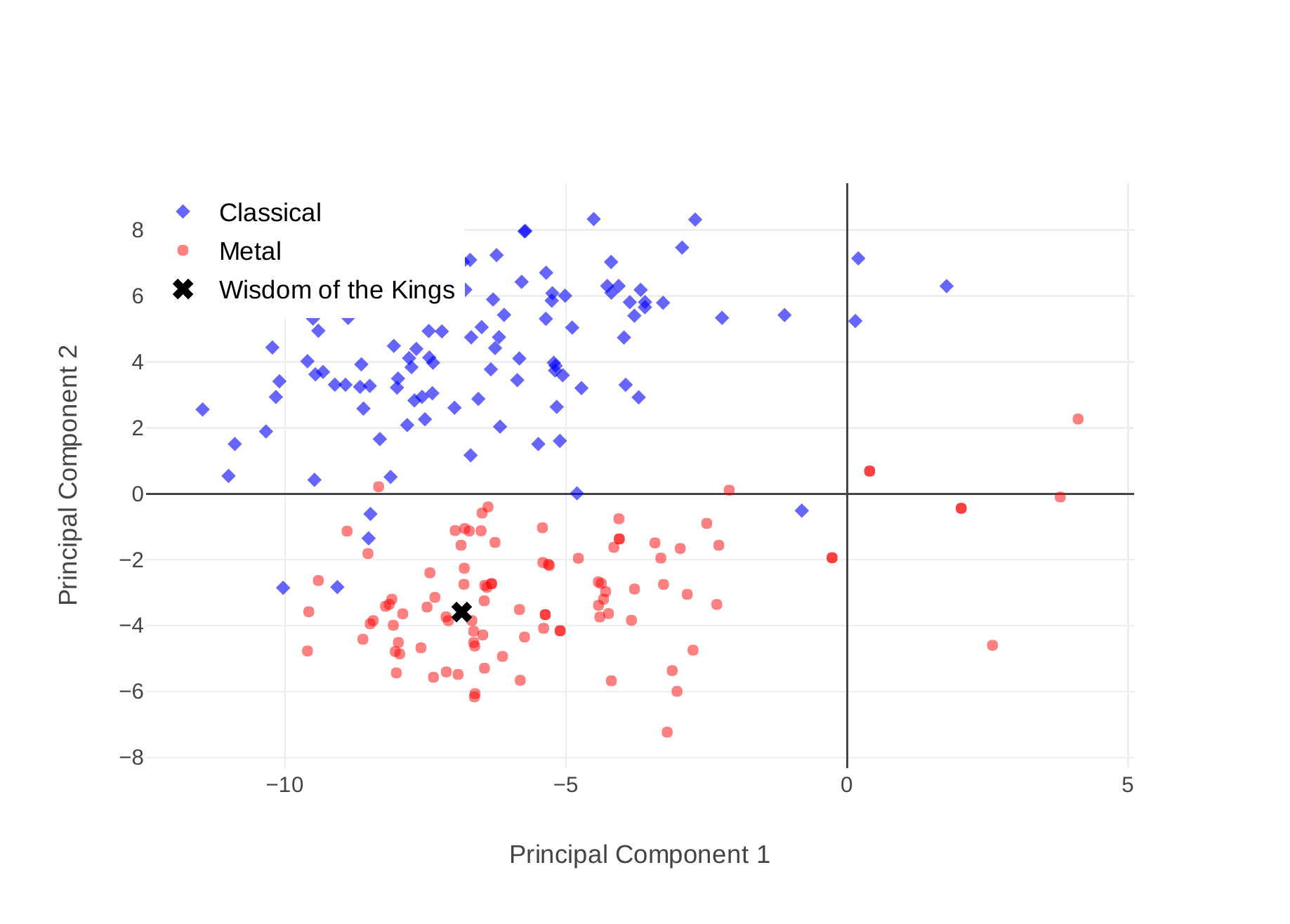}
        \label{fig:pca_fts}
    }
    \\
    \vspace{-10pt}
    \subfloat[KMEANSC 5]{
        \includegraphics[scale=0.43]{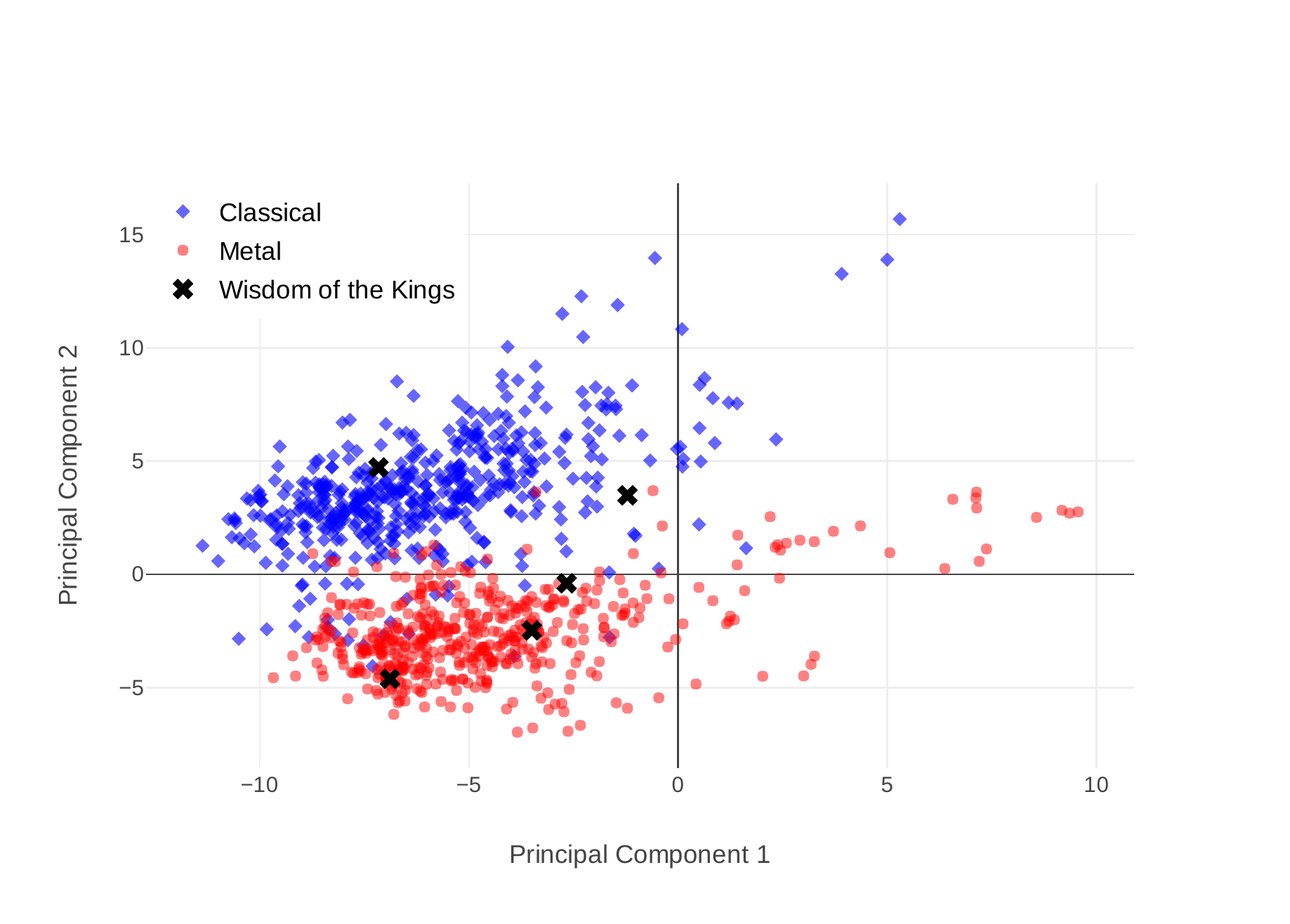}
        \label{fig:pca_kmeansc5}
    }
    \hspace{-10pt}
    \subfloat[LINSPACE 5]{
        \includegraphics[scale=0.43]{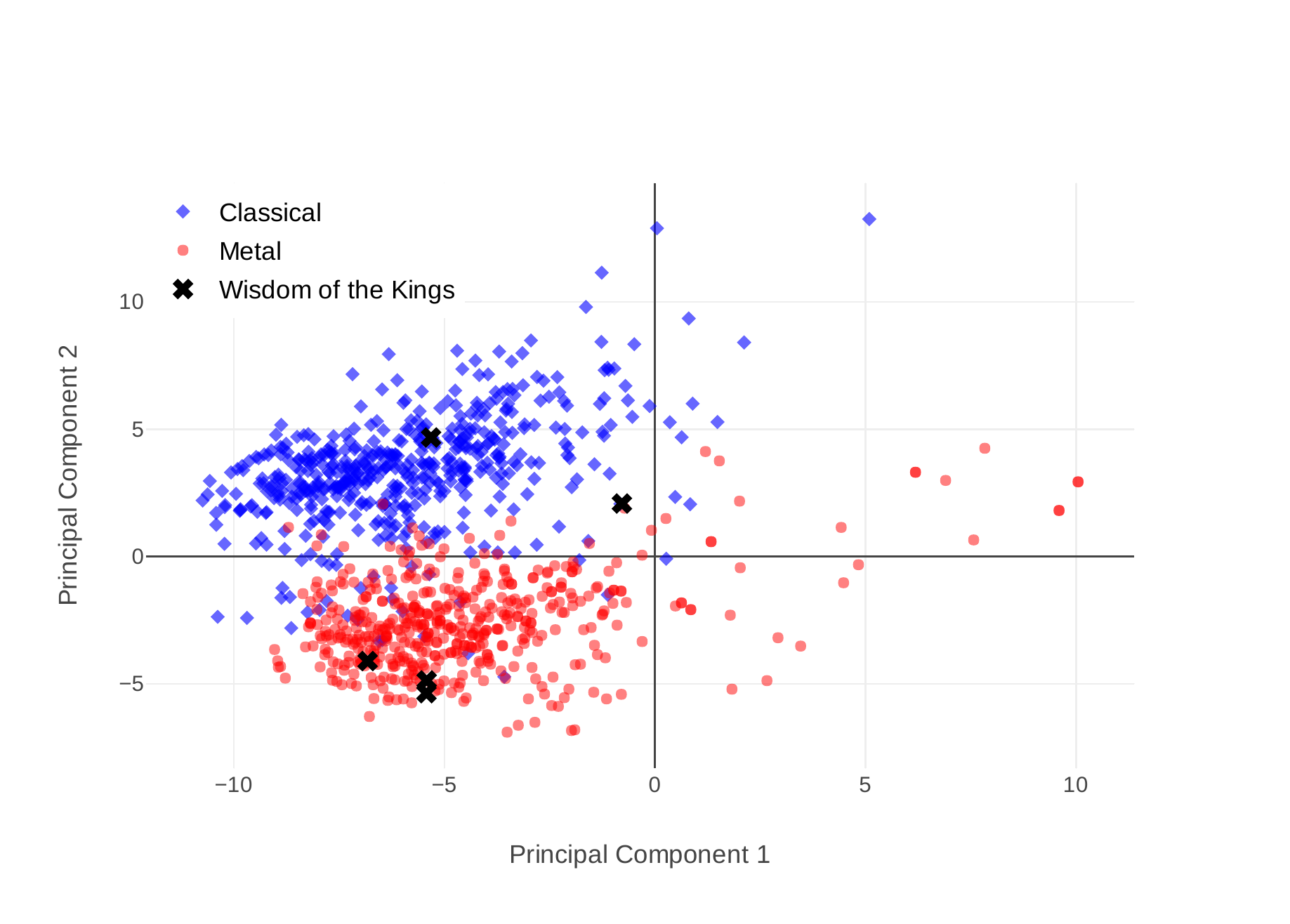}
        \label{fig:pca_linspace5}
    }    
    \caption{PCA projections of the Texture Selection Approaches over the GTZAN dataset (HANDCRAFTED features)} \label{fig:pca_ts}
\end{figure}

In this section, we show the effects of feature selection in the representation of single tracks. For such, we used PCA projections of textures textures calculated from the GTZAN dataset using the handcrafted feature set. This allows visualizing the projections related to a single track over a genre-related point cloud, as shown in Figure \ref{fig:pca_ts}.

The discussions conducted here use the track ``Wisdom of the Kings'', as played by the Italian Heavy Metal band Rhapsody. This track was chosen because it has both typical Heavy Metal parts, with fast drums and guitars, and interludes composed in Classical style. As shown in Figure \ref{fig:pca_all}, the textures of the track span over both the Metal and the Classical clouds.

The FTS representation shown in Figure \ref{fig:pca_fts} correctly positions the track close to the Metal genre. However, it discards the information related to the Classical parts. This indicates that FTS representations can fail on representing genre fusion within musical tracks, regardless of their effectiveness for the classification task.

This genre fusion information was preserved both for the K-Means selection (Figure \ref{fig:pca_kmeansc5})  and the linear downsampling (Figure \ref{fig:pca_linspace5}) representations. It is interesting to note that the linear downsampling selected more textures that are part of the most used textures, whereas the K-Means selection selected textures that are more distributed along the space. This indicates that K-Means texture selection is more effective to represent a texture palette that was used in each track.

This means that K-Means texture selection can be used to summarize songs. For such, it is possible to select one texture from each of the K clusters and then concatenate them in a single track. This resulted in Audio Sample 1 \footnote{\url{http://bit.ly/2Xabeob}}, presented as supplementary material.  It is possible to use linear downsampling to generate another track summary, as shown in Audio Sample 2 \footnote{\url{http://bit.ly/2Qsgl0m}}. However, this result is generated independently of audio content. This means that the chosen textures are probably those that are used more often in the track, not those that best represent the track diversity.

The next section discusses the effects of using different texture selection methods in the whole dataset.

\subsection{Dataset Texture Selection}
\label{sec:tsne}
\begin{figure}[h!]
    \centering
    \subfloat[KMEANSC 5 (full)]{
        \includegraphics[scale=0.43]{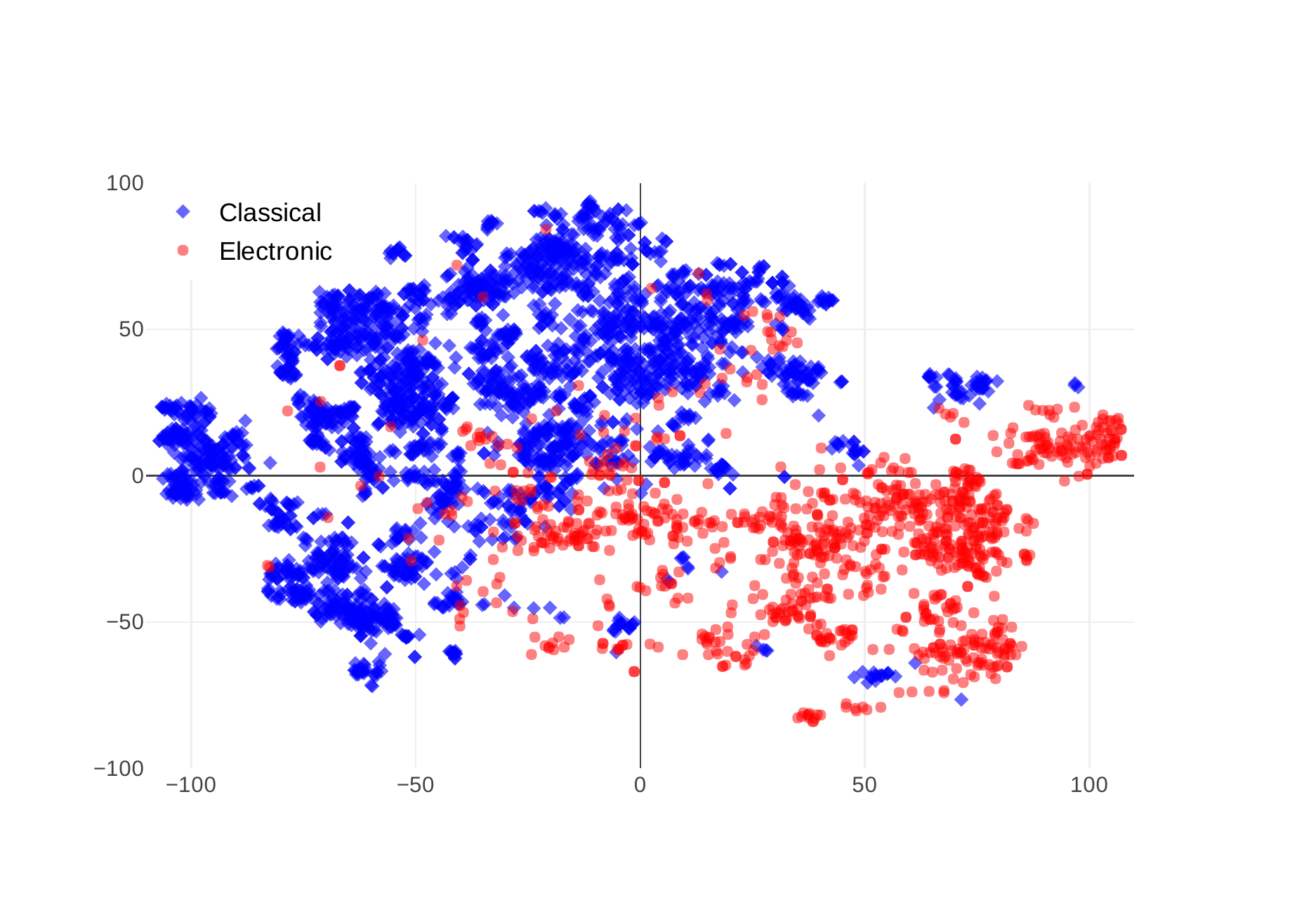}
        \label{fig:tsne_k5f}
    }
    \hspace{-10pt}
    \subfloat[LINSPACE 5 (full)]{
        \includegraphics[scale=0.43]{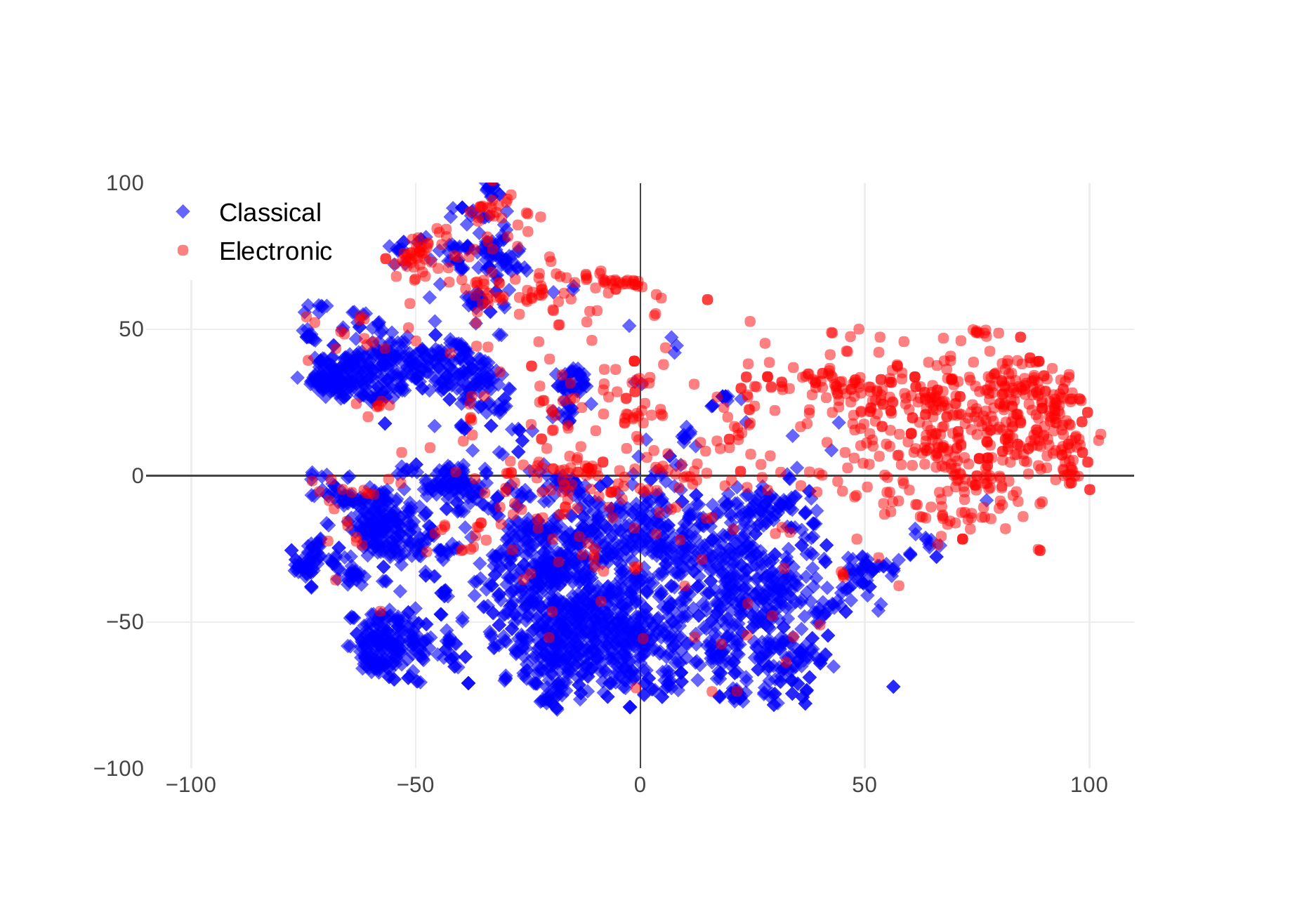}
        \label{fig:tsne_l5f}
    }
    
    \subfloat[KMEANSC 5 (10s)]{
        \includegraphics[scale=0.43]{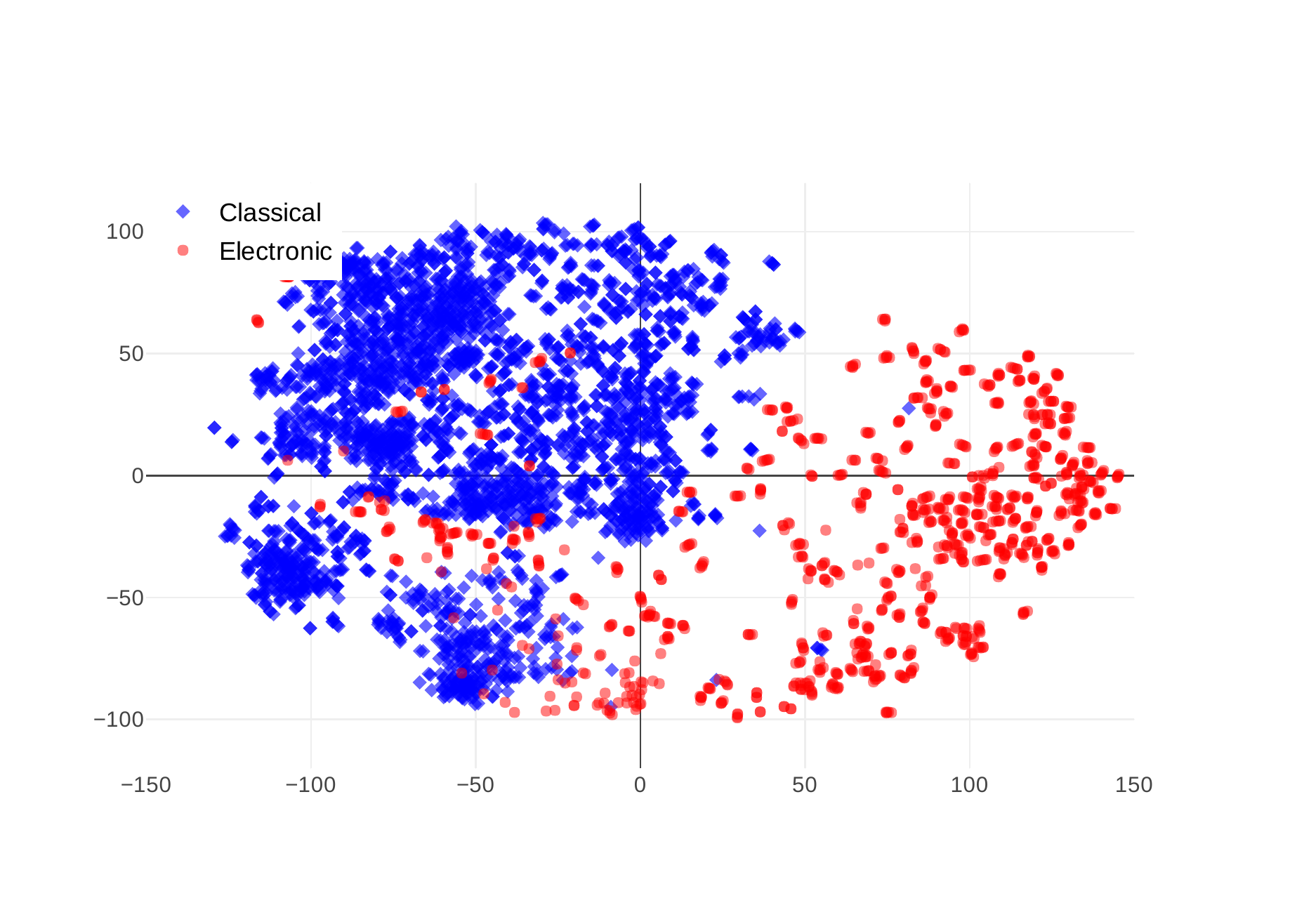}
        \label{fig:tsne_k510}
    }
    \hspace{-10pt}
    \subfloat[LINSPACE 5 (10s)]{
        \includegraphics[scale=0.43]{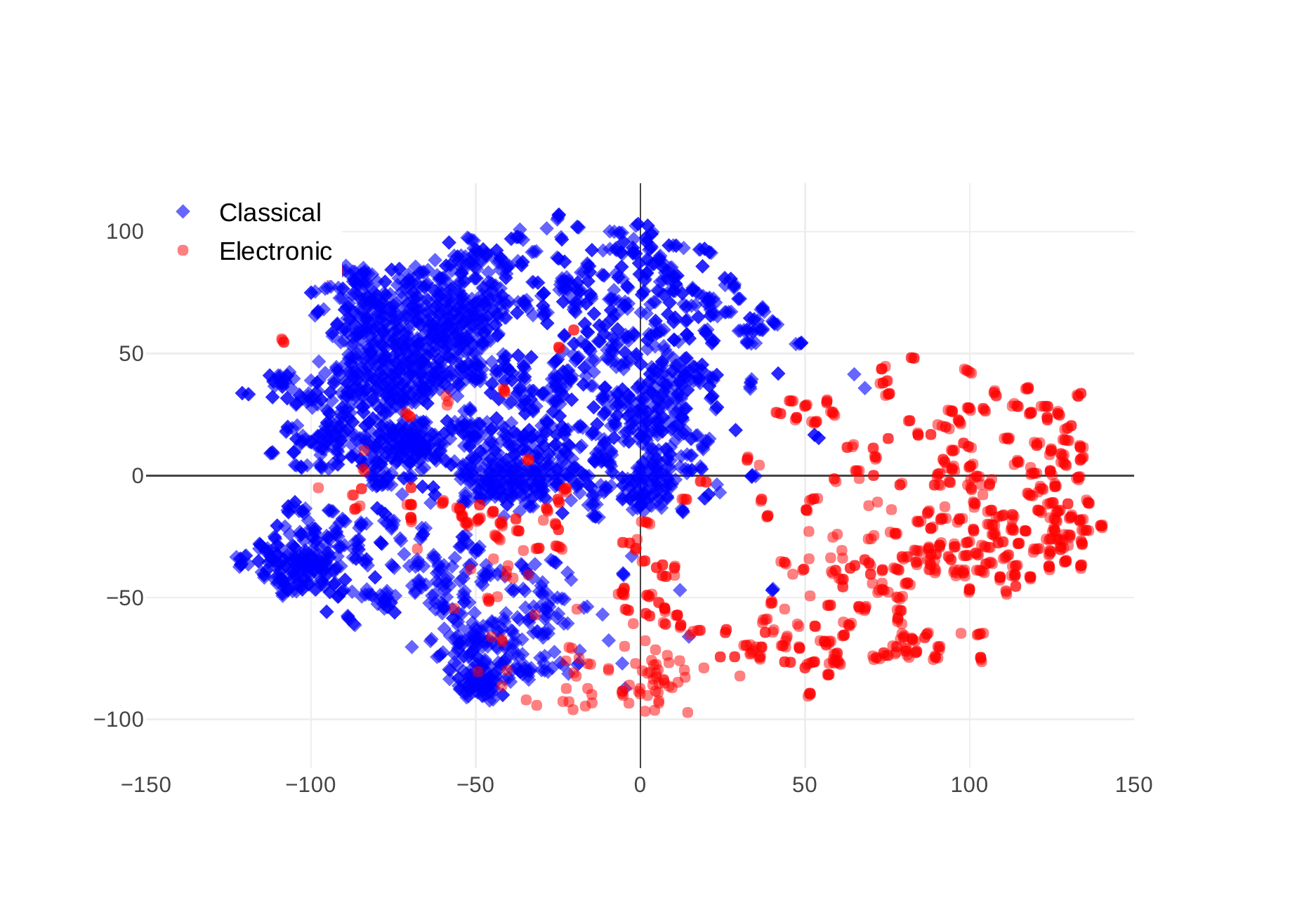}
        \label{fig:tsne_l510}
    }        
    
    \subfloat[KMEANSC 20 (full)]{
        \includegraphics[scale=1.1]{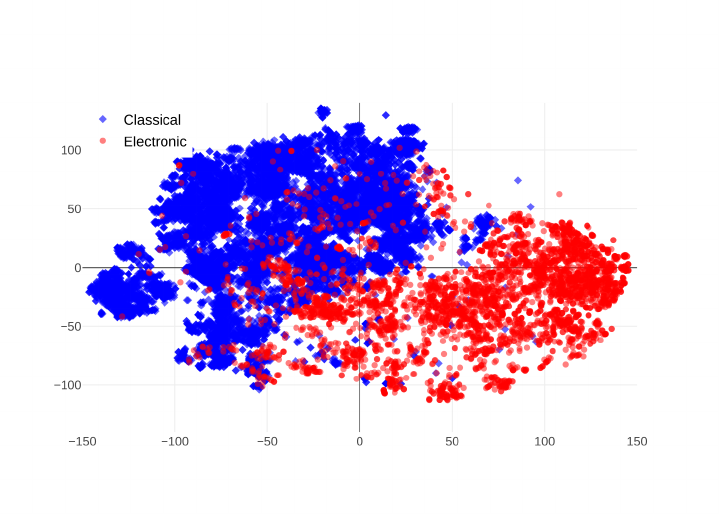}
        \label{fig:tsne_k20f}
    }
    \hspace{-10pt}
    \subfloat[LINSPACE 20 (full)]{
        \includegraphics[scale=1.1]{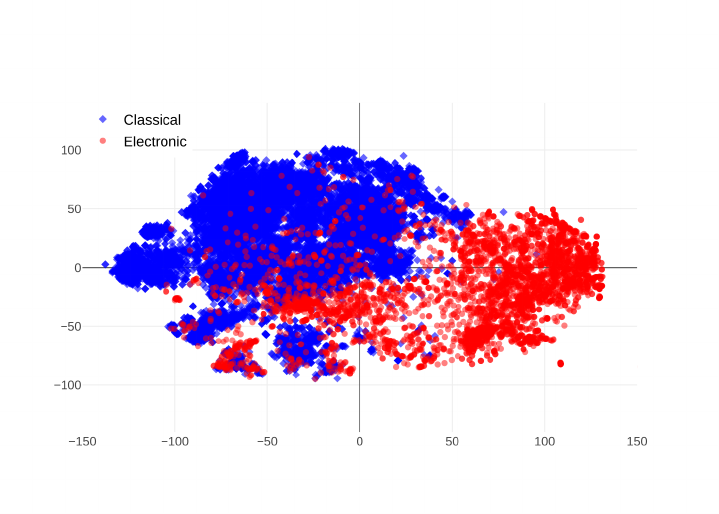}
        \label{fig:tsne_l20f}
    }    
    \caption{t-SNE embeddings depicting feature spaces derived from texture selection over the ISMIR dataset (HANDCRAFTED features)} \label{fig:tsne}
\end{figure}

This section discusses the effects of feature selection in the representation of datasets. For such, we used t-SNE projections of textures calculated from the ISMIR dataset using the handcrafted feature set. We visualize the projection related to the Classical and Electronic genres, as shown in Figure \ref{fig:tsne}.

A comparison between the spaces spanned by selecting 5 textures with K-Means (Figure \ref{fig:tsne_k5f}) and with linear downsampling (Figure \ref{fig:tsne_l5f}) shows that the K-Means selection leads to a higher amount of observable clusters. This means that the selected textures are more diverse, which suggests that they can be more informative when training classifiers. This difference disappears when 10s excerpts are used instead of full tracks. This behavior, shown in figures \ref{fig:tsne_k510} and \ref{fig:tsne_l510}, indicate that the texture selection method is less relevant when the audio tracks contain less information. As a result, there is no observable difference between the results yielded by each method in datasets comprising short excerpts, as discussed in Section \ref{sec:results}.

When more textures are selected from each track, the corresponding clusters tend to grow and, eventually, merge. This is shown in figures \ref{fig:tsne_k20f} and \ref{fig:tsne_l20f}. Nevertheless, the clusters related to linear texture downsampling seem to merge faster than the ones related to K-Means selection.

These results indicate that texture variety is more relevant than the number of textures used to build classification models. The lack of textural variety can be a result of either the texture selection method, such as a small $k$ in LINSPACE, or textural homogeneity within tracks, such as datasets consisting of short clips. Both of these causes lead to less variety, which has a harmful effect to system generalization.

The next section presents conclusive remarks.

\section{Conclusion}
\label{sec:conclusion}
% * One paragraph per research question, answering it using data gathered in this work
% * Another paragraph with closing remarks, noting the limitations of the work
% * Paragraph discussing other possible applications for the method and indicating directions for future work (ex: music summarization)

In this work we presented a study about the effect of texture selection in a variety of machine learning setups. We presented a novel texture selection method based on K-Means clustering, aiming to select diverse typical textures from each track. We evaluated four feature sets at different levels of abstraction, as well as a univariate feature selection strategy. Four publicly available datasets were used in the expriments. Our results show that texture collections improve classification results on the datasets comprising full-length tracks when compared to a full-track statistics (FTS) vector representation. %These texture collections may be the full set of textures, as well as a subset chosen by a texture selection algorithm. 

The results showed that our K-Means texture selection is able to achieve significant improvements over the FTS baseline using only 5 textures per track. The baseline strategy of selecting linearly-spaced textures required at least 20 textures to achieve similar results. K-Means texture selection has shown to reduce the number of textures needed to train models  while maintaining the same performance as using all textures from every track. %Our results also show that texture selection is complimentary to feature selection, although the effects of texture selection seems to be higher in the datasets evaluated.

We also showed that texture variety in the training set is key to improving classification performance. Through a qualitative analysis based on PCA an t-SNE projections, we showed that KMEANSC extracts diverse textures from each track, while LINSPACE is more likely to select textures similar to more common sounds within the track. As a result, LINSPACE needs more textures to be able to capture texture variety and achieve a significant performance improvement.

The results discussed in this article can be used for building automatic music summaries. Also, they can be used as basis for further exploration in texture selection methods for audio and music classification. 

\section*{Acknowledgements}
The authors would like to thank the University of Campinas and the Federal University of Technology -- Paraná for supporting this research through the DINTER agreement. Thanks to Prof. Dr. Rogerio Aparecido Gonçalves for lending GPUs to support large-scale SVM training.

This research did not receive any specific grant from funding agencies in the public, commercial, or not-for-profit sectors.

Declarations of interest: none.

\bibliographystyle{unsrt}  
\bibliography{refs}  %%% Remove comment to use the external .bib file (using bibtex).

\clearpage
\appendix
%\newpage
\section{KNN and SVM Results for All Datasets}
\label{app:results}

\begin{figure}[h!]
    \centering
    \subfloat[KNN]{
        \includegraphics[scale=0.37]{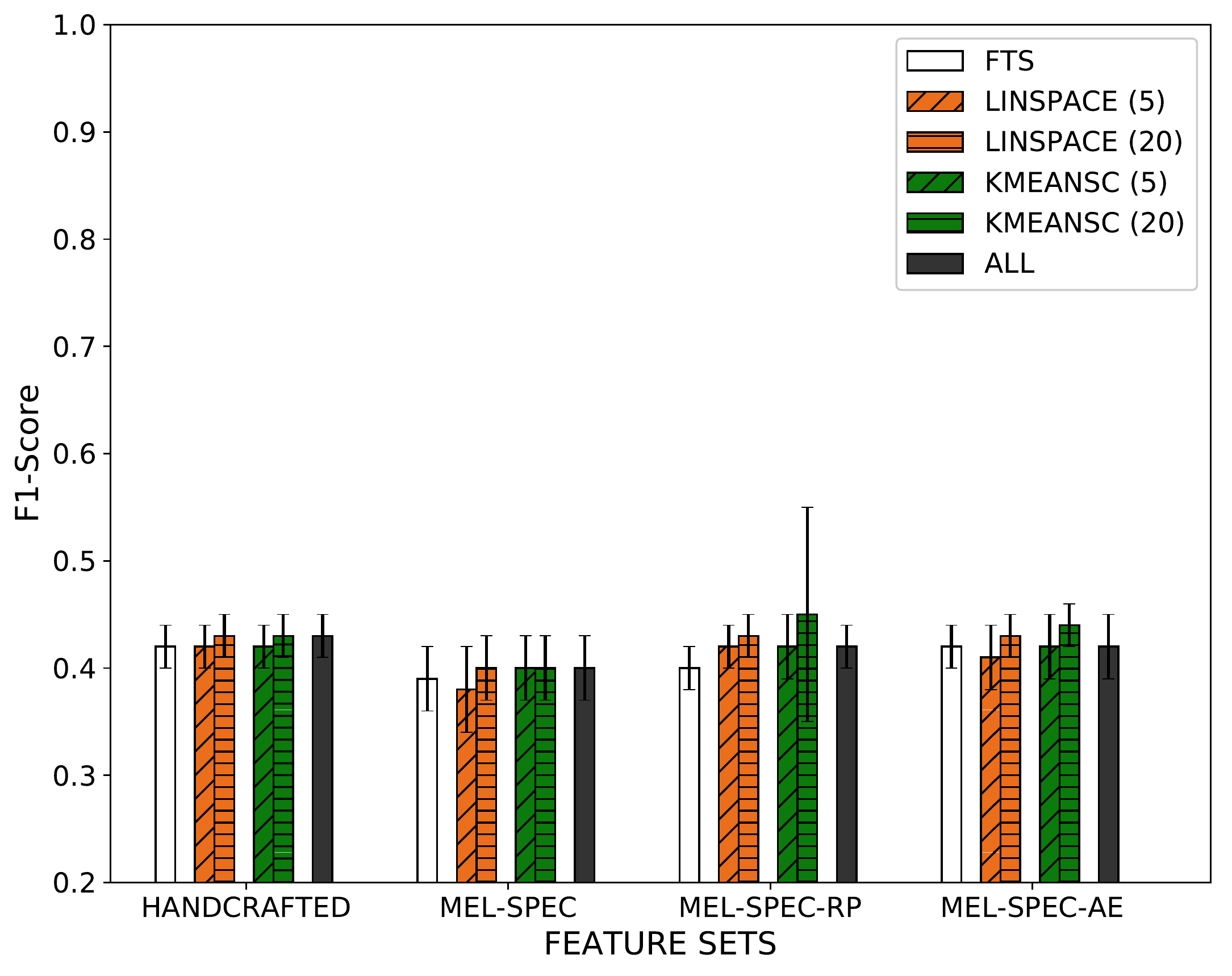}
    }
    \hspace{-10pt}
    \subfloat[KNN+ANOVA]{
        \includegraphics[scale=0.37]{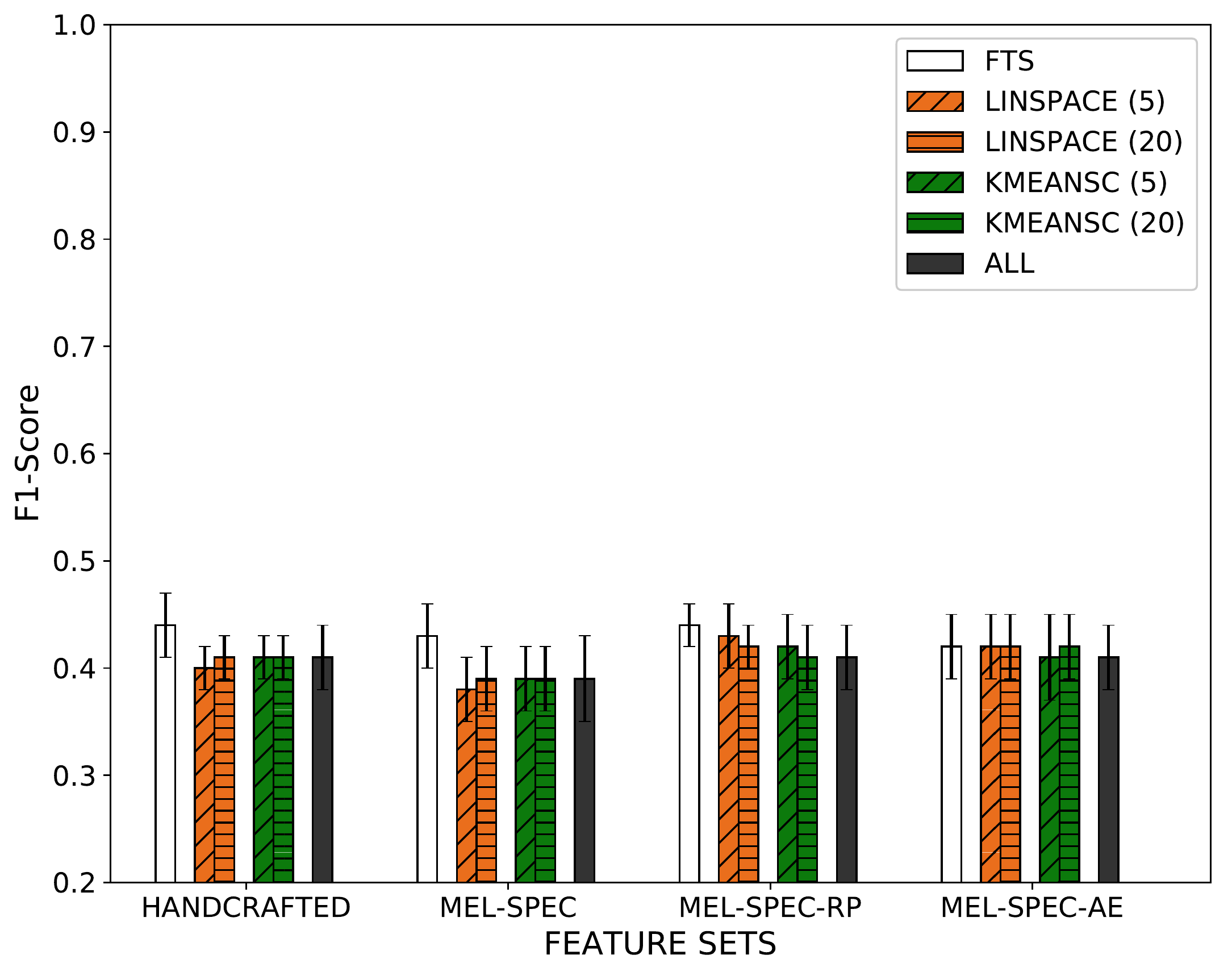}
    }
    
    \subfloat[SVM]{
        \includegraphics[scale=0.37]{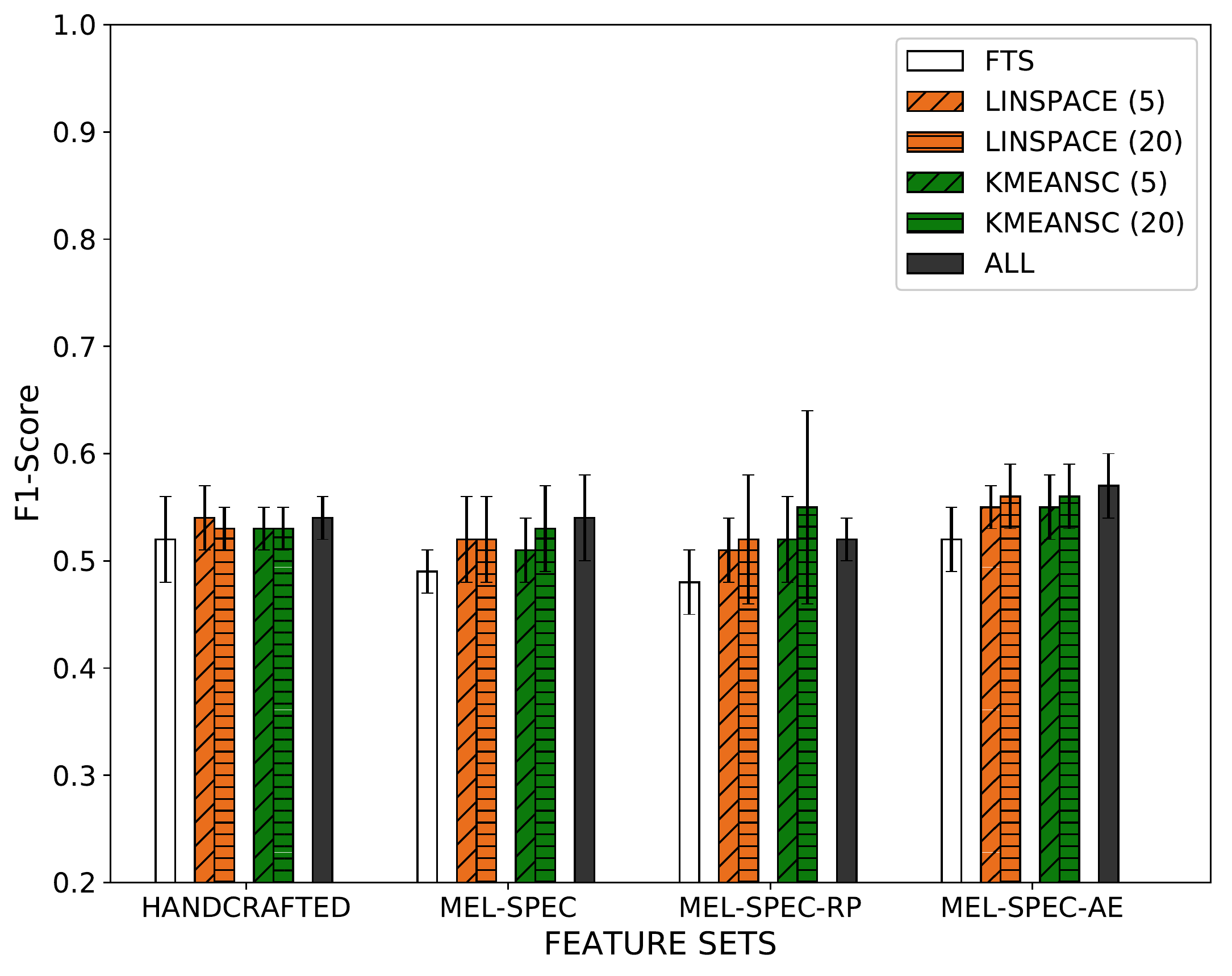}
    }
    \hspace{-10pt}
    \subfloat[SVM+ANOVA]{
        \includegraphics[scale=0.37]{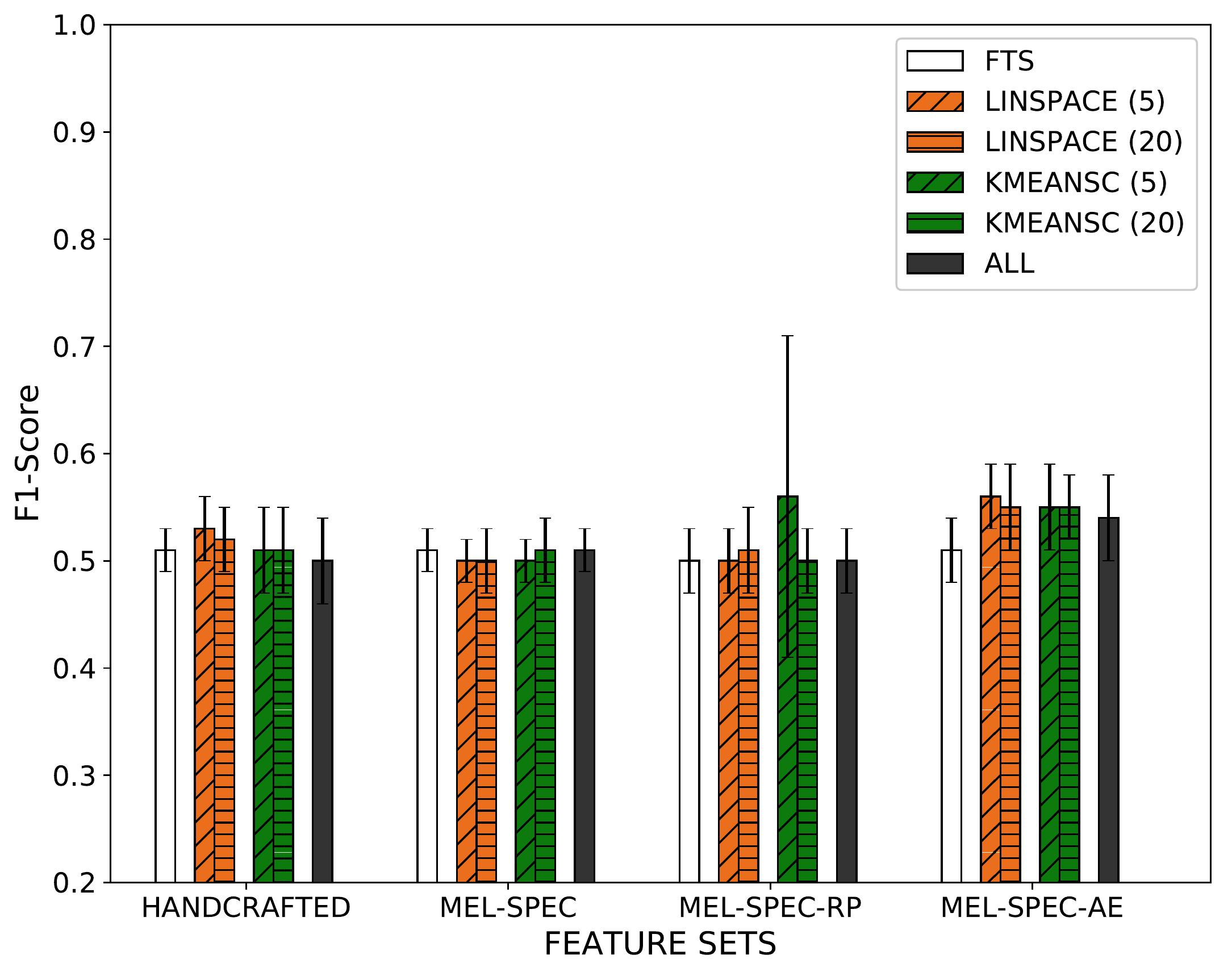}
    }
    \caption{Average F1-score across 10 RANDOM Folds for the HOMBURG Dataset} \label{fig:homburg_f1}
\end{figure}

\begin{figure}[h!]
    \centering
    \subfloat[KNN]{
        \includegraphics[scale=0.37]{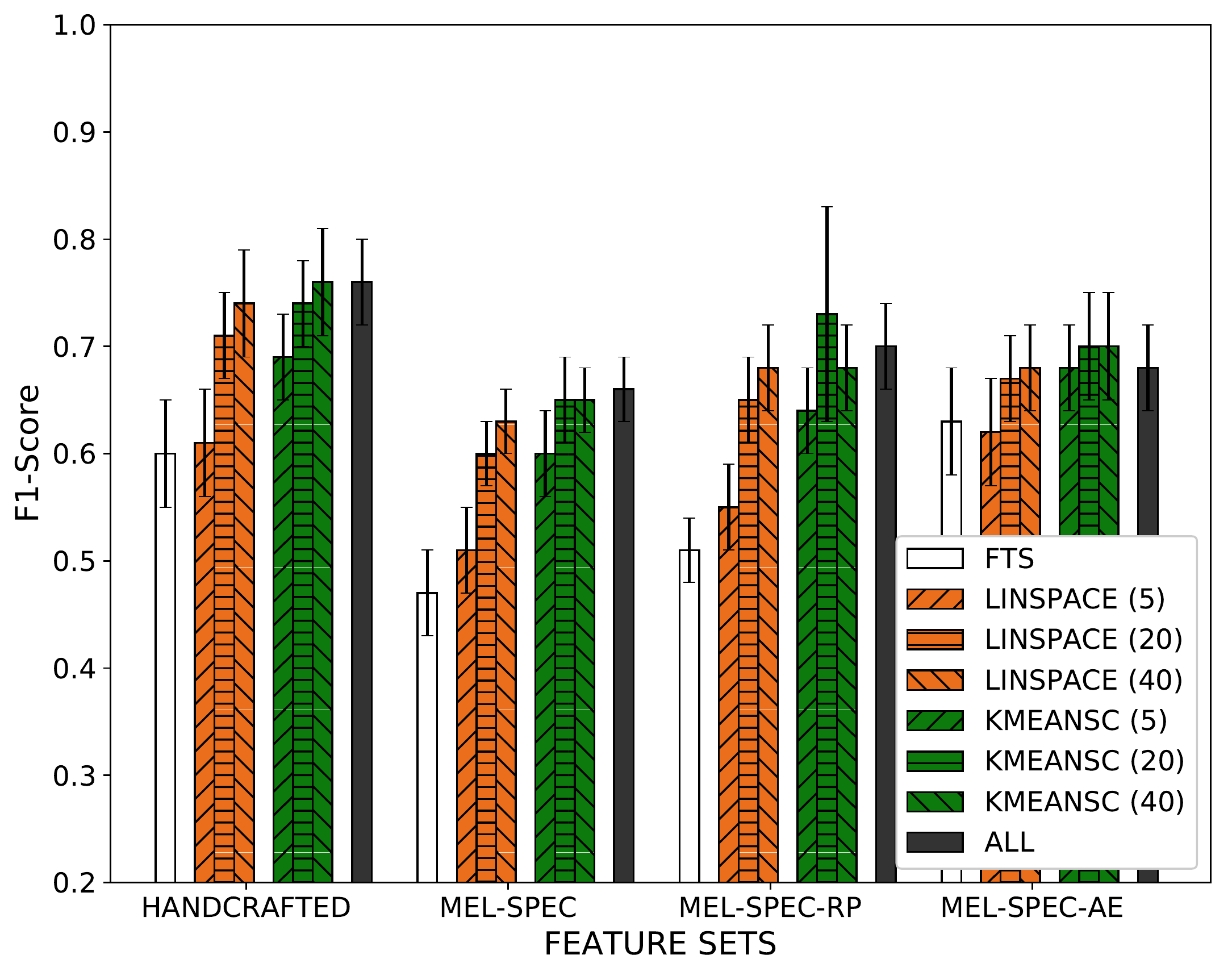}
    }
    \hspace{-10pt}
    \subfloat[KNN+ANOVA]{
        \includegraphics[scale=0.37]{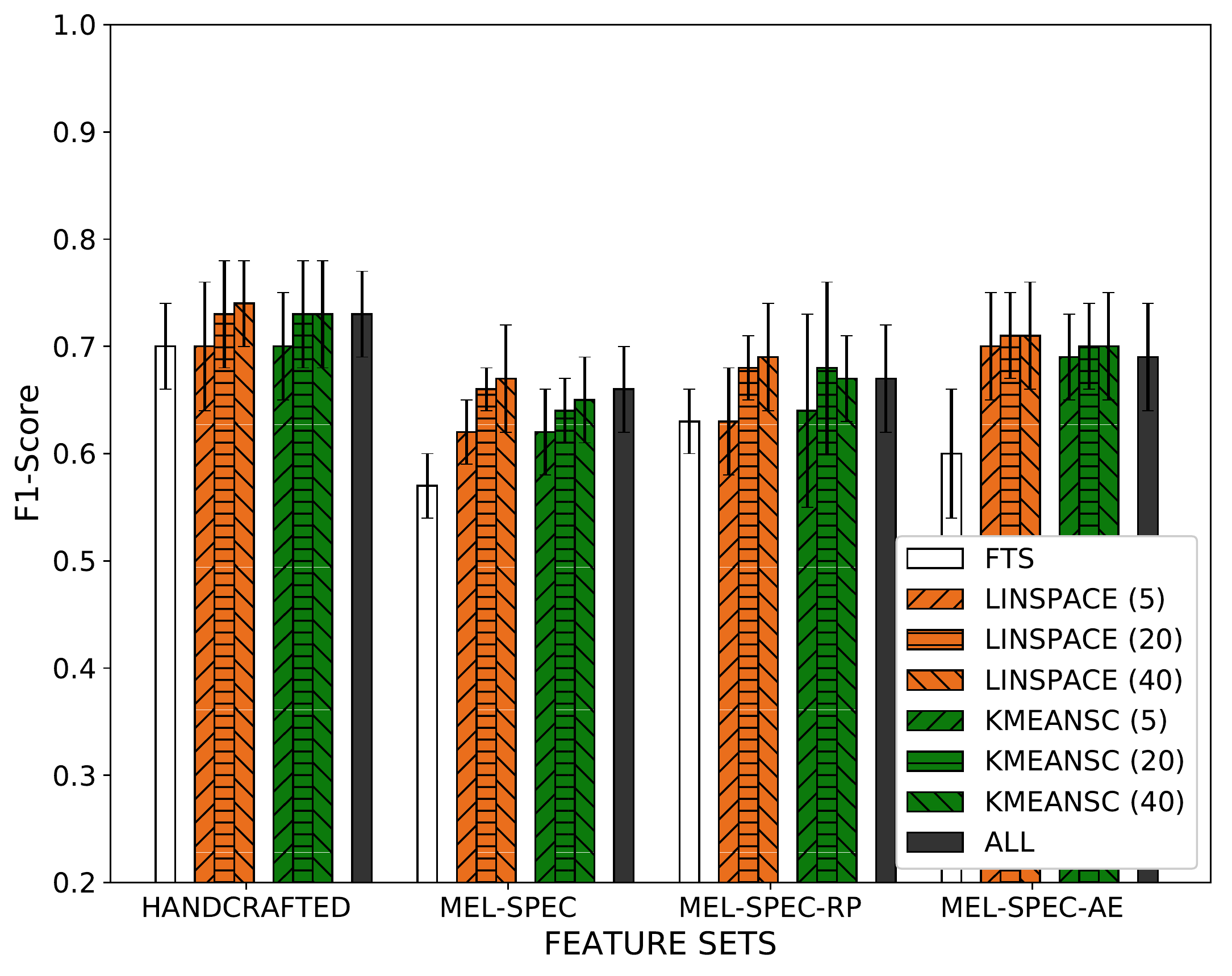}
    }
    
    %\vspace{-10pt}
    \subfloat[SVM]{
        \includegraphics[scale=0.37]{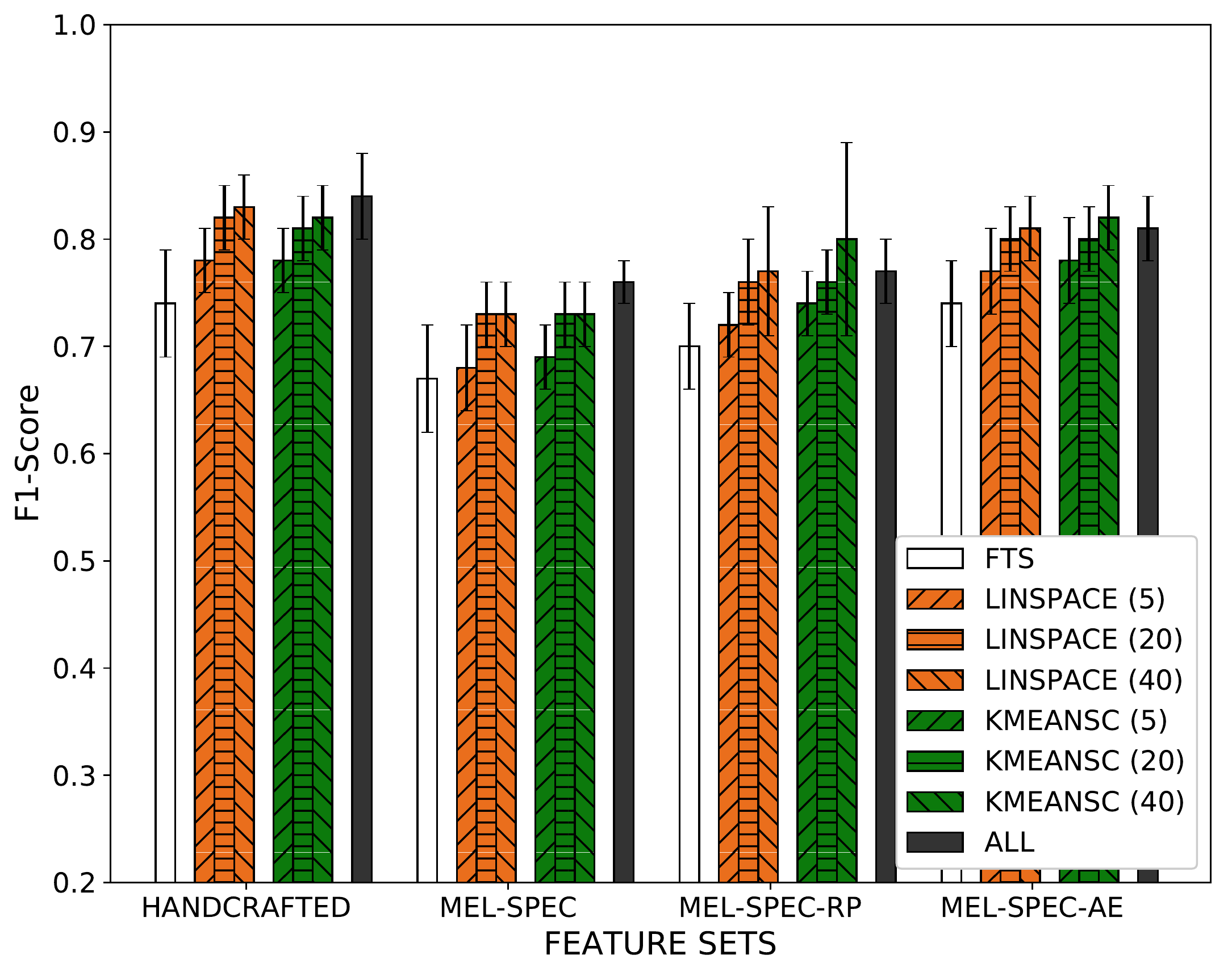}
    }
    \hspace{-10pt}
    \subfloat[SVM+ANOVA]{
        \includegraphics[scale=0.37]{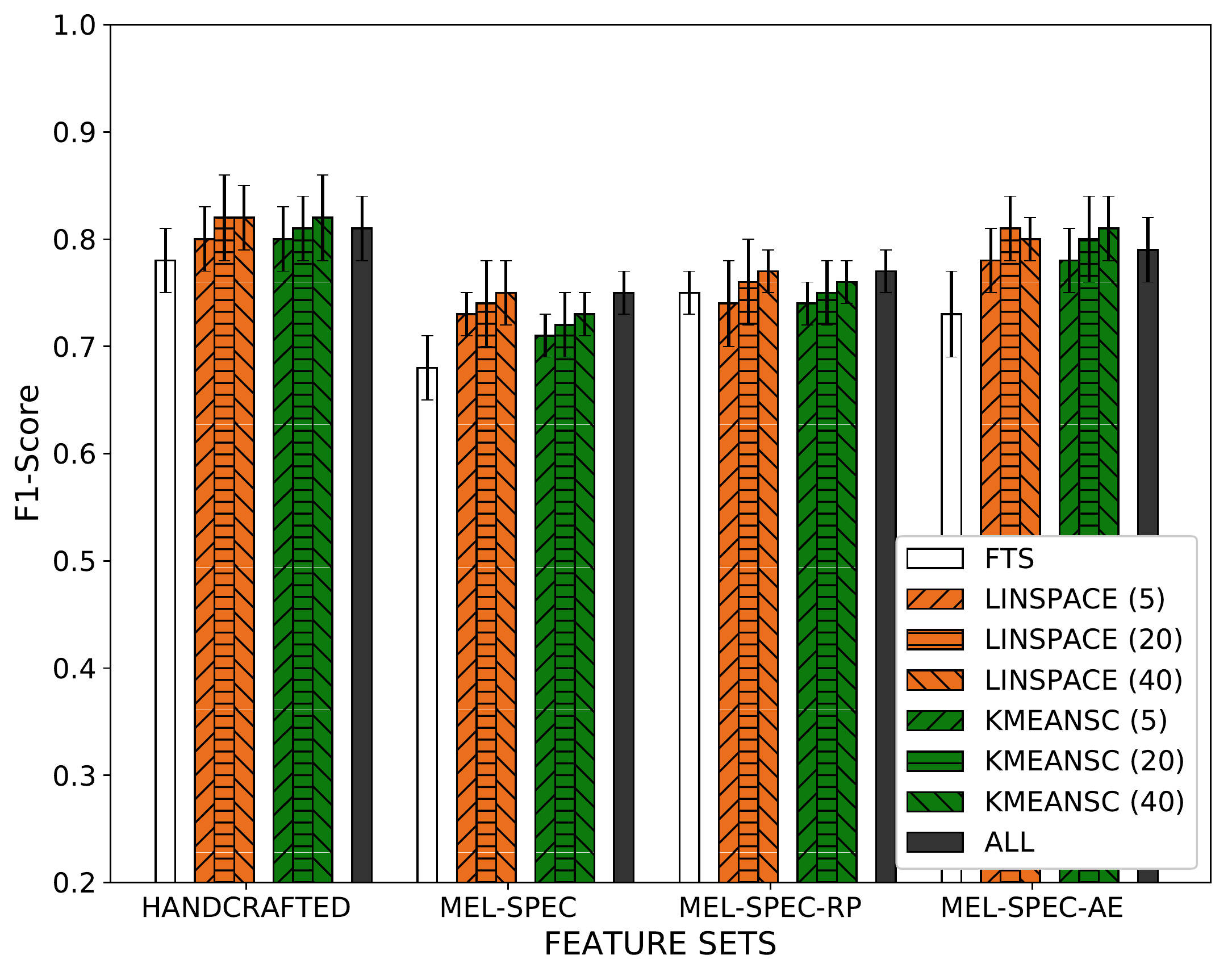}
    }
    \caption{Average F1-score across 10 RANDOM Folds for the GTZAN Dataset} \label{fig:gtzan_random_f1}
\end{figure}

\begin{figure}[h!]
    %\ContinuedFloat
    \centering
    \subfloat[KNN]{
        \includegraphics[scale=0.37]{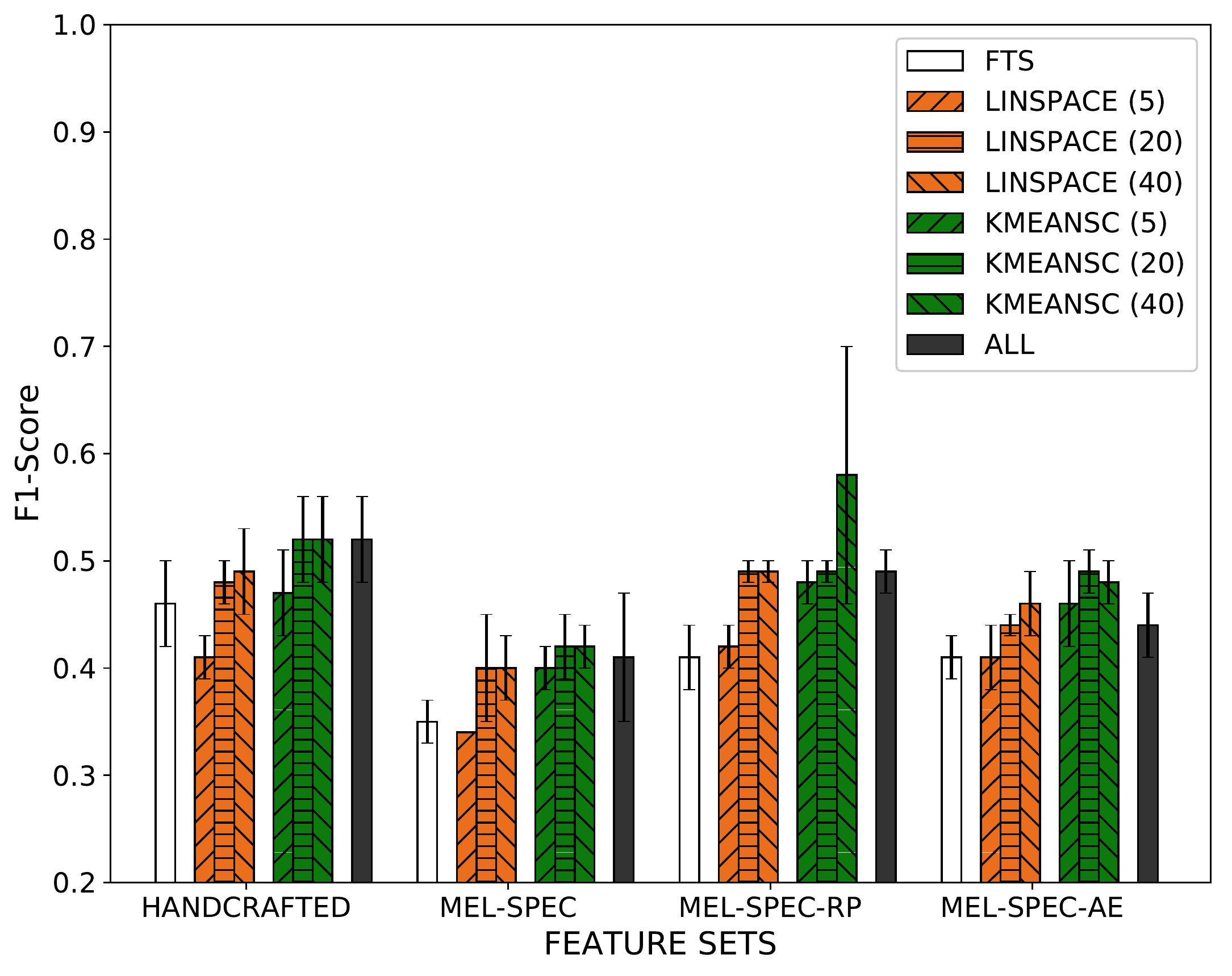}
    }
    \hspace{-10pt}
    \subfloat[KNN+ANOVA]{
        \includegraphics[scale=0.37]{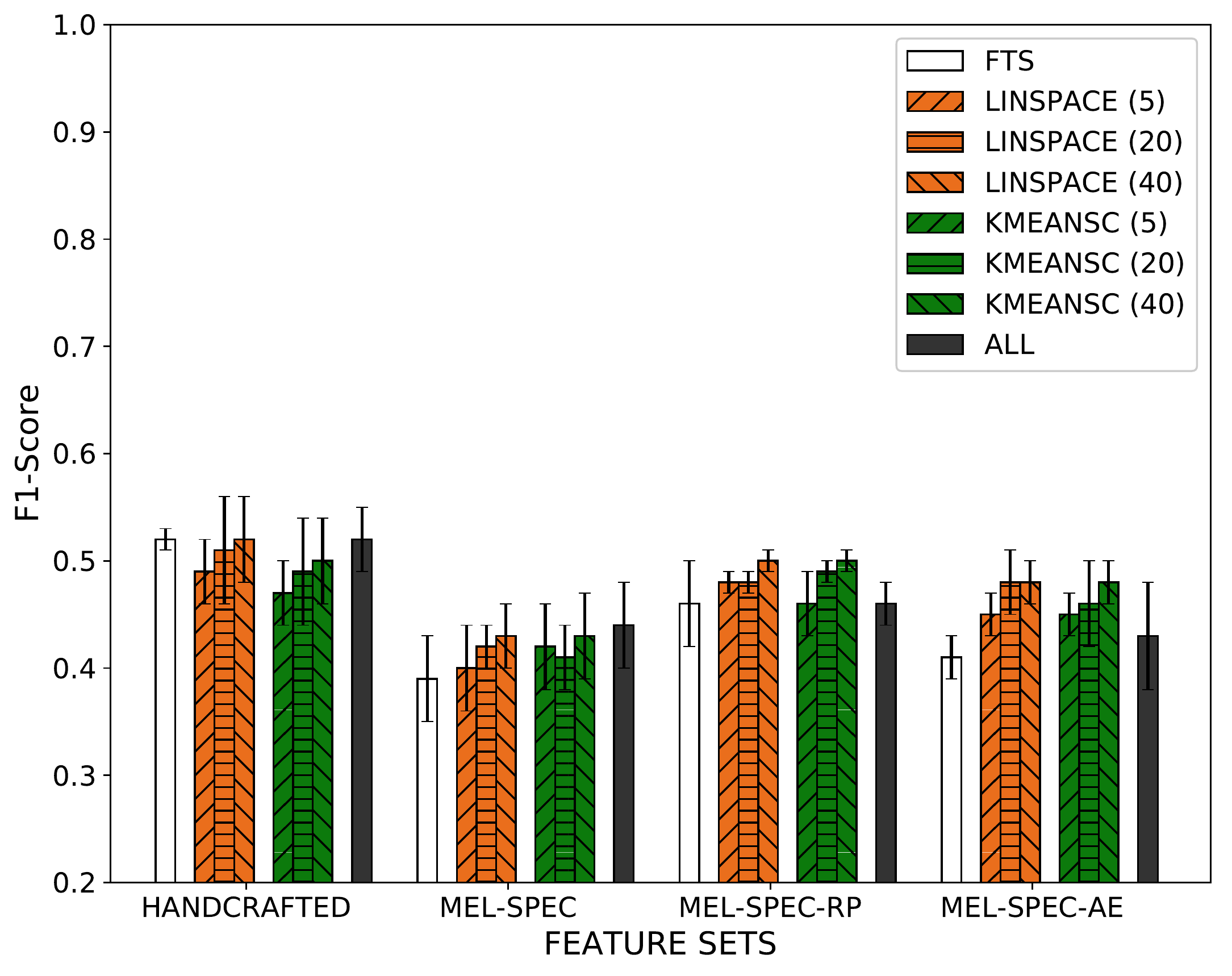}
    }
    
    \subfloat[SVM]{
        \includegraphics[scale=0.37]{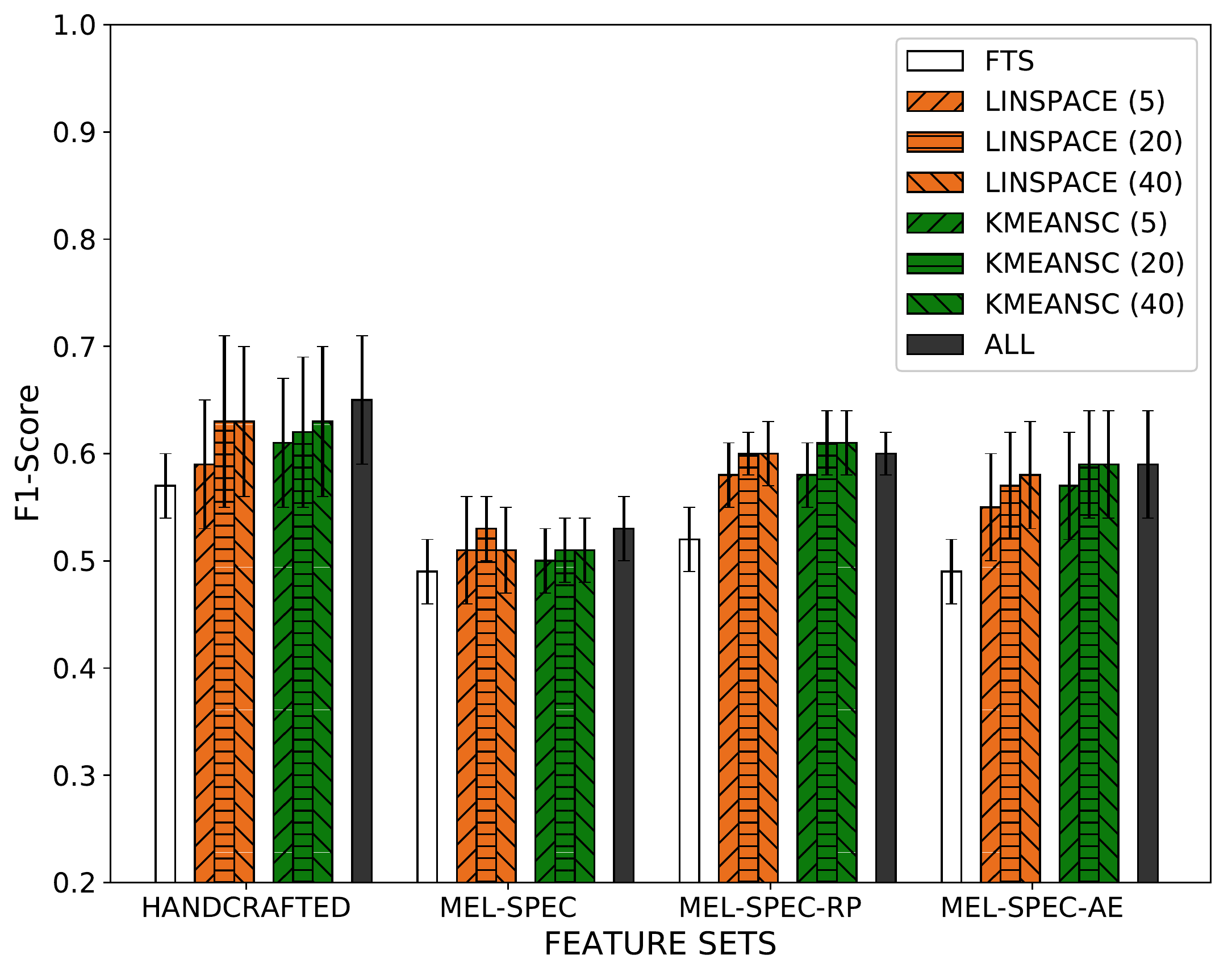}
    }
    \hspace{-10pt}
    \subfloat[SVM+ANOVA]{
        \includegraphics[scale=0.37]{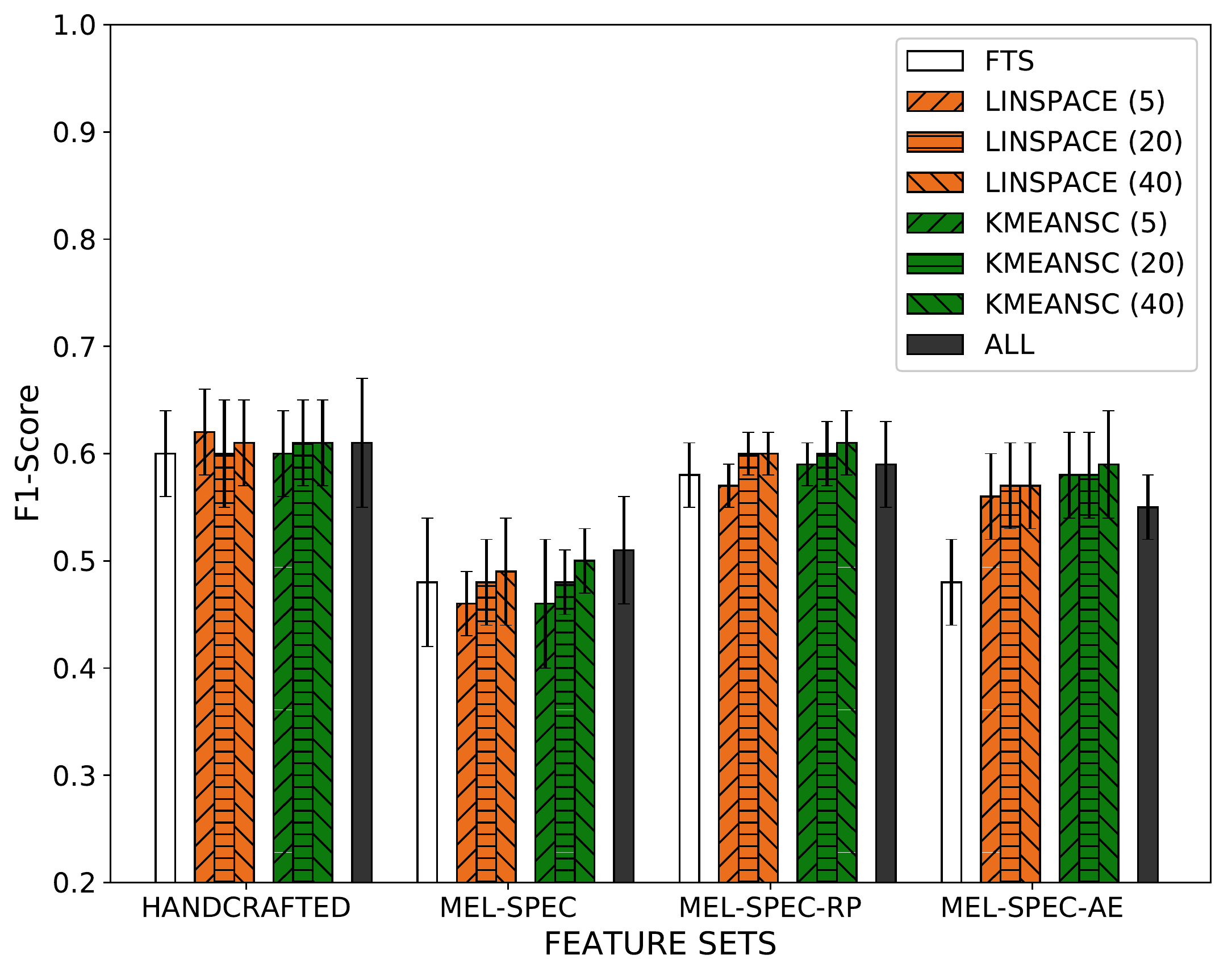}
    }
    \caption{Average F1-score across 3 ARTF Folds for the GTZAN Dataset} \label{fig:gtzan_artf_f1}
\end{figure}

\begin{figure}[h!]
    \centering
    \subfloat[KNN]{
        \includegraphics[scale=0.28]{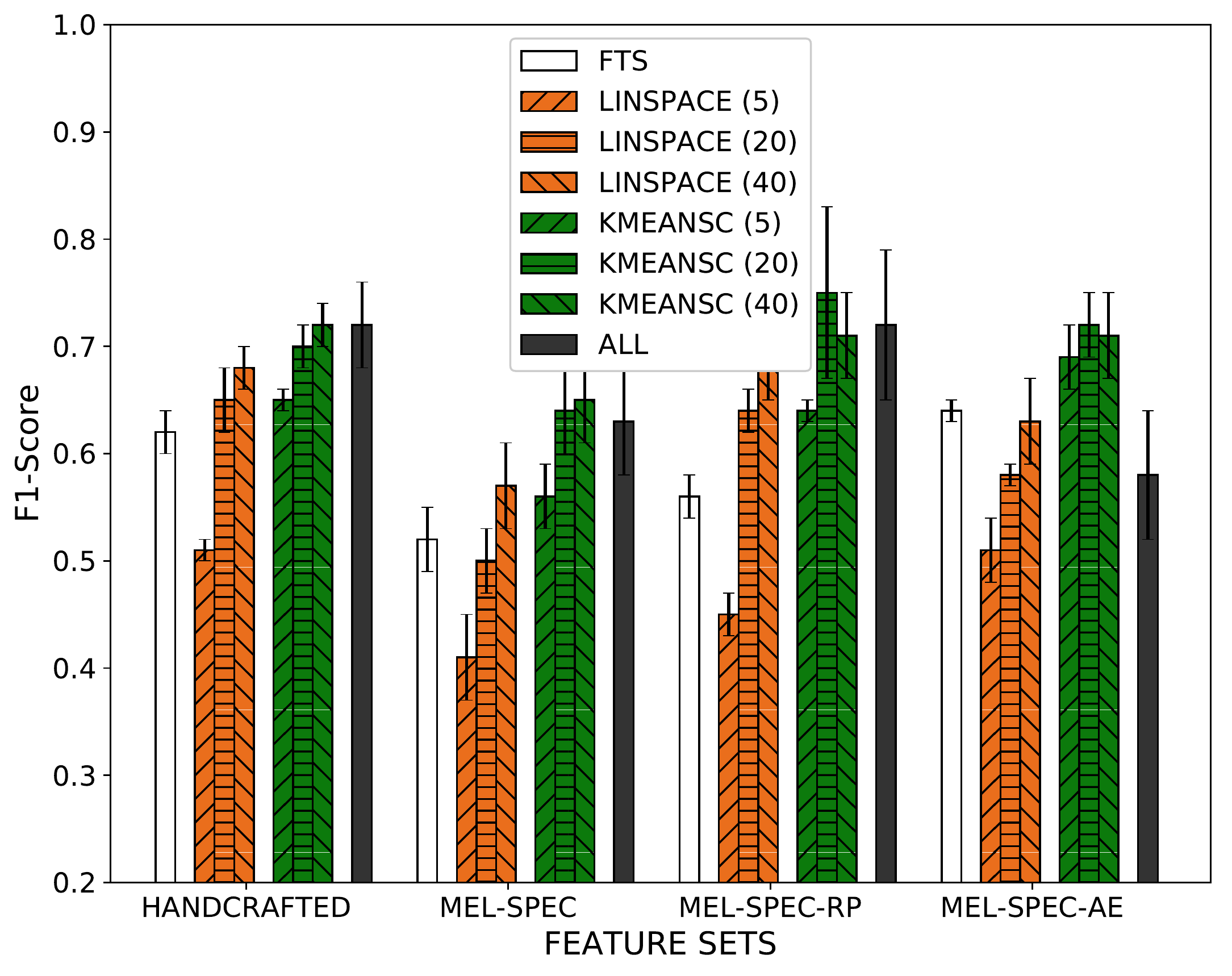}
    }
    \hspace{-10pt}
    \subfloat[KNN+ANOVA]{
        \includegraphics[scale=0.28]{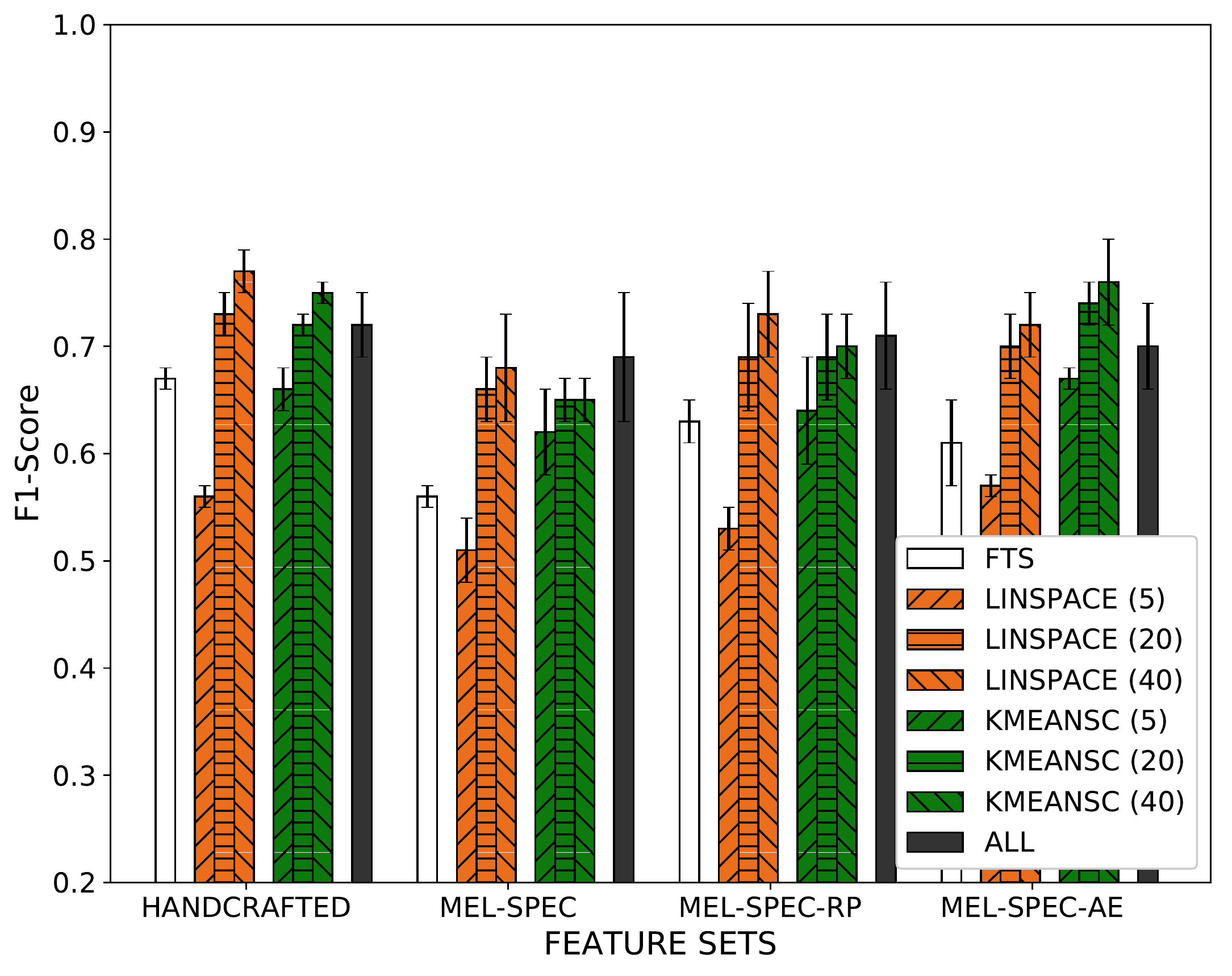}
    }
    ~\\
    \subfloat[SVM]{
        \includegraphics[scale=0.28]{figures/LMD-SVM.pdf}
    }
    \hspace{-10pt}
    \subfloat[SVM+ANOVA]{
        \includegraphics[scale=0.28]{figures/LMD-SVMANOVA.pdf}
    }
    ~\\
    \subfloat[KNN (10s)]{
        \includegraphics[scale=0.28]{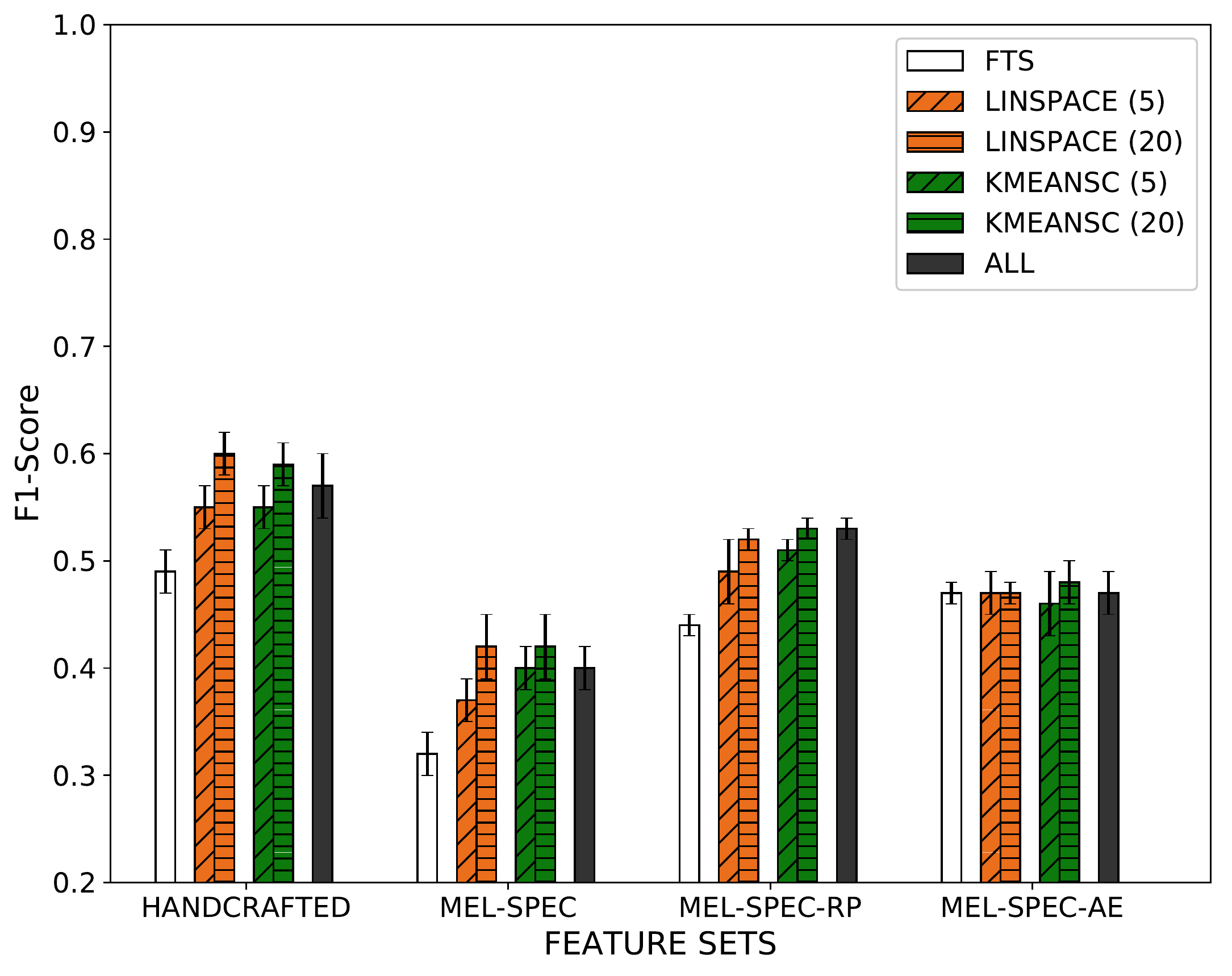}
    }
    \hspace{-10pt}
    \subfloat[KNN+ANOVA (10s)]{
        \includegraphics[scale=0.28]{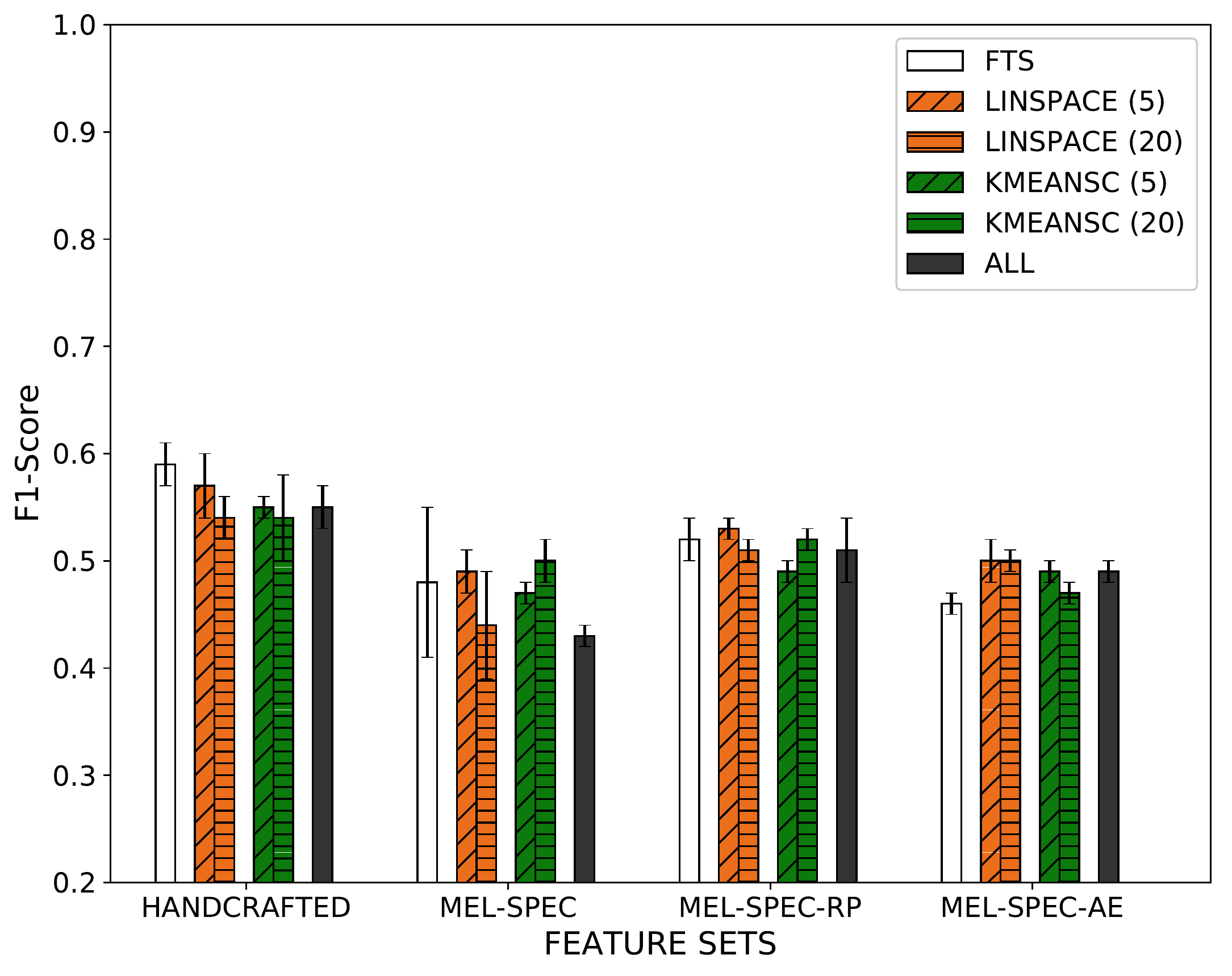}
    }
    ~\\
    \subfloat[SVM (10s)]{
        \includegraphics[scale=0.28]{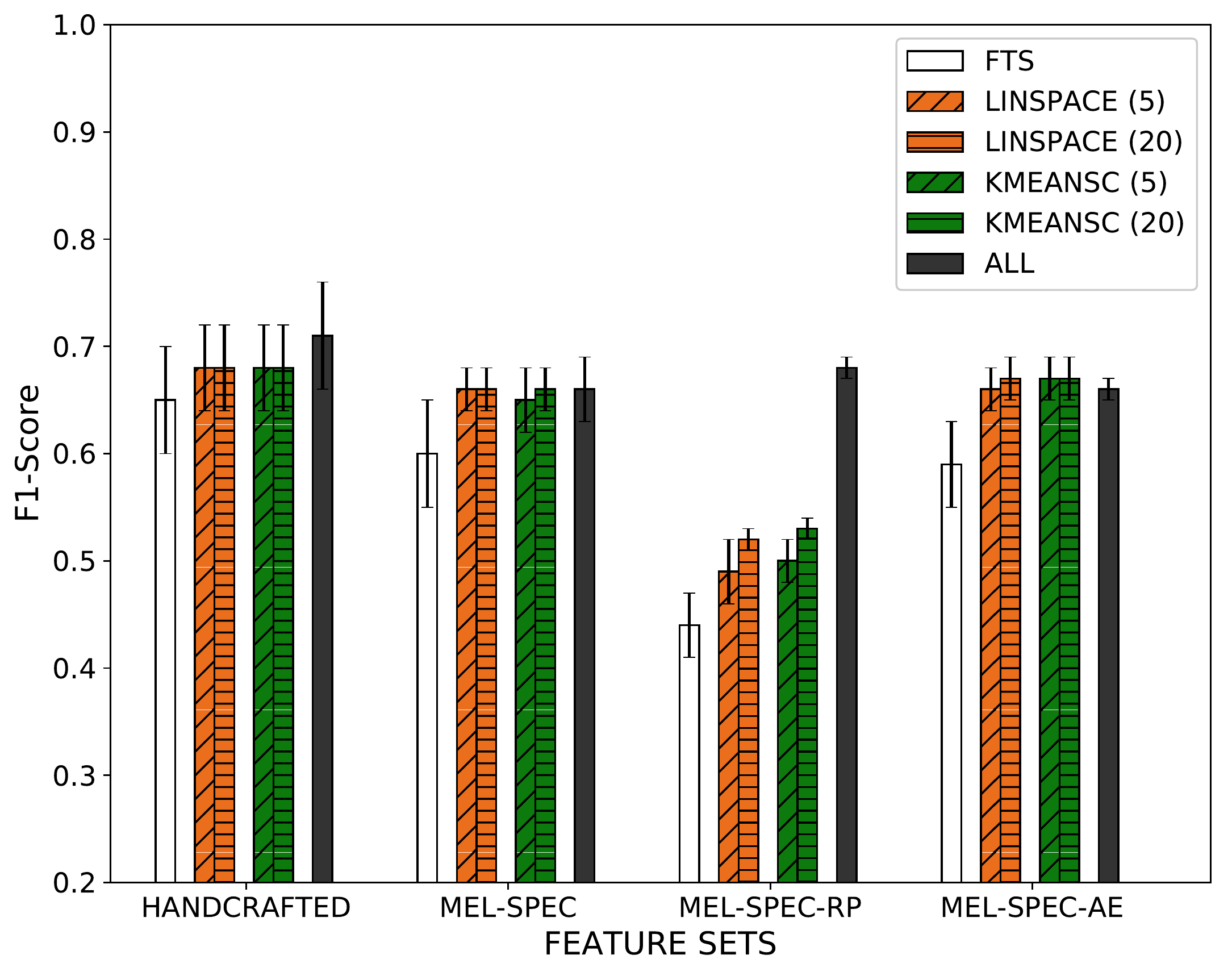}
    }
    \hspace{-10pt}
    \subfloat[SVM+ANOVA (10s)]{
        \includegraphics[scale=0.28]{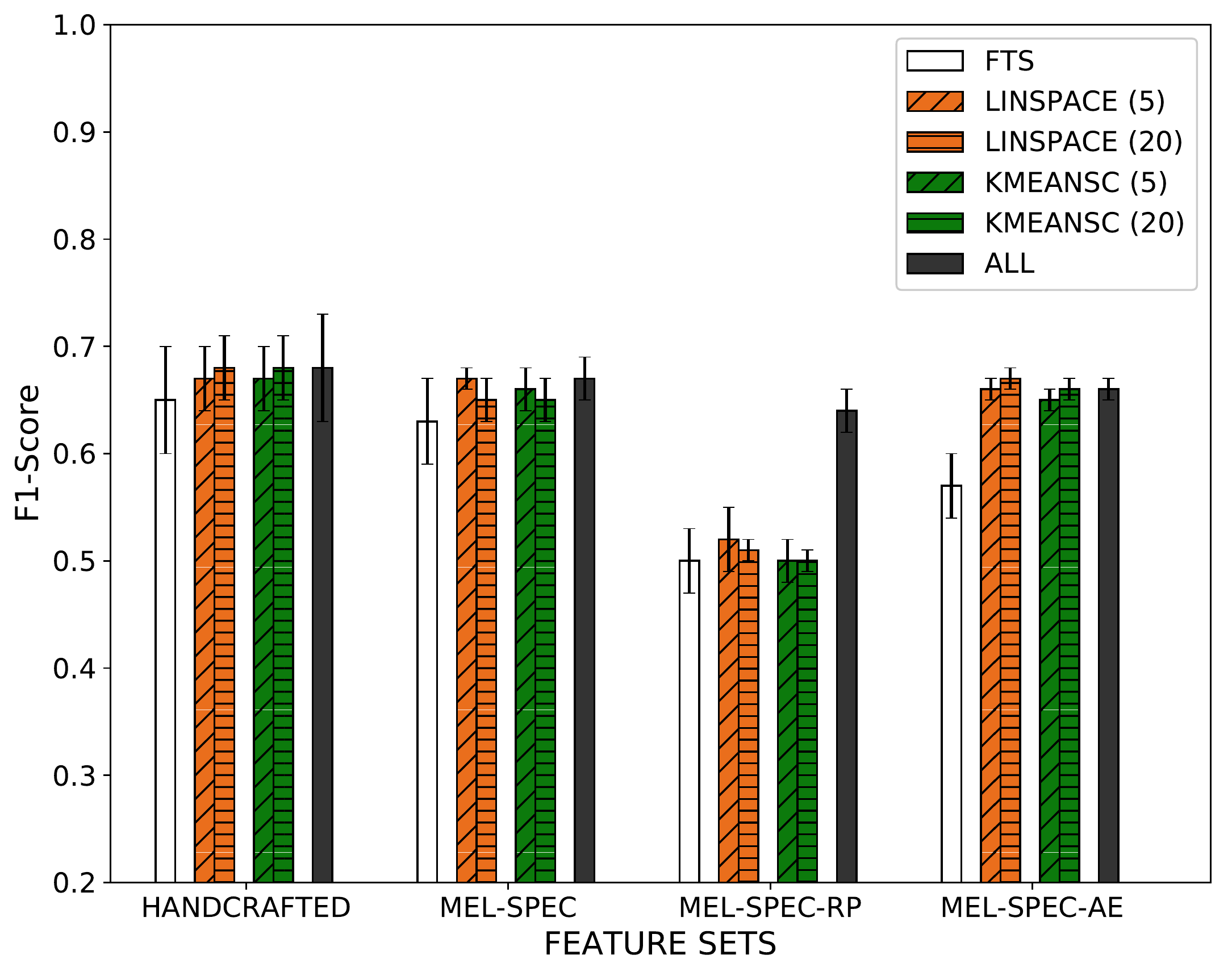}
    }
    \caption{Average F1-score across 3 Folds for the LMD Dataset} \label{fig:lmd_f1}
\end{figure}

\begin{figure}[h!]
    \centering
    \subfloat[KNN]{
        \includegraphics[scale=0.28]{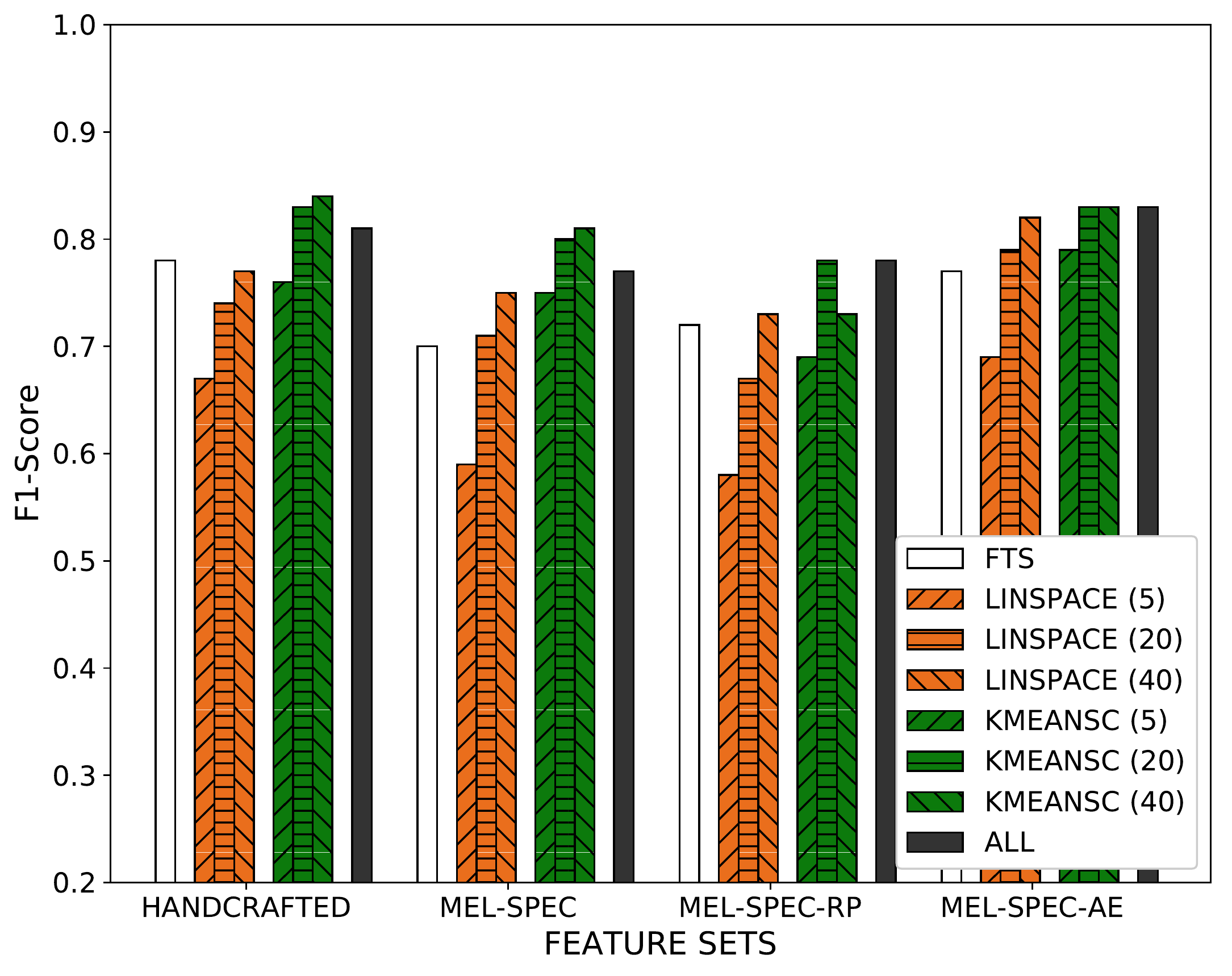}
    }
    \hspace{-10pt}
    \subfloat[KNN+ANOVA]{
        \includegraphics[scale=0.28]{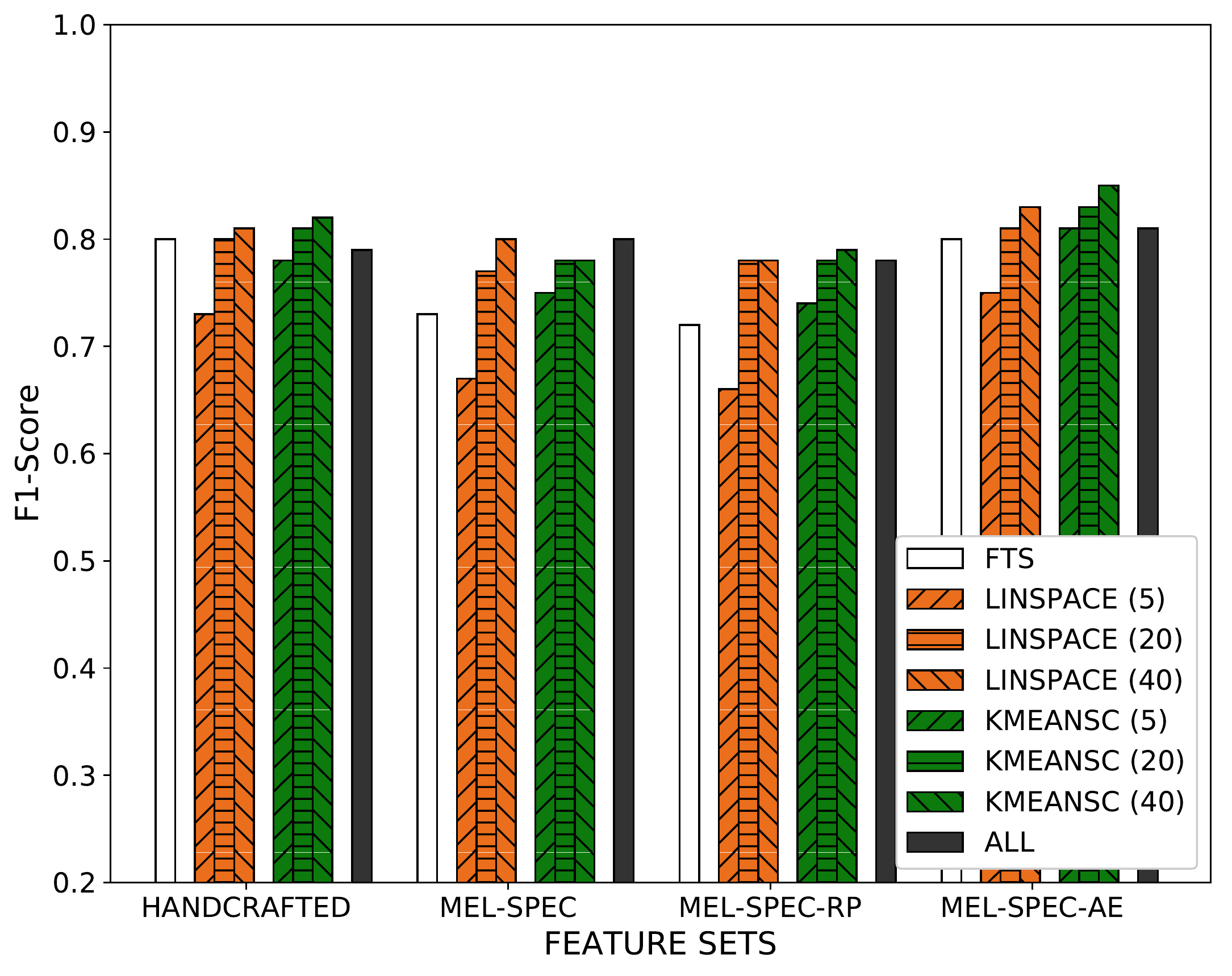}
    }
    ~\\
    \subfloat[SVM]{
        \includegraphics[scale=0.28]{figures/ISMIR-SVM.pdf}
    }
    \hspace{-10pt}
    \subfloat[SVM+ANOVA]{
        \includegraphics[scale=0.28]{figures/ISMIR-SVMANOVA.pdf}
    }
    ~\\
    \subfloat[KNN (10s)]{
        \includegraphics[scale=0.28]{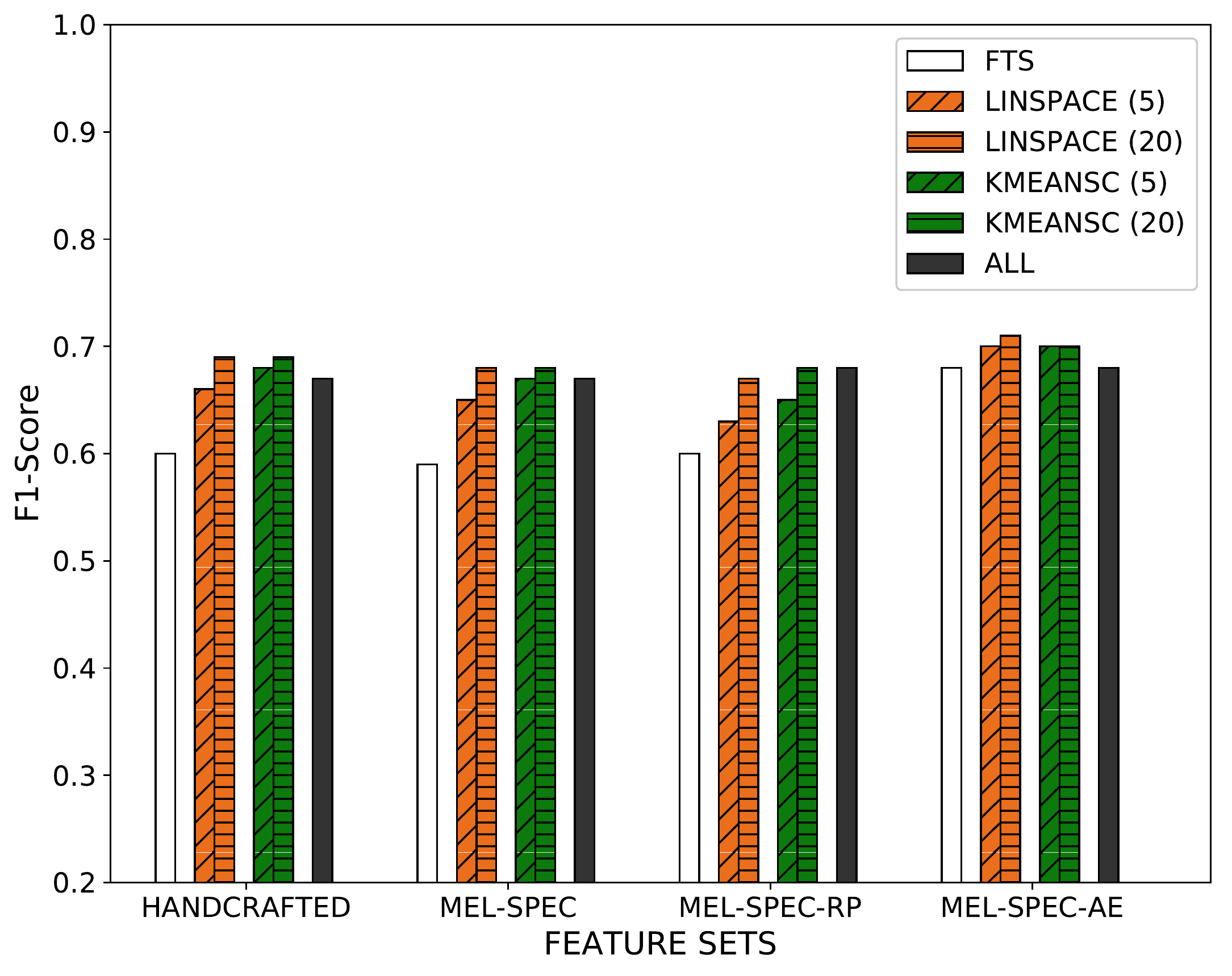}
    }
    \hspace{-10pt}
    \subfloat[KNN+ANOVA (10s)]{
        \includegraphics[scale=0.28]{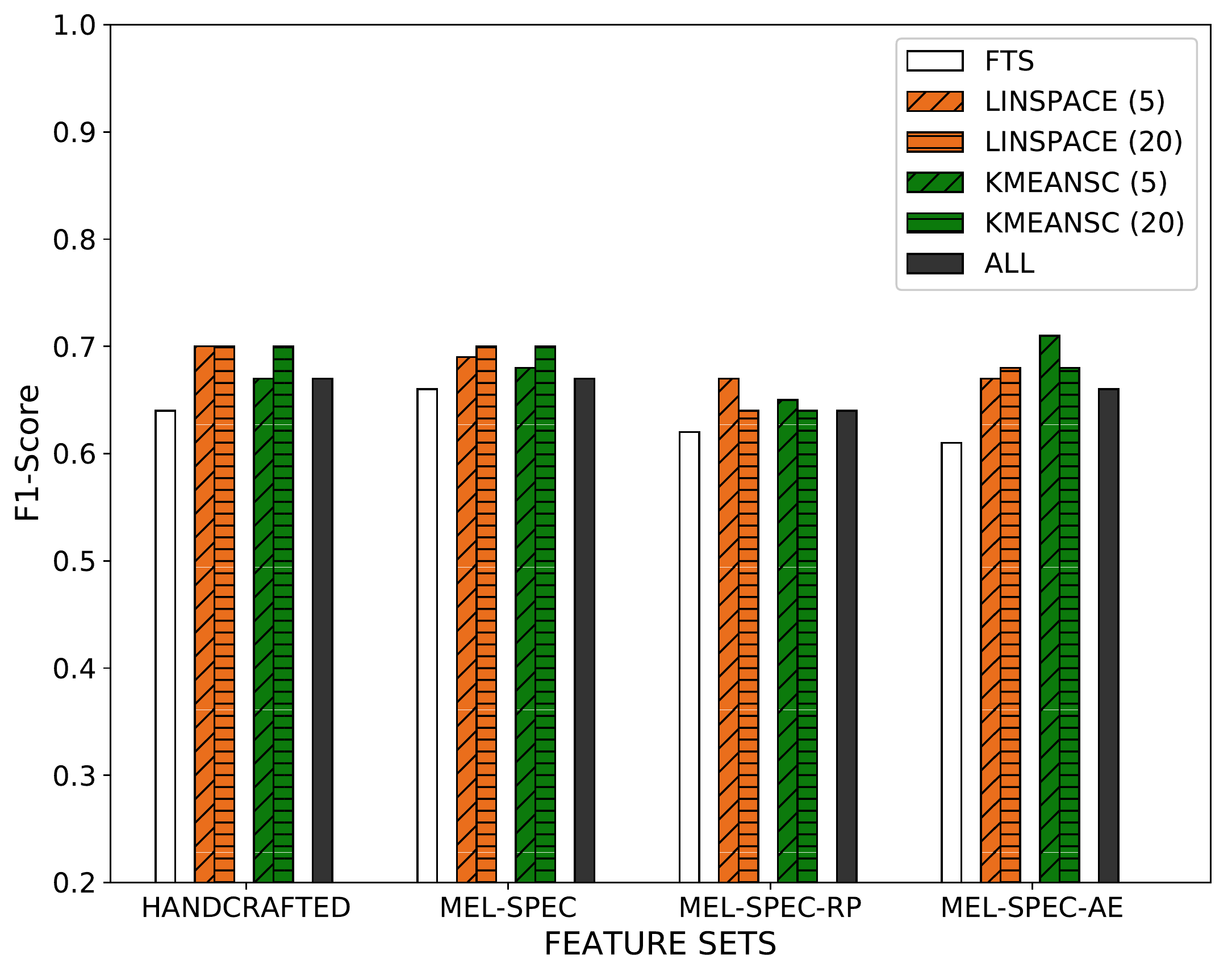}
    }
    ~\\
    \subfloat[SVM (10s)]{
        \includegraphics[scale=0.28]{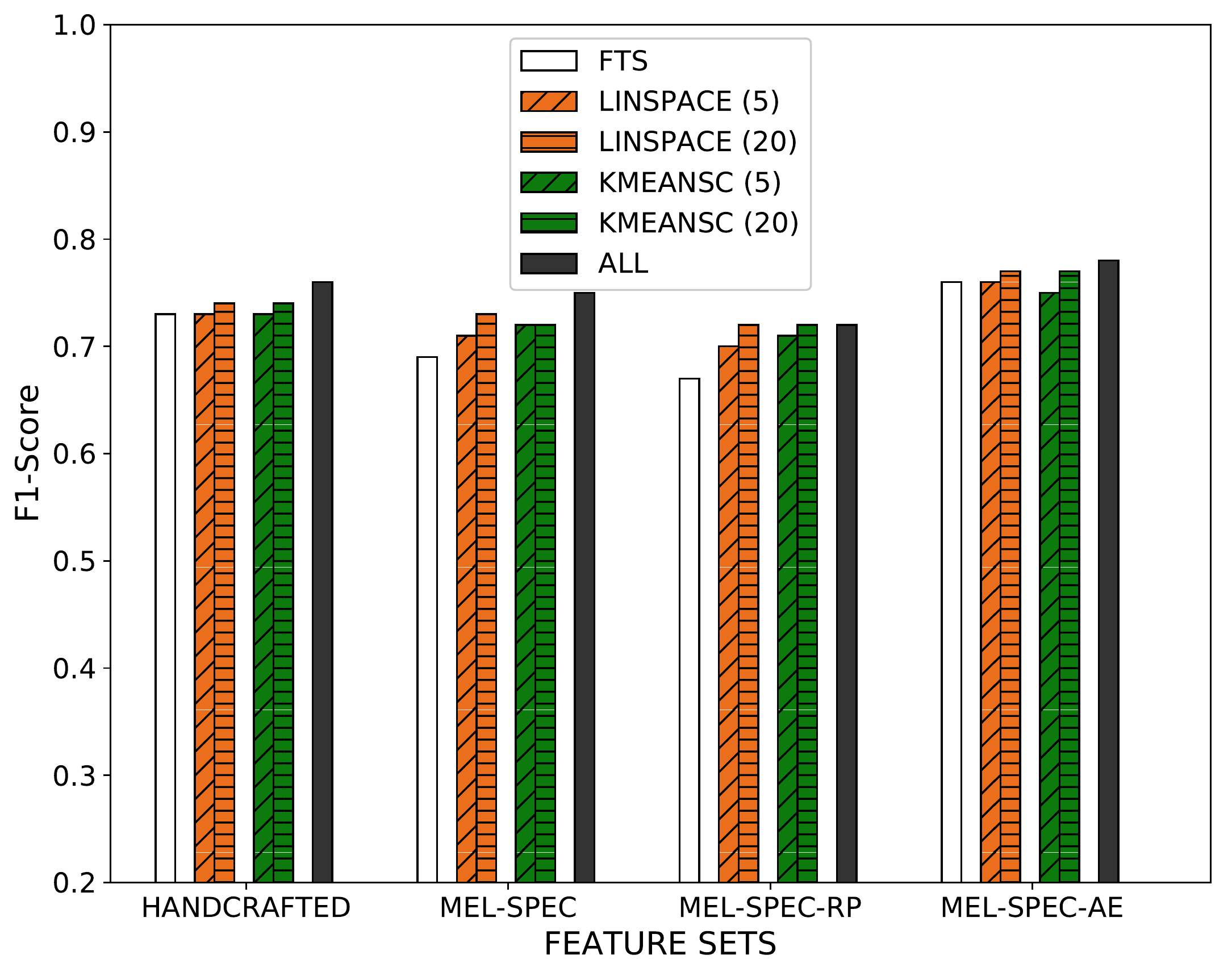}
    }
    \hspace{-10pt}
    \subfloat[SVM+ANOVA (10s)]{
        \includegraphics[scale=0.28]{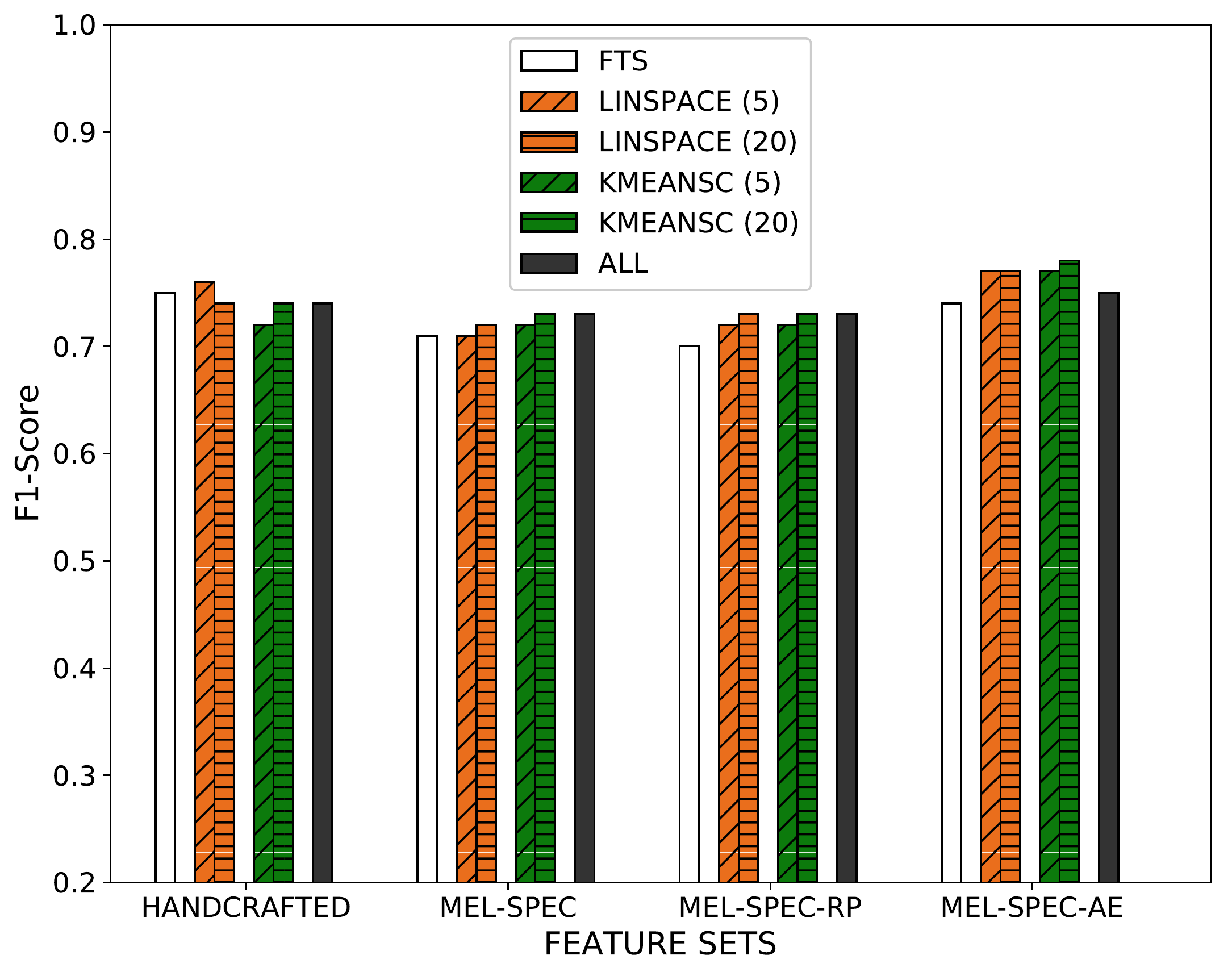}
    }
    \caption{F1-scores for the ISMIR Dataset} \label{fig:ismir_f1}
\end{figure}

\end{document}